\RequirePackage{amsmath}
\documentclass[twocolumn,twocolappendix]{aas}
\usepackage[utf8]{inputenc}
\usepackage[varg]{txfonts}
\usepackage{gensymb}
\usepackage{nameref}
\usepackage{multirow}
\usepackage{listings}
\usepackage{lipsum}
\usepackage{lmodern}
\usepackage{breqn}
\usepackage{outlines}
\usepackage{soul}
\usepackage{colortbl}
\usepackage{booktabs}
\usepackage{dashrule}
\usepackage{array}
\usepackage[table]{xcolor}
\usepackage{tabularx}

\hypersetup{linkcolor=blue,citecolor=blue,filecolor=blue,urlcolor=blue}

\newcommand{\Mfid}{\mathcal{M}_{\mathrm{fiducial}}}
\newcommand{\Mlow}{\mathcal{M}_{\mathrm{low\,SSN}}}
\newcommand{\Mpol}{\mathcal{M}_{\mathrm{polar}}}
\newcommand{\Mroc}{\mathcal{M}_{\mathrm{rocket}}}
\newcommand{\Mlowroc}{\mathcal{M}_{\mathrm{low\,SSN+rocket}}}

\begin{document}

\title{\Large The Kick Velocities of Neutron Stars in Binary Systems}
\shorttitle{The Kick Velocities of Neutron Stars in Binary Systems}

\author[0000-0002-0492-4089]{\normalsize Paul Disberg}
\affiliation{School of Physics and Astronomy, Monash University, Clayton, Victoria 3800, Australia}
\affiliation{The ARC Centre of Excellence for Gravitational Wave Discovery---OzGrav, Australia}
\email[show]{\href{mailto:paul.disberg@monash.edu}{paul.disberg@monash.edu}}

\author[0000-0002-6134-8946]{\normalsize Ilya Mandel}
\affiliation{School of Physics and Astronomy, Monash University, Clayton, Victoria 3800, Australia}
\affiliation{The ARC Centre of Excellence for Gravitational Wave Discovery---OzGrav, Australia}
\email{}

\author[0000-0002-8032-8174]{\normalsize Ryosuke Hirai}
\affiliation{School of Physics and Astronomy, Monash University, Clayton, Victoria 3800, Australia}
\affiliation{The ARC Centre of Excellence for Gravitational Wave Discovery---OzGrav, Australia}
\email{}

\shortauthors{Disberg, Mandel, \& Hirai}

\received{August 25, 2026}
\revised{October 12, 2026}
\accepted{October 14, 2026}

\begin{abstract}
\noindent Neutron stars (NSs) receive natal kicks on their formation in supernovae (SNe). We consider constraints placed on the natal kick magnitudes by NSs in different classes of binary systems. We compare observed systems to predictions from the \lstinline{COMPAS} rapid population synthesis code, where we apply kick models with varied natal kick prescriptions. Specifically, we compare binary orbits (i.e., periods and eccentricities) and systemic kick estimates of (1) \textit{Gaia} observations of NS-harboring binaries (\textit{Gaia} NSs), (2) NS low-mass X-ray binaries (LMXBs), (3) NS-white dwarf binaries (NSWDs), (4) NS high-mass X-ray binaries (HMXBs) and in particular Be X-ray binaries (BeXBs), and (5) double NSs (DNSs). In this comparison, we find that we can reproduce most of the observed properties of the \textit{Gaia} NSs, LMXBs, and NSWDs with natal kicks calibrated to the velocities of young isolated pulsars, although we need a small amount of ``rocket" kicks to explain the \textit{Gaia} NS eccentricities. The HMXBs and DNSs, in contrast, show evidence of significantly reduced NS natal kicks. In particular, we find that an apparent correlation between eccentricity and systemic kick for DNSs can be explained by Blaauw kicks, if the natal kicks are ${\lesssim}\,10\,$km\,s$^{-1}$. Although our model does not align well with low-metallicity \textit{Gaia} NSs, high-eccentricity BeXBs, and DNS mass estimates, we provide alternative hypothetical explanations for these systems. We conclude that a model in which NSs that are formed in binaries with high-mass companions receive significantly reduced natal kicks can provide a relatively consistent explanation for the observed NSs in binary systems.
\end{abstract}

\section{Introduction}
\label{sec1}
\noindent Neutron stars (NSs) are compact objects that are formed when massive stars explode in supernovae (SNe). These explosions are likely not isotropic \citep{Janka_1994,Muller_2019,Burrows_2024,Shishkin_2025,Soker_2025}, which results in the NS obtaining a natal kick velocity of several hundred of km\,s$^{-1}$ \citep{Lyne_1994,Hobbs_2005,Faucher_2006,Verbunt_2017,Igoshev_2020,Disberg_2025b}. If this happens in a binary system it affects the binary orbit and can even unbind the system \citep{Brandt_1995}. If the system remains bound, its center of mass receives a systemic kick velocity relative to its local standard of rest (LSR). Isotropic mass loss can also change the binary's orbit and deliver a systemic kick \citep{Blaauw_1961}. Therefore, post-SN binary orbits and systemic velocities hold clues to the SN mass loss and kick magnitude. Even for binaries that are relatively old, whose present-day systemic velocities have been affected by migration through the Galactic potential \citep{Disberg_2024a}, the systemic kicks can be inferred by analyzing their Galactic trajectories \citep{Atri_2019,Disberg_2024b,Disberg_2025a}.

Based on their analysis of isolated NSs, \citet{Disberg_2025b} estimate their kick velocities to have a median magnitude of ${\sim}\,250$\,km\,s$^{-1}$. However, there are reasons to expect kicks in binary systems to differ from this because binary interactions can affect stellar structure, including the carbon-oxygen (CO) core mass, which is thought to be relevant for kick magnitudes \citep[see][and references therein]{Mandel_2020}. In the case of double neutron stars, for example, \citet{Grichener_2026} suggest that the orbits can be explained by low natal kicks, and \citet{Disberg_2024b} indeed find that they have likely experienced relatively low systemic kicks \citep[see also][]{Gaspari_2024a}. In contrast, the systemic kicks of low-mass X-ray binaries containing NSs \citep{ODoherty_2023} appear to indicate larger natal kicks that are similar to the kicks of isolated NSs \citep{Disberg_2026}. The connection between binary evolution and kick magnitude is currently not well understood, and this raises the question of what kick velocities can explain observed NSs in different kinds of binary systems \citep[see also, e.g.,][]{Richards_2023,Shariat_2026}.

In order to address this question, we apply different kick models to simulated binaries from population synthesis, and compare the results to several categories of NS-harboring binaries. We use the rapid population synthesis code \lstinline{COMPAS}\footnote{\url{https://github.com/TeamCOMPAS/COMPAS}} v3.29.02 \citep{Stevenson_2017,Vigna_2018,COMPAS_2022a,COMPAS_2022b,COMPAS_2025} to estimate the pre-SN binary properties. Then, we determine the post-SN properties using a kick model in which we vary the magnitude and directions of the natal kicks. We also consider the possibility of post-SN acceleration through a ``rocket" kick \citep{Harrison_1975,Lai_2001,Hirai_2024,Baibhav_2026}. In this comparison, we do not account for different SN mechanisms, although ultra-stripped SNe (USSNe), electron-capture SNe (ECSNe), and accretion-induced collapse (AIC) are thought to yield different kick magnitudes than the standard core-collapse SNe (CCSNe). This way, we estimate the kick magnitudes while remaining agnostic about their physical causes. 

We compare the results of different kick models to observational samples of several classes of binary systems:
\begin{itemize}
    \item \textit{Gaia} binaries with an NS candidate (\textit{Gaia} NSs). In the third \textit{Gaia} data release \citep{Gaia_2016,Gaia_2023} there are $21$ (wide) binaries with a possible NS companion \citep{El-Badry_2024b}, which have constrained orbital periods, eccentricities, and systemic velocities.
    \item NS low-mass X-ray binaries (LMXBs). Based on the catalog of \citet{Arnason_2021}, \citet{ODoherty_2023} selected a sample of $19$ NS LMXBs and determined their systemic kicks through the method of \citet{Atri_2019}.
    \item NS--white dwarf systems (NSWDs). This sample consists of millisecond pulsars (MSPs) with a white dwarf (WD) companion, where in $33$ systems the companion is a helium white dwarf (He WD) and in $8$ systems it is a carbon-oxygen white dwarf (CO WD), as listed in the Australian National Telescope Facility (ATNF) Pulsar Catalog\footnote{\url{https://www.atnf.csiro.au/research/pulsar/psrcat/}} v2.7.0 \citep{Manchester_2005}.
    \item NS high-mass X-ray binaries (HMXBs). In particular, we use the Be X-ray binary (BeXB) sample of \citet{Valli_2025}, which contains $23$ systems, and compare our model to their orbital periods and eccentricities. Moreover, we use the HMXB sample of \citet{Fortin_2022b}, which contains $44$ binaries, and consider their peculiar velocity estimates.
    \item Double neutron stars (DNSs). \citet{Grichener_2026} list $29$ (potential) DNS systems \citep[see also][]{Chattaraj_2026}. They estimate the post-second-SN properties by accounting for the effects of gravitational-wave (GW) emission. Furthermore, \citet{Disberg_2024b} estimated the systemic kicks of $11$ DNSs that have constrained proper motions.
\end{itemize}
Our comparison makes use of the orbital periods and eccentricities of \textit{Gaia} NSs, BeXBs, and DNSs, since the orbits of LMXBs and NSWDs have been affected by post-SN binary evolution. We also use the systemic kick estimates for LMXBs \citep{ODoherty_2023}, HMXBs \citep{Fortin_2022b}, and DNSs \citep{Disberg_2024b}, and apply the method of \citet{Disberg_2024b,Disberg_2025a} to constrain the systemic kicks of the \textit{Gaia} NSs and NSWDs.

In Section \ref{sec2}, we derive the effects of SN kicks and mass loss on (eccentric) binary systems, mainly based on the work of \citet{Brandt_1995}, and define the kick models that we consider in this work. We compare our modeled \lstinline{COMPAS} binaries to the observed binary categories listed above: \textit{Gaia} NSs (Section \ref{sec3}), LMXBs (Section \ref{sec4}), NSWDs (Section \ref{sec5}), HMXBs (Section \ref{sec6}), and DNSs (Section \ref{sec7}). We discuss the implications and caveats of our model in Section \ref{sec8}, and summarize our conclusions in Section \ref{sec9}.

\section{Model}
\label{sec2}
\noindent In order to model the effects of supernova kicks on binary systems, we extend the analytical framework of \citet{Brandt_1995}. They derived the post-SN binary properties for binaries with no initial eccentricity, which we generalize to eccentric binaries. In part, this overlaps with the works of \citet{Hills_1983} and \citet{Pfahl_2002}, who also discussed kicks in eccentric binaries, but here we give a comprehensive derivation in which we define several quantities that are useful in our comparison to observations. In Sections \ref{sec2.1} and \ref{sec2.2}, we derive the post-SN orbital properties and systemic kicks, respectively. Then, in Section \ref{sec2.3}, we describe the effects of rocket kicks on the binary system, based on the formalism of \citet{Hirai_2024}. We define prescriptions for the NS kick magnitudes \citep[based on the model of][]{Mandel_2020} and directions in Section \ref{sec2.4}, and describe how we apply our kick model to results from \lstinline{COMPAS} in Section \ref{sec2.5}. Finally, in Section \ref{sec2.6}, we summarize our model for estimating systemic kicks \citep[based on the work of][]{Disberg_2024b,Disberg_2025a}.

\subsection{Natal Kicks}
\label{sec2.1}
\noindent We consider a binary system with an SN progenitor and a companion star with masses $M_p$ and $M_c$, respectively, in an orbit with a semi-major axis $a$ and an eccentricity $e$. After the SN, the star has a mass $M_p'=M_p-\Delta M_p$ and the orbit changes to a semi-major axis $a'$ and an eccentricity $e'$. The progenitor has a pre-SN velocity $\vec{v}$ relative to the companion when it receives an instantaneous kick with velocity $\vec{v}_k$, after which its velocity equals $\vec{v}\hspace{.5mm}'$. The distance $r$ between the objects at the moment of the SN depends on where in the orbit the SN occurs (if $e>0$), and is determined by the phase angle $\omega$ (i.e., the true anomaly) in the elliptical orbit through
\begin{equation}
    \label{eq_r}
    r=\dfrac{a(1-e^2)}{1+e\cos\omega}=\dfrac{h^2}{G(M_p+M_c)(1+e\cos\omega)}, 
\end{equation}
where $G$ is the gravitational constant and $h$ is the specific angular momentum ($h=r^2\dot{\omega}=rv_t$). The probability of the SN occurring at a phase $\omega$ equals
\begin{equation}
    \label{eq_omega}
    p(\omega)\propto\dfrac{1}{\dot{\omega}}=\dfrac{r^2}{h}\propto\left(1+e\cos\omega\right)^{-2},
\end{equation}
which allows for sampling multiple values of $\omega$ and thus $r$ to determine the posteriors of the post-SN orbital properties. 

The pre-SN velocity vector is a combination of the transverse ($v_t$) and radial ($v_r$) components relative to the companion. If $v_{\text{orb}}$ equals the Keplerian orbital velocity of the binary if it were circular (i.e., $v_{\text{orb}}=\sqrt{G(M_p+M_c)/a}$), the transverse component is described by
\begin{equation}
    \label{eq_vt}
    \left(\dfrac{v_t}{v_{\text{orb}}}\right)^2=\dfrac{r^2\dot{\omega}^2}{v_{\text{orb}}^2}=\dfrac{\left(1+e\cos\omega\right)^2}{1-e^2},
\end{equation}
whereas the radial component is given by
\begin{equation}
    \label{eq_vr}
    \left(\dfrac{v_r}{v_{\text{orb}}}\right)^2=\left(\dfrac{dr}{d\omega}\dfrac{\dot{\omega}}{v_{\text{orb}}}\right)^2=\dfrac{e^2\sin^2\omega}{1-e^2}.
\end{equation}
This means that the magnitude of the total pre-SN velocity can be determined through
\begin{equation}
    \label{eq_v}
    \left(\dfrac{v}{v_{\text{orb}}}\right)^2=\dfrac{v_t^2+v_r^2}{v_{\text{orb}}^2}=\dfrac{1+2e\cos\omega+e^2}{1-e^2},
\end{equation}
where $v=v_{\text{orb}}$ for $e=0$. We choose axes such that the pre-SN orbit is in the $x,y$-plane with (1) the stars located on the $y$-axis, (2) the companion at the origin, and (3) the SN progenitor having a positive velocity in the $x$-direction. The angle $\zeta$ between $\vec{v}$ and the $x$-axis is given by
\begin{equation}
    \label{eq_zeta}
    \tan\zeta=\dfrac{v_r}{v_t}=\dfrac{e\sin\omega}{1+e\cos\omega},
\end{equation}
such that $\vec{v}=\left(\vec{\hat{x}}\cos\zeta+\vec{\hat{y}}\sin\zeta\right)v$. When an instantaneous kick velocity $\vec{v}_k$ is added to this vector, which makes an angle $\phi$ in the pre-SN orbital plane and an angle $\theta$ with the positive $z$-axis, the post-SN velocity $\vec{v}\hspace{.5mm}'$ becomes
\begin{equation}
    \label{eq_v_prime_vec}
    \vec{v}\hspace{.5mm}'=\begin{pmatrix}v\cos\zeta+v_k\cos\phi\sin\theta\\v\sin\zeta+v_k\sin\phi\sin\theta\\v_k\cos\theta\end{pmatrix},
\end{equation}
which has a magnitude of
\begin{equation}
    \label{eq_v_prime}
    \left(\dfrac{v'}{v}\right)^2=1+2\tilde{v}\cos(\phi-\zeta)\sin\theta+\tilde{v}^2,
\end{equation}
where we follow \citet{Brandt_1995} in defining $\tilde{v}=v_k/v$, although we generalize to eccentric pre-SN orbits with $v\neq v_{\text{orb}}$.

The energy of the post-SN binary can be expressed as
\begin{dmath}
    \label{eq_energy_prime}
    E'=\dfrac{\mu'}{2}v'{^2}-\dfrac{GM_p'M_c}{r}=-\dfrac{GM'_pM_c}{2a}\left[\dfrac{2a}{r}-\dfrac{M_p+M_c}{M_p'+M_c}\left(\dfrac{v'}{v_{\text{orb}}}\right)^2\right],
\end{dmath}
where $\mu^{(\prime)}=M_p^{(\prime)}M_c/(M_p^{(\prime)}+M_c)$ is the pre(post)-SN reduced mass \citep[cf.\ equation 2.4 of][]{Brandt_1995}. The binary remains bound if $E'<0$, meaning that in the absence of a natal kick, a circular binary can lose up to 50\% of its mass without being disrupted \citep[e.g.,][]{Nelemans_1999}. Since the post-SN orbit (with a semi-major axis $a'$) is again Keplerian, its energy can also be expressed as $E'=-GM_p'M_c/(2a')$, which means that the change in semi-major axis is described by
\begin{equation}
    \label{eq_a_prime}
    \dfrac{a'}{a}=\left[\dfrac{2a}{r}-\dfrac{M_p+M_c}{M_p'+M_c}\left(\dfrac{v'}{v_{\text{orb}}}\right)^2\right]^{-1}.
\end{equation}
This equation shows that with no natal kick the binary semi-major axis can still change if there is mass-loss (i.e., $M_p'\neq M_p$) due to a recoil-induced Blaauw kick \citep{Blaauw_1961}. The post-SN orbital period follows from combining Equation \ref{eq_a_prime} with Kepler's third law. The post-SN specific angular momentum vector $\vec{h}\hspace{.2mm}'$, in turn, depends on $\vec{r}=r\vec{\hat{y}}$ and $\vec{v}\hspace{.3mm}'_t=\vec{v}\hspace{.3mm}'_x+\vec{v}\hspace{.3mm}'_z$ and is given by
\begin{equation}
    \label{eq_h_prime_vec}
    \vec{h}\hspace{.2mm}'={\vec{r}\times \vec{v}\hspace{.3mm}'_t=rv\begin{pmatrix}\tilde{v}\cos\theta\\0\\-\cos\zeta-\tilde{v}\cos\phi\sin\theta\end{pmatrix}}.
\end{equation}
In combination with Equation \ref{eq_r}, this allows for determining the post-SN eccentricity through
\begin{equation}
    \label{eq_e_prime}
    1-e'{^2}={\dfrac{h'{^2}}{a'G(M'_p+M_c)}=\left(\dfrac{h'}{a'v'_{\text{orb}}}\right)^2},
\end{equation}
where $h'=|\vec{h}\hspace{.2mm}'|$ and $v'_{\text{orb}}=\sqrt{G(M_p'+M_c)/a'}$ \citep[cf.\ equation B7 of][]{Pfahl_2002}. The inclination $\iota$ relative to the pre-SN orbit equals the angle between -$\vec{h}\hspace{.2mm}'$ and $\vec{\hat{z}}$, since $\vec{h}$ is aligned with the negative $z$-axis, which gives
\begin{equation}
    \label{eq_iota}
    \cos\iota=\dfrac{\text{-}\vec{h}\hspace{.2mm}'\cdot\vec{\hat{z}}}{h'}=\dfrac{\cos\zeta+\tilde{v}\cos\phi\sin\theta}{\left[\left(\cos\zeta+\tilde{v}\cos\phi\sin\theta\right)^2+\tilde{v}^2\cos^2\theta\right]^{1/2}},
\end{equation}
where $\iota=0$ if $\tilde{v}=0$ because the Blaauw kick is restricted to the orbital plane.

\subsection{Systemic Kicks}
\label{sec2.2}
\noindent The natal and Blaauw kicks do not only change the orbital parameters (i.e., the period and eccentricity), but they also cause the center of mass (CM) of the binary to obtain a systemic kick. In the CM frame the pre-SN velocities for the supernova progenitor and the companion are $\vec{v}_p=\vec{v}\mu/M_p$ and $\vec{v}_c=-\vec{v}\mu/M_c$, respectively, assuming no pre-SN systemic velocity, $v_{\text{sys},0}=(M_p\vec{v}_p+M_c\vec{v}_c)/(M_p+M_c)=0$. The systemic kick imparted on the binary, then, can be described by $\Delta\vec{v}_{\text{sys}}=(M_p'(\vec{v}_p+\vec{v}_k)+M_c \vec{v}_c)/(M_p'+M_c)$. Combining this with $\vec{v}$ and $\vec{v}_k$ as described in Equation \ref{eq_v_prime_vec} gives
\begin{equation}
    \label{eq_dv_sys_vec}
    \Delta\vec{v}_{\text{sys}}=\dfrac{\mu'}{M_c}\begin{pmatrix}v_k\cos\phi\sin\theta-\dfrac{\mu\Delta M_p}{M_pM_p'}v\cos\zeta\\v_k\sin\phi\sin\theta-\dfrac{\mu\Delta M_p}{M_pM_p'}v\sin\zeta\\v_k\cos\theta\end{pmatrix},
\end{equation}
which has a magnitude of
\begin{dmath}
    \label{eq_dv_sys}
    \left(\dfrac{\Delta v_{\text{sys}}}{v}\right)^2=\left(\dfrac{\mu'}{M_c}\right)^2\left[\left(\dfrac{\mu\Delta M_p}{M_pM_p'}\right)^2-2\dfrac{\mu\Delta M_p}{M_pM_p'}\tilde{v}\cos\left(\phi-\zeta\right)\sin\theta+\tilde{v}^2\right].
\end{dmath}
We define $\Delta\lambda$ as the angle between $\Delta\vec{v}_{\text{sys}}$ and -$\vec{h}\hspace{.2mm}'$, similarly to Equation \ref{eq_iota}, such that $\cos\Delta\lambda=(\Delta\vec{v}_{\text{sys}}\cdot\text{-}\vec{h}\hspace{.5mm}')/(\Delta v_{\text{sys}}h')$, which results in
\begin{equation}
    \label{eq_dlambda}
    \cos\Delta\lambda=\dfrac{v_k}{\Delta v_{\text{sys}}}\dfrac{rv}{h'}\dfrac{\mu'}{M_c}\left(\dfrac{\mu\Delta M_p}{M_pM_p'}+1\right)\cos\zeta\cos\theta.
\end{equation}
This angle may be of interest when, for example, analyzing the offsets of short-duration gamma-ray bursts (SGRBs) from their host galaxies \citep[e.g.,][]{Gaspari_2025}, since these signals are thought to result from DNS mergers \citep[which might also be the case for a significant fraction of long-duration GRBs, see][]{Levan_2026} and the jets are aligned with the post-SN orbital angular momentum. However, the systemic velocities of DNSs are affected by two systemic kicks at the formation of both NSs. In order to combine two subsequent systemic kicks, we rotate the first systemic kick $\Delta\vec{v}_{\text{sys,0}}$ into the frame of the second kick. We define $\omega'$ as the true anomaly at the moment of the first SN in the post-SN orbit, such that $r(\omega|a,e)=r(\omega'|a',e')$ as defined in Equation \ref{eq_r}. If the first natal kick had an angle $\theta=\theta_0$, occurred at a post-SN phase $\omega'_0$, and induced an inclination $\iota_0$, then the total systemic kick equals
\begin{equation}
    \label{eq_v_sys}
    \vec{v}\hspace{.2mm}'_{\text{sys}}=\Delta\vec{v}_{\text{sys}}+\mathcal{R}_z(\Delta\omega)\mathcal{R}_y(\iota_0\,\text{sgn}\left(\cos\theta_0\right))\Delta\vec{v}_{\text{sys,0}},
\end{equation}
where $\Delta\omega=\omega-\omega_0'$ and $\mathcal{R}_i$ is the rotation matrix around the $i$-axis. The combination of $\Delta\vec{v}_{\text{sys}}$ and $\Delta\vec{v}_{\text{sys,0}}$ in this equation assumes that the systemic velocity is only affected by the SN and neglects interaction with, for example, the Galactic potential. The angle $\lambda$ between the final systemic velocity and the orbital angular momentum comes down to $\cos\lambda=(\vec{v}\hspace{.2mm}'_{\text{sys}}\cdot\text{-}\vec{h}\hspace{.5mm}')/(\vec{v}\hspace{.2mm}'_{\text{sys}}h')$, similar to Equation \ref{eq_dlambda}.

\subsection{Rocket Kicks}
\label{sec2.3}
\noindent It has been theorized that asymmetric spin-down radiation can accelerate NSs after their formation \citep{Harrison_1975,Lai_2001}. This acceleration is aligned with the NS spin and takes place on the spin-down timescale, which generally exceeds the binary orbital period. Due to its secular nature, the effect on the binary orbit is qualitatively different from that of instantaneous natal kicks. \citet{Hirai_2024} show that rocket kicks induce oscillations in eccentricity, while keeping the semi-major axis constant. In order to quantify this, they define a dimensionless orbital angular momentum vector $\vec{l}$ with $l=\sqrt{1-e^2}$. The pre-rocket orbit is then defined through an eccentricity vector $\vec{e}$ and the orthogonal $\vec{l}$. If the rocket direction is described by the angles $\phi_r$ and $\theta_r$, which are defined similar to $\phi$ and $\theta$ in Equation \ref{eq_v_prime_vec}, we rotate the $z$-axis to be aligned with the rocket such that $\vec{e}$ and $\vec{l}$ are given by
\begin{equation}
    \label{eq_e_l_vec}
    \vec{e}=e\begin{pmatrix}\cos\phi_r\cos\theta_r\\ -\sin\phi_r\\\cos\phi_r\sin\theta_r\end{pmatrix}\quad\text{and}\quad\vec{l}=\sqrt{1-e^2}\begin{pmatrix}-\sin\theta_r\\0\\\cos\theta_r\end{pmatrix},
\end{equation}
which are obtained through $\vec{e}=\mathcal{R}_y(-\theta_r)\mathcal{R}_z(-\phi_r)e\hat{\vec{x}}$ and $\vec{l}=\mathcal{R}_y(-\theta_r)\mathcal{R}_z(-\phi_r)l\hat{\vec{z}}$. If the rocket acceleration $a_{\text{roc}}$ takes place over a period $\tau_{\text{roc}}$ then it applies a velocity equal to $\Delta v_{\text{roc}}=\int_0^{\tau_{\text{roc}}}a_{\text{roc}}dt$. In this model $\tau_{\text{roc}}$ exceeds the orbital period of the binary, but is expected to be small enough that all observed binaries are observed at $t>\tau_{\text{roc}}$. \citet{Hirai_2024} define the vectors $\vec{\mathcal{H}}_{\pm}=\vec{l}\pm\vec{e}$ and show that the rocket kick effectively rotates these vectors around the rocket axis by an angle equal to $\pm3/2\cdot\Delta v_{\text{roc}}/v_{\text{orb}}$. This means that, in the case where the rocket is aligned with the $z$-axis, the post-rocket vectors $\vec{\mathcal{H}}_{\pm}'=\mathcal{R}_{z}(\pm3/2\cdot\Delta v_{\text{roc}}/v_{\text{orb}})\vec{\mathcal{H}}_{\pm}$ are described by
\begin{equation}
    \label{eq_roc_vec}
    \vec{\mathcal{H}}_{\pm}'=\begin{pmatrix}\mathcal{H}_{\pm,x}\cos\left(\pm\dfrac{3\tilde{v}_{\text{roc}}}{2}\right)-\mathcal{H}_{\pm,y}\sin\left(\pm\dfrac{3\tilde{v}_{\text{roc}}}{2}\right)\\\mathcal{H}_{\pm,x}\sin\left(\pm\dfrac{3\tilde{v}_{\text{roc}}}{2}\right)+\mathcal{H}_{\pm,y}\cos\left(\pm\dfrac{3\tilde{v}_{\text{roc}}}{2}\right)\\\mathcal{H}_{\pm,z}\end{pmatrix},
\end{equation}
where we define $\tilde{v}_{\text{roc}}=\Delta v_{\text{roc}}/v_{\text{orb}}$. The post-rocket angular momentum and eccentricity vectors are then determined through $\vec{l}\hspace{.5mm}'=(\vec{\mathcal{H}}_+'+\vec{\mathcal{H}}_-')/2$ and $\vec{e}\hspace{.5mm}'=(\vec{\mathcal{H}}_+'-\vec{\mathcal{H}}_-')/2$, where the latter is equal to
\begin{equation}
    \label{eq_e_roc_vec}
    \vec{e}\hspace{.5mm}'=e\begin{pmatrix}a\cos\left(3\tilde{v}_{\text{roc}}/2\right)\\ b\sin\left(3\tilde{v}_{\text{roc}}/2\right)+c\cos\left(3\tilde{v}_{\text{roc}}/2\right)\\d\end{pmatrix},
\end{equation}
with $a=\cos\phi_r\cos\theta_r$, $b=-(l/e)\sin\theta_r$, $c=-\sin\phi_r$, and $d=\cos\phi_r\sin\theta_r=\sqrt{1-a^2-c^2}$. This means that the change in eccentricity due to the rocket kick, as a function of $\tilde{v}_{\text{roc}}$, can be expressed as
\begin{equation}
    \label{eq_e_roc}
    \left(\dfrac{e'}{e}\right)^2=1+A\sin\left(3\tilde{v}_{\text{roc}}\right)+B\left[\cos\left(3\tilde{v}_{\text{roc}}\right)-1\right],
\end{equation}
where $A=bc$ and $B=(a^2-b^2+c^2)/2$. However, the eccentricity oscillation described by this equation can reach values of $e'\simeq1$, which would induce a merger and stop the oscillation. In Appendix \ref{appA} we determine the maximum rocket kick a binary can obtain before inducing a contact merger. Lastly, we note that \citet{Baibhav_2026} show that a rocket acceleration is not possible in force-free electrodynamics, meaning it can only occur in ``weak" pulsars with suppressed pair production around the light cylinder, and they estimate the possible rocket magnitude for these pulsars to be approximately ${\lesssim}\,30$\,km\,s$^{-1}$.

\subsection{Kick Prescriptions}
\label{sec2.4}
\begin{figure}
    \centering
    \resizebox{\hsize}{!}{\includegraphics{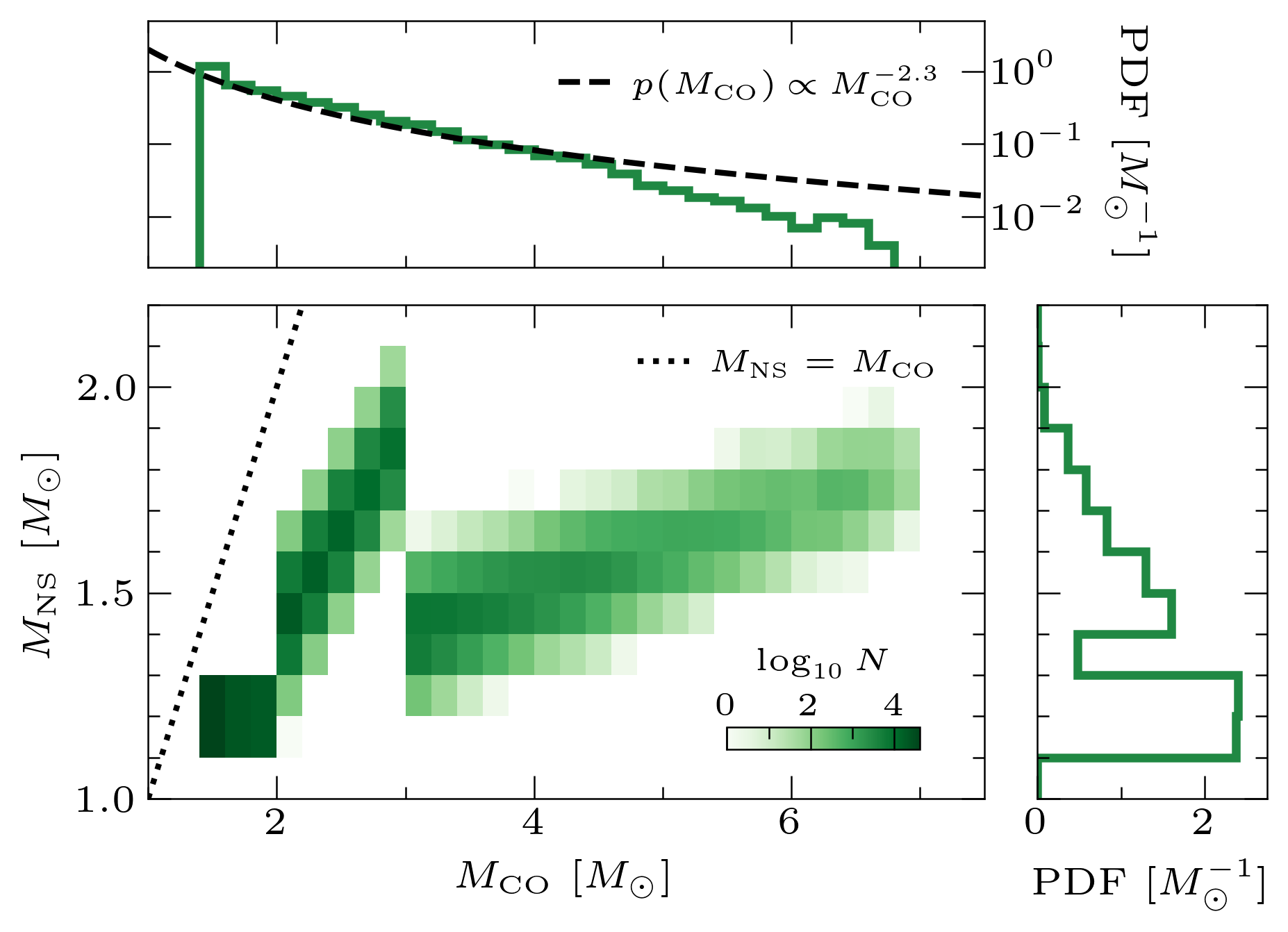}}
    \caption{Relationship between CO core mass ($M_{\text{CO}}$) and NS mass ($M_{\text{NS}}$), following the prescription of \citet{Mandel_2020} as implemented in the SSE \lstinline{COMPAS} model (bottom left panel). The density is shown on a logarithmic color scale, but we note that in all subsequent figures the color scales are linear. The black dotted line shows $M_{\text{NS}}=M_{\text{CO}}$. The top panel shows the CO core mass distribution (green histogram) and a line defined by $p(M_{\text{CO}})\propto M_{\text{CO}}^{-2.3}$, which aligns with the IMF (black dashed line). We note that the histogram starts to diverge from the IMF at $M_{\text{CO}}>4M_{\odot}$, since for these masses an increasing fraction of SNe form black holes in the prescription of \citet{Mandel_2020}. The bottom right panel shows the NS mass distribution.}
    \label{Fig_SSE}
\end{figure}
\begin{figure*}
    \centering
    \includegraphics[width=18cm]{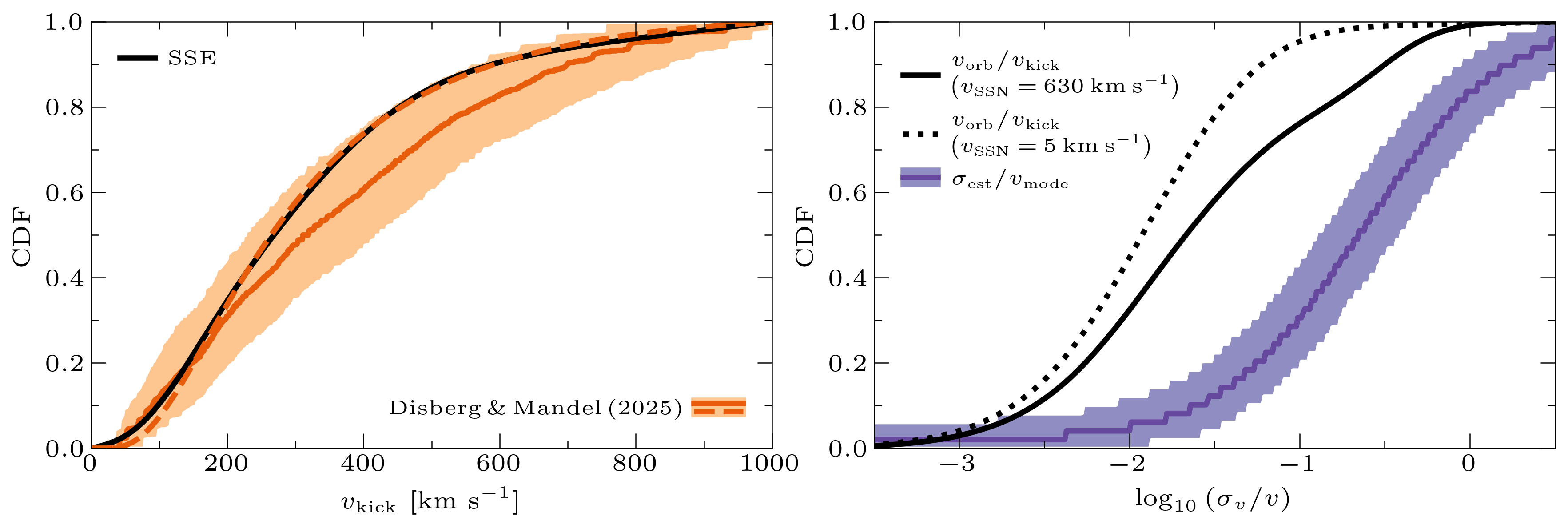}
    \caption{Natal kicks of isolated NSs. The left panel shows the kicks of NSs in the SSE \lstinline{COMPAS} model (black line) calibrated to the kick estimates of \citet{Disberg_2025b}, with $v_{\text{NS}}=630$\,km\,s$^{-1}$ and $f_{\text{NS}}=0.45$ as defined in Equation \ref{eq_mu_sig}. The kick distributions are normalized between $0$ and $1000$\,km\,s$^{-1}$. The orange solid line and light orange shaded region correspond to the median distribution and the bootstrapped $95\%$ uncertainty interval of the velocity estimates for observed pulsars younger than 10 Myr of \citet{Disberg_2025b}, whereas the orange dashed line shows the kick distribution: a lognormal model with $\mu=5.60(12)$ and $\sigma=0.68(10)$. The figure shows that the calibrated SSE \lstinline{COMPAS} model indeed aligns well with the lognormal model, which was fitted to the same observational sample. The right panel shows the distributions of the velocity ratios, as labeled ($\sigma_v/v$), which include the values of $v_{\text{orb}}/v_{\text{kick}}$ for binaries in the binary stellar evolution (BSE) \lstinline{COMPAS} model that become unbound due to the natal kick. We show simulated binaries for $v_{\text{SSN}}=630$\,km\,s$^{-1}$ (black solid line) and $v_{\text{SSN}}=5$\,km\,s$^{-1}$ (black dashed line), where in both cases $v_{\text{SN}}=630$\,km\,s$^{-1}$ as defined in Equations \ref{eq_mu_sig} and \ref{eq_v_NS}. In order to compare these curves to the accuracy of the observational natal kick estimates of \citet[][see also their appendix B]{Disberg_2025b}, we consider their individual kick posteriors $\{v_{\text{post}}\}$ and define $\sigma_{\text{est}}=|v_{\text{post}}-v_{\text{mode}}|$, where $v_{\text{mode}}$ is the most likely value given their lognormal model. The purple solid line and light purple shaded region correspond to the median distribution and bootstrapped $95\%$ interval of $\sigma_{\text{est}}/v_{\text{mode}}$, which describes the relative width of their kick posteriors and exceeds the values of $v_{\text{orb}}/v_{\text{kick}}$ by approximately an order of magnitude.}
    \label{Fig_Calibration}
\end{figure*}
\noindent In order to estimate the kick velocities of NSs in binary systems, we apply kick models to results from population synthesis simulations with \lstinline{COMPAS} and compare them to observations. In particular, we model the NS natal kicks following the model of \citet{Mandel_2020}, in which the natal kick velocity is drawn from a normal distribution with mean $\mu_k$ and standard deviation $\sigma_k$ described by
\begin{equation}
    \label{eq_mu_sig}
    \mu_{k}=v_{\text{NS}}\dfrac{M_{\text{CO}}-M_{\text{NS}}}{M_{\text{NS}}}\quad\text{and}\quad\sigma_{k}=f_{\text{NS}}\,\mu_k,
\end{equation}
where $M_{\text{CO}}$ and $M_{\text{NS}}$ are the pre-SN CO-core mass and NS mass, and $v_{\text{NS}}$ and $f_{\text{NS}}$ are parameters that determine the magnitude and spread of the kick distribution, respectively. When drawing kicks for the normal distribution described by this equation, we require the sampled kick velocity to be positive. Moreover, \citet{Mandel_2020} describe the relationship between $M_{\text{CO}}$ and $M_{\text{NS}}$ using three mass ranges. In Figure \ref{Fig_SSE} we show this relationship for a single-star evolution (SSE) \lstinline{COMPAS} run with 1 million stars. Before the SN, the CO core masses approximately track the initial mass function (IMF), but there is a non-monotonic relation between $M_{\text{CO}}$ and $M_{\text{NS}}$ which shapes the NS mass distribution and the kick distribution through Equation \ref{eq_mu_sig} \citep[see also][]{Karim_2026}.

Recently, \citet{Grichener_2026} argued that USSNe in DNS progenitor systems may impart negligible natal kicks. In order to distinguish between the kicks connected to normal SNe and stripped SNe (SSNe), we separate the $v_{\text{NS}}$ parameter into $v_{\text{SSN}}$ for stripped SNe where the progenitor has lost its envelope through binary interactions (which includes but is not limited to USSNe) and $v_{\text{SN}}$ for SNe in progenitors that retained their envelope. We implement this as
\begin{equation}
    \label{eq_v_NS}
    v_{\text{NS}}=\left\{\begin{matrix}v_{\text{SN}}&\text{if }M_p-M_{\text{CO}}\geq4M_{\odot}\\v_{\text{SSN}}&\text{if }M_p-M_{\text{CO}}<4M_{\odot}\end{matrix}\right.,
\end{equation}
because we find that progenitors with $M_p-M_{\text{CO}}<4M_{\odot}$ have lost their hydrogen envelope and result in a stripped SN. We do not change $f_{\text{NS}}$ because for significantly reduced SSN kicks this parameter has a negligible effect.

\begin{figure*}
    \centering
    \includegraphics[width=18cm]{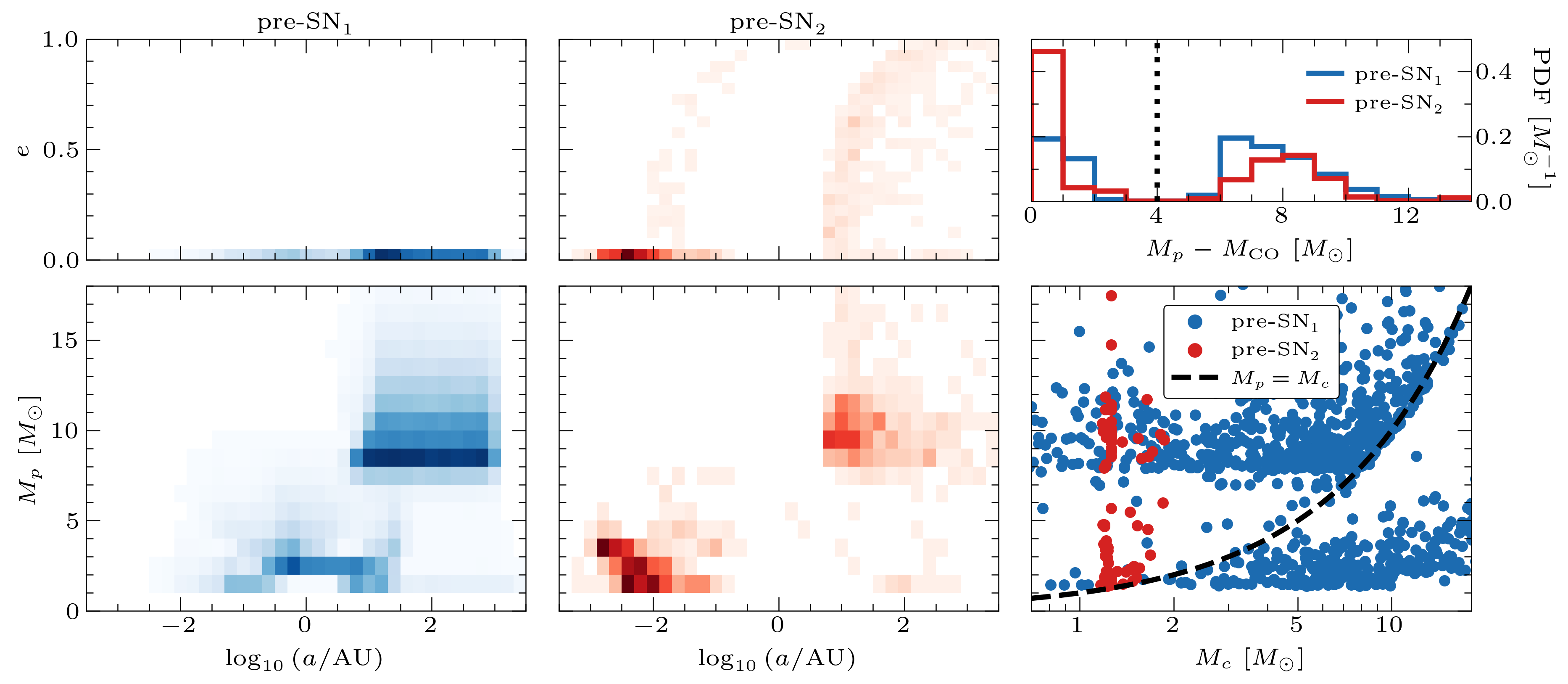}
    \caption{Pre-supernova binary parameters from \lstinline{COMPAS}, for the first SN in a binary (SN$_1$, blue) and the second SN when the companion is already an NS (SN$_2$, red), only including SNe that form NSs. The left and central panels show the semi-major axes $a$ as a function of the progenitor mass $M_p$ and eccentricity $e$, for SN$_1$ (left) and SN$_2$ (center). The bottom right panel shows the relationship between progenitor mass and companion mass for $10^3$ randomly selected SN$_1$ points and $10^2$ randomly selected SN$_2$ points, together with a black dashed line corresponding to $M_p=M_c$. The top right panel shows the distribution of envelope masses $M_p-M_{\text{CO}}$. The black dotted line shows $M_p-M_{\text{CO}}=4M_{\odot}$ since we classify SNe below this limit as SSNe in Equation \ref{eq_v_NS}. For SN$_1$, $33\%$ of SNe are SSNe and for SN$_2$ this number increases to $54\%$.}
    \label{Fig_BSE}
\end{figure*}

We estimate $v_{\text{SN}}$ by applying Equation \ref{eq_mu_sig} to the SSE simulation shown in Figure \ref{Fig_SSE} and fitting the resulting kick distribution to the kick estimates of \citet{Disberg_2025b}, which are based on the velocities of young isolated pulsars. This results in fitted values of $v_{\text{NS}}=(630\pm100)$\,km\,s$^{-1}$ and $f_{\text{NS}}=0.45\pm0.20$ (as also used in the analyses of \citealt{Disberg_2026} and \citealt{Grichener_2026}). In Figure \ref{Fig_Calibration} we show that the cumulative distribution function (CDF) of the fitted kick distribution aligns well with the lognormal model of \citet{Disberg_2025b}. Moreover, \citet{Chrimes_2026} find that the distribution of \citet{Disberg_2025b} is relatively consistent with the observed transverse velocities of magnetars, although they argue that there may be evidence for a dearth of high-velocity magnetars in their sample. However, in Appendix \ref{appB} we show that the lognormal kick distribution of \citet{Disberg_2025b} can indeed explain the observed proper motions of young isolated pulsars as well as the magnetars in the sample of \citet{Chrimes_2026}. We thus choose $v_{\text{SN}}=630$\,km\,s$^{-1}$ and consider different values for $v_{\text{SSN}}$, such as ultra-low kicks with $v_{\text{SSN}}=5$\,km\,s$^{-1}$.

The kick distribution shown in Figure \ref{Fig_Calibration} predicts lower kick velocities than the widely used distribution of \citet{Hobbs_2005}, but \citet{Disberg_2025b} show that the Maxwellian distribution of \citet{Hobbs_2005} is based on an erroneous histogram interpretation that misses a Jacobian needed to correct for its logarithmic bin sizes. Correcting for this missing Jacobian reconciles the distribution of \citet{Hobbs_2005} with the lognormal distribution shown in Figure \ref{Fig_SSE}. Moreover, \citet{Disberg_2025b} conclude that their distribution is consistent with the results of \citet{Verbunt_2017} and \citet{Igoshev_2020}, but that the bimodality that they find is likely not statistically significant.

Previously, \citet{Kapil_2023} fitted these parameters to a dataset of pulsar velocities, but found different values because (1) their pulsar sample is not age-limited, meaning it incorporates dynamically old pulsars whose velocities have been decreased by the Galactic potential \citep{Disberg_2024a}, and (2) they estimate the line-of-sight component of the velocity vector by assuming isotropy in the observer frame instead of the LSR frame of the pulsar. For these reasons, we are more confident in the values calibrated to the results of \citet{Disberg_2025b}.

This calibration compares the SSE \lstinline{COMPAS} model to the pulsar kick estimates of \citet{Disberg_2025b}. However, a significant fraction of their pulsar sample can be expected to have a binary origin, which would affect their velocities \citep[e.g.,][]{Tauris_1998}. For this reason, we apply the same natal kicks to a binary stellar evolution (BSE) \lstinline{COMPAS} simulation. The binary origin affects the kick estimates by a scatter of order $v_{\text{orb}}$ \citep[e.g.,][]{Mandel_2026}. In Figure \ref{Fig_Calibration} we show that the median value of $v_{\text{orb}}/v_{\text{kick}}$ for binaries that become unbound is ${\sim}\,1\%$, because most NS-forming binaries are relatively wide (meaning $v_{\text{orb}}$ is low compared to $v_{\text{kick}}$). Figure \ref{Fig_Calibration} also shows that the scatter caused by $v_{\text{orb}}$ is approximately an order of magnitude smaller than the uncertainty in the pulsar velocity estimates in the sample of \citet{Disberg_2025b}. We therefore conclude that a binary origin is unlikely to significantly affect kick estimates based on isolated pulsars, as also found by \citet{Kapil_2023}.

Besides the kick magnitudes, we also consider the kick directions relative to the binary system, since an alignment between the kick and the orbital angular momentum has been proposed \citep[e.g.,][]{Bear_2025,Valli_2025}. Relevantly, an alignment has been observed between the spins and velocities of young pulsars \citep[e.g.,][]{Noutsos_2012,Biryukov_2025}. Although \citet{Muller_2019} find no evidence of spin-kick alignment in their SN simulations, \citet{Janka_2022} propose that this alignment may be caused by fallback onto the newly-formed NS (but see also \citealt{Mueller_2023}). If the progenitor spin is aligned with the orbital angular momentum, because they were formed aligned (which is not obviously true, as e.g.\ shown by \citealt{Offner_2016} and \citealt{Lee_2019}) or they were aligned due to tides and mass transfer \citep[e.g.,][]{Repetto_2014}, and if the NS spin is aligned with the progenitor spin, then the kick directions in the binary system may indeed be constrained. Even though there is little observational or theoretical evidence for these strong alignments, we explore the effects of kick-orbit alignment. In particular, we choose the kick angle $\phi$ in the orbital plane to be isotropic (i.e., $\phi$ is uniformly distributed between $0$ and $2\pi$) whereas we either sample $\theta$ with $p(\theta)\propto\sin\theta$ for isotropic kicks or set $\theta=0$ for polar kicks.

\subsection{Application}
\label{sec2.5}
\begin{table}
\centering
\caption{Variations in the different kick models ($\mathcal{M}$), for the stripped SN kick magnitude ($v_{\text{SSN}}$, defined in Equations \ref{eq_mu_sig} and \ref{eq_v_NS}), the natal kick directions (either isotropic or polar), and the rocket kick magnitude (defined in Section \ref{sec2.3}).}
\label{tab_models}
\begin{tabular}{l|c|c|c}
    \hline\hline\\[-13pt]
     model & $v_{\text{SSN}}$ & natal kick & $\Delta v_{\text{roc}}$ \\
    & [km\,s$^{-1}$] & directions & [km\,s$^{-1}$]\\
[1pt]\hline\\[-13pt]
    $\Mfid$ & 630 & isotropic & 0\\
    $\Mlow$ & 5 & isotropic & 0\\
    $\Mpol$ & 630 & polar & 0\\
    $\Mroc$ & 630 & isotropic & 30\\
    $\Mlowroc$ & 5 & isotropic & 30\\
[1pt]\hline
\end{tabular}
\tablecomments{\footnotesize In all models we set $v_{\text{SN}}=630$\,km\,s$^{-1}$ and $f_{\text{NS}}=0.45$, as fitted to the velocities of isolated pulsars (see also Figure \ref{Fig_Calibration}).}
\end{table}
\noindent We use the prescription for natal kick magnitudes given in Equation \ref{eq_mu_sig}, and apply it to \lstinline{COMPAS} binaries through the model described in Sections \ref{sec2.1}, \ref{sec2.2}, and \ref{sec2.3}. In particular, we take the pre-SN progenitor masses, CO-core masses, companion masses, semi-major axes, and eccentricities from a \lstinline{COMPAS} v$3.29.02$ simulation of $1$ million binaries at solar metallicity, in which we use the main sequence core-mass prescription of \citet{Brcek_2026}. Then, we compute the post-SN orbital properties by sampling $100$ kick magnitudes from Equation \ref{eq_mu_sig} and choose the kick directions to be distributed either isotropically or polar. For the DNSs, we determine the systemic kicks by applying the kick model to both SNe, but for the binary orbits we only consider the second SN (we elaborate on this in Section \ref{sec7}).

We show these pre-SN orbital parameters in Figure \ref{Fig_BSE}, for the first SN in a binary and the second SN where the companion is an NS. There is a clear bimodality between non-interacting binaries and interacting binaries that have reduced orbital separations and lower progenitor masses due to envelope stripping. This is also reflected in the distribution of $M_p-M_{\text{CO}}$, which has one population with $M_p-M_{\text{CO}}<4M_{\odot}$ and one population with $M_p-M_{\text{CO}}>4M_{\odot}$, reflecting the difference between stripped and non-stripped SNe, respectively. The figure also shows that for most interacting binaries mass transfer onto the companion star has resulted in $M_c>M_p$. 

We create several different kick models $\mathcal{M}$, in which we assume $v_{\text{SN}}=630$\,km\,s$^{-1}$ and $f_{\text{NS}}=0.45$, while choosing (1) different values of $v_{\text{SSN}}$, (2) natal kicks to be either isotropic or polar, and (3) whether or not the NSs obtain a post-SN rocket kick. In particular, we consider the following five kick models:
\begin{itemize}
    \item $\Mfid$: all kicks are calibrated to isolated NSs and are isotropic, with no rocket kick.
    \item $\Mlow$: SSN kicks are significantly reduced to $v_{\text{SSN}}=5$\,km\,s$^{-1}$ \citep[motivated by the results of][]{Grichener_2026}, all kicks are isotropic with no rocket kick.
    \item $\Mpol$: all kicks are calibrated to isolated NSs and aligned with the pre-SN orbital angular momentum (i.e., $\theta=0$), with no rocket kick.
    \item $\Mroc$: all kicks are calibrated to isolated NSs and are isotropic, with an isotropically distributed rocket kick of $\Delta v_{\text{roc}}=30$\,km\,s$^{-1}$ \citep[motivated by the results of][]{Baibhav_2026}. The natal kick and rocket kick directions are sampled from independent isotropic distributions.
    \item $\Mlowroc$: SSN kicks are significantly reduced similarly to $\Mlow$, all kicks are isotropic with the same rocket kick as $\Mroc$.
\end{itemize}
In Table \ref{tab_models} we summarize the parameters of the different kick models. We take the pre-SN binaries from \lstinline{COMPAS} and apply the kick models following the equations listed in the above sections.

\subsection{Systemic Kick Estimates}
\label{sec2.6}
\noindent After applying our kick models to the \lstinline{COMPAS} binaries, we compare them to observations of the orbits (i.e., periods and eccentricities) and systemic velocities of several populations of NS-harboring binaries. The comparison with periods and eccentricities is relatively straightforward, given that the orbit has not significantly changed since the SN (e.g., for BeXBs, see discussion in \citealt{Valli_2025}) or the orbital evolution is well understood (e.g., for gravitational-wave emission from DNSs). However, the comparison with systemic velocities requires a more elaborate analysis.

For binaries that are relatively young, the present-day systemic velocity is approximately equal to the systemic kicks, but for binaries older than a few tens of Myr the velocities have been affected by their motion through the Galactic potential \citep{Disberg_2024a}. In order to estimate the systemic kicks of the dynamically old binaries for which no estimates exist in the literature, we employ the method of \citet[][see also \citealt{Disberg_2025a} and \citealt{Disberg_2025b}]{Disberg_2024b}, which entails integrating the Galactic trajectories of objects back through the Galaxy and inferring the systemic kicks through a simulated relationship between kick magnitude and eccentricity of the Galactic orbit. The fact that this relationship is relatively stable over time \citep{Disberg_2025a} allows for kick estimates of old objects \citep[cf.\ the method of][]{Atri_2019}. The resulting kick estimate for a given object effectively corresponds to the systemic kick needed to change a circular Galactic orbit into its current Galactic trajectory.

We apply this method to kinematically constrain the systemic kicks of \textit{Gaia} NSs, LMXBs, NSWDs, and DNSs, since these are all likely older than $40$\,Myr and we thus consider them dynamically old (for HMXBs, which may be relatively young, we adopt the systemic velocity estimates of \citealt{Fortin_2022b}). We use bootstrapping to determine the confidence intervals and also fit lognormal distributions to these systemic kick estimates. In Appendix \ref{appC} we explain this method in more detail and list relevant caveats.

For the different types of binary systems, we want to compare our model to these confidence intervals and the lognormal fits. In order to do this, we select systems from our simulated population that show post-SN characteristics (such as orbital period or companion mass) that are comparable to the observed systems. The systemic kicks of these objects, determined through Equation \ref{eq_v_sys}, correspond to the intrinsic systemic kick distribution as predicted through our model.

However, in the observed distribution there is a kinematic bias induced by the fact that larger systemic kicks lead to Galactic trajectories with bigger apocenters and vertical offsets from the Galactic disc \citep[e.g.,][]{Song_2025}, and thus larger distances between the object and the Sun, leading to a decreased probability of being observed. This means that the same intrinsic systemic kick distribution will lead to different observed distributions for objects located at different distances. \citet{Disberg_2026} model this distance-dependent kinematic bias and describe a method for inferring the predicted observed systemic kick distribution based on an intrinsic distribution and a simulated relationship between systemic kick magnitude and the probability of the object obtaining a certain distance from the Sun. We combine this method with our model results and the distances of the objects in order to compare them to the systemic kick estimates.

\begin{figure*}
    \centering
    \includegraphics[width=18cm]{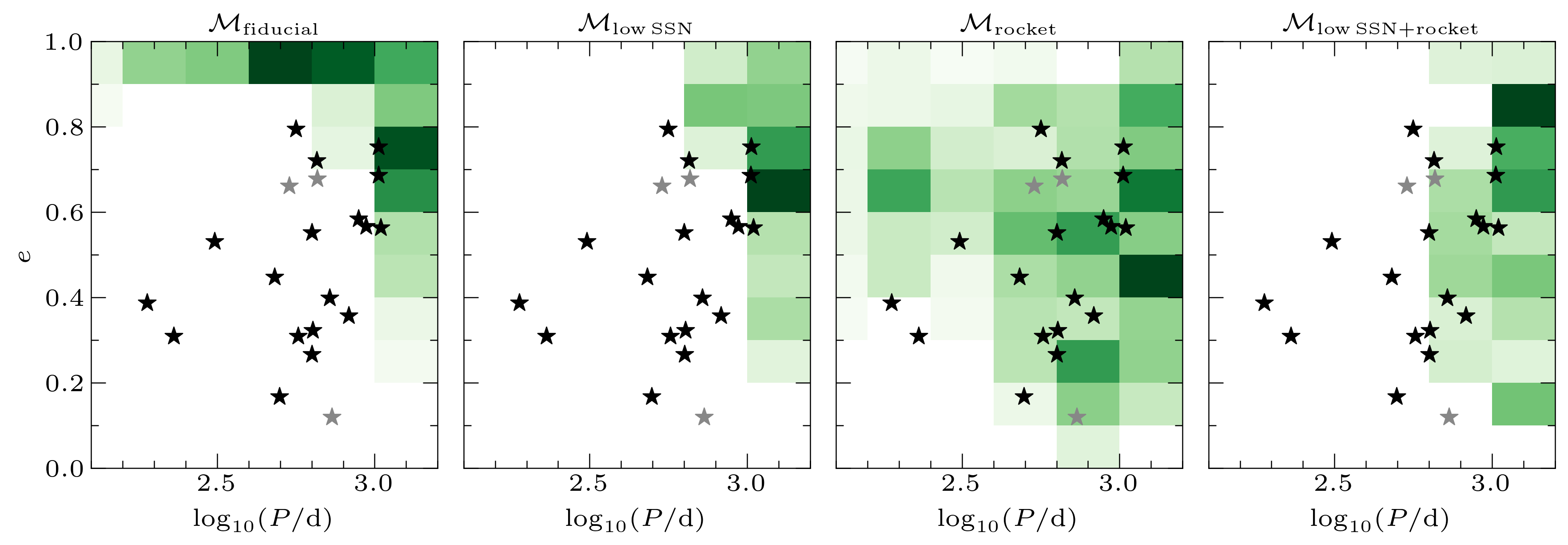}
    \caption{Binary orbital properties of the \textit{Gaia} NSs in the sample of \citet{El-Badry_2024b}, containing systems with disc-like Galactic orbits (black stars) and three systems on halo-like Galactic orbits (grey stars). The green distributions show the density of the simulated post-SN binaries as a function of the orbital period $P$ and the eccentricity $e$, which are weighted by the \textit{Gaia} detection probability function of \citet{Baibhav_2026}. These distributions depend on the different kick models $\mathcal{M}$ (Table \ref{tab_models}), and are re-normalized for each panel.}
    \label{Fig_Gaia_Orbits}
\end{figure*}

\section{Gaia Neutron Star Binaries}
\label{sec3}
\noindent The \textit{Gaia} mission \citep{Gaia_2016} has provided invaluable data on the positions, distances, and velocities of millions of stars \citep{Gaia_2023}. \citet{El-Badry_2024b} find $21$ binary systems among the \textit{Gaia} observations that consist of main-sequence stars, with masses between ${\sim}\,0.7M_{\odot}$ and ${\sim}\,1.3M_{\odot}$, and companions with masses between ${\sim}\,1.25M_{\odot}$ and ${\sim}\,1.90M_{\odot}$ that are theorized to be NSs. Although they note that it is difficult to decisively determine the nature of the companion star, we assume that these binaries indeed harbor an NS and compare them to our different kick models in order to determine what kick model (if any) could explain their orbits (Section \ref{sec3.1}) and systemic velocities (Section \ref{sec3.2}). 

\subsection{Gaia NS Orbits}
\label{sec3.1}
\noindent From our different kick models, we select the binaries that remain bound after the SN and have main-sequence companions with masses below $1.5M_{\odot}$. Then, we describe the \textit{Gaia} detection probability, as a function of orbital period and eccentricity, through the analytical formulae of \citet{Baibhav_2026}, which are based on the framework of \citet{El-Badry_2024a}. In Figure \ref{Fig_Gaia_Orbits} we show the binary orbits of the resulting simulated binaries for kick models $\Mfid$, $\Mlow$, $\Mroc$, and $\Mlowroc$, compared to the observational sample of \citet{El-Badry_2024b}. The results for $\Mpol$ are similar to the results for $\Mfid$. For $\Mfid$ and $\Mlow$, there are binaries in relatively eccentric orbits at $P\gtrsim1000\,$d. The progenitors of these systems are the non-interacting binaries with the shortest periods ($P_0\approx1000\,\text{d}$) that have retained their orbital period. For $\Mfid$, however, the interacting binaries also obtain significant natal kicks, which results in a population of highly eccentric ($e\gtrsim0.95$) systems in the period range of the \textit{Gaia} NSs. Neither of these models is able to explain the observed binary orbits, which are dominated by systems with eccentricity between $\sim 0.2$ and $\sim 0.8$ and orbital periods $P \lesssim 1000\,$d.

These results align well with the argument of \citet{Hirai_2024} that it is difficult to explain the observed wide \textit{Gaia} NSs with moderate eccentricities because progenitor systems with these orbital periods will interact and reach significantly tighter orbits before the SN \citep[contrasting the argument favoring low kicks of][which requires pre-SN orbital separations in this regime]{Shariat_2026}. \citet{Hirai_2024} argue that rocket kicks might explain these systems, since the eccentricity oscillations induced by rocket accelerations can reduce the eccentricity of the highly eccentric post-SN systems originating from these tight pre-SN orbits. Moreover, they show that for wide orbits similar to those of the \textit{Gaia} NSs only relatively small rocket kicks (i.e., $\Delta v_{\text{roc}}\lesssim30$\,km\,s$^{-1}$) are required to reduce the eccentricity significantly. \citet{Baibhav_2026}, in turn, find that the period--eccentricity distribution of \textit{Gaia} NSs can be reproduced as long as the rocket magnitude is $\Delta v_{\text{roc}}\gtrsim30$\,km\,s$^{-1}$, but note that they can explain these systems without rocket kicks if the pre-SN orbits are not significantly reduced by binary interactions and if they use a natal kick distribution that peaks at zero. Indeed, \citet{VanSon_2026} point out that the periods and eccentricities of the \textit{Gaia} NSs are not necessarily different from other post-mass-transfer binaries, although (1) they note that post-mass-transfer systems do have significantly greater mass ratios and (2) most post-mass-transfer binaries do not contain an NS whose formation can significantly affect the orbits.

\begin{figure}
    \centering
    \resizebox{\hsize}{!}{\includegraphics{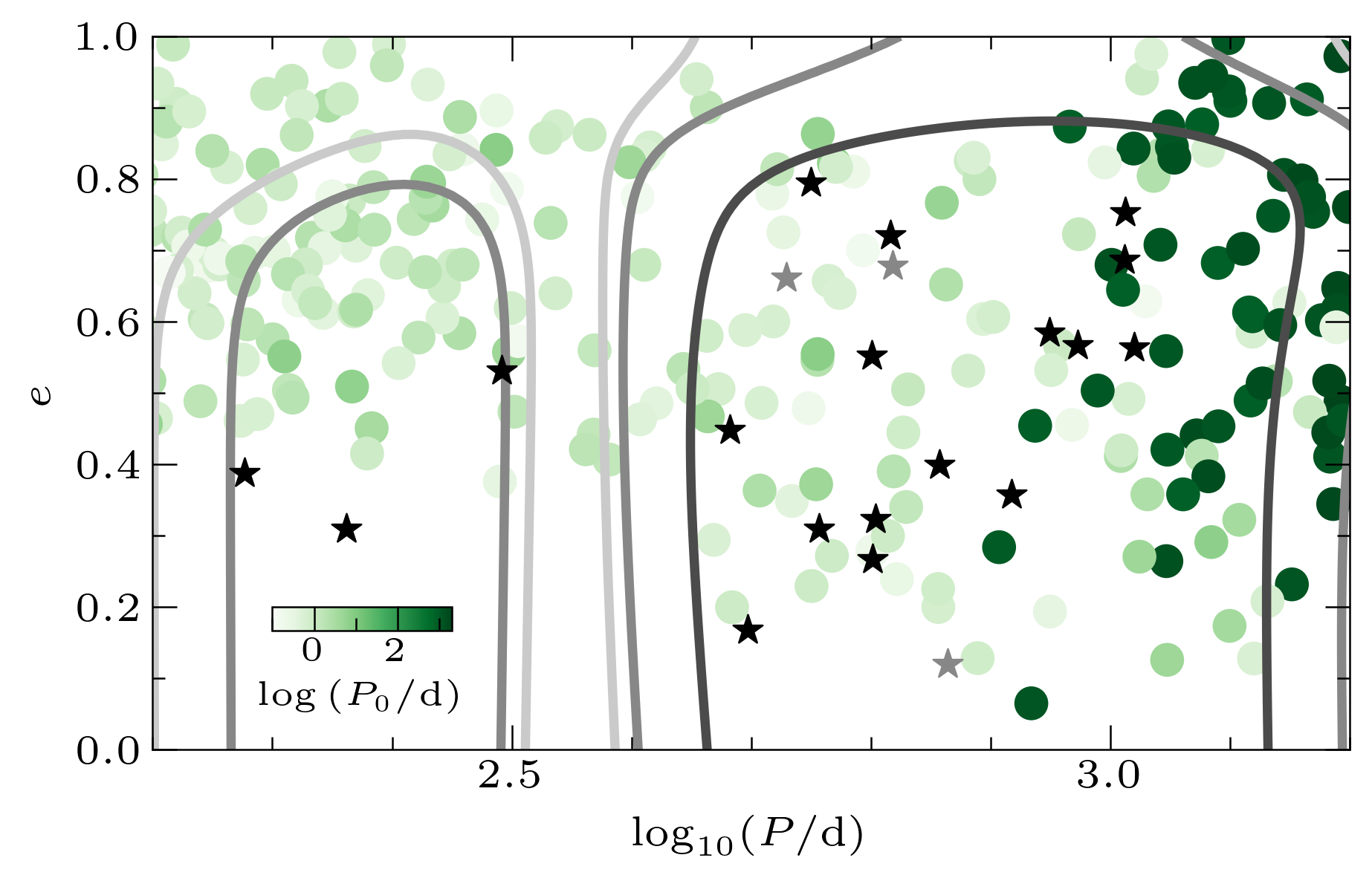}}
    \caption{Simulated systems in the \textit{Gaia} detection range for $\Mroc$. The black and grey stars are the systems in the sample of \citet{El-Badry_2024b}, with the latter corresponding to the systems on halo-like Galactic orbits. The color-scale shows the pre-SN orbital period $P_0$, where there is a clear difference between systems with small $P_0$ that went through a CE event and systems with large $P_0$ that did not. The lines correspond to the $68\%$, $90\%$, and $95\%$ contours of the \textit{Gaia} detection probability distribution (dark grey, grey, and light grey lines, respectively), as formulated by \citet{Baibhav_2026} based on the framework of \citet{El-Badry_2024a}.}
    \label{Fig_Gaia_Progenitors}
\end{figure}

The models $\Mroc$ and $\Mlowroc$, which contain a rocket kick of $30\,$km\,s$^{-1}$, indeed yield reduced eccentricities. However, for $v_{\text{SSN}}=5$\,km\,s$^{-1}$ there are no systems at $P\lesssim1000$\,d, which means that the lower orbital periods of the \textit{Gaia} NSs are difficult to explain with $\Mlowroc$. The $\Mroc$ model, in contrast, reduces the eccentricities of the highly eccentric systems predicted by $\Mfid$ and fills the parameter space of the observations. In Figure \ref{Fig_Gaia_Progenitors} we show the systems in the $\Mroc$ distribution and their pre-SN orbital periods. There is a clear distinction between the non-interacting binaries, which are also present in the results for $\Mlow$ and $\Mlowroc$, and the binaries that have significantly tighter pre-SN orbits and are kicked by SSNe onto highly eccentric orbits with periods in the sensitivity range of \textit{Gaia}, with eccentricity subsequently reduced by rocket kicks.

\begin{figure*}
    \centering
    \includegraphics[width=18cm]{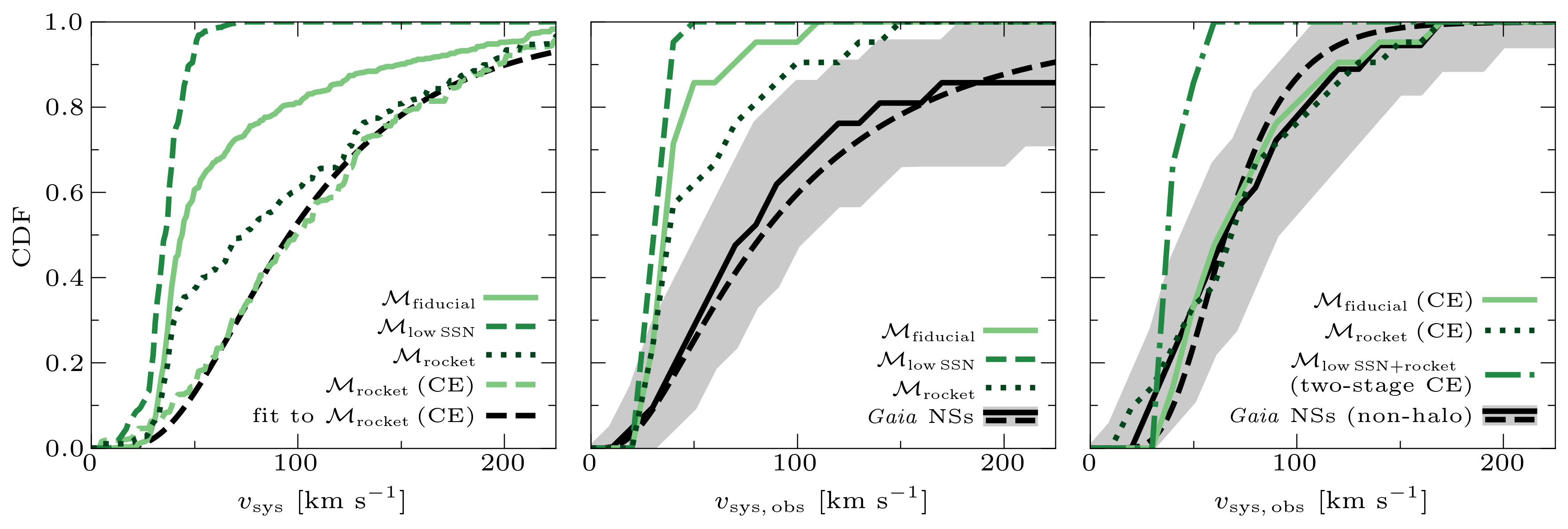}
    \caption{Systemic kicks of the \textit{Gaia} NSs. The left panel shows the intrinsic kick distributions as predicted through our models $\Mfid$, $\Mlow$, and $\Mroc$ (light green solid line, green dashed line, and dark green dotted line, respectively), where we sampled the systems from our results following the \textit{Gaia} detection probability function of \citet{Baibhav_2026}. We also show the distribution for the systems in $\Mroc$ that experienced a CE (light green dotted line), which in practice corresponds to the systems with $P_0<100\,$d (as shown in Figure \ref{Fig_Gaia_Progenitors}), and fit a lognormal distribution (see Eq.~\ref{eq_lognormal}) to these velocities which results in $\mu=4.60$ and $\sigma=0.55$. In the central panel we show the median distribution of the systemic kick estimates for the observed systems in the sample of \citet{El-Badry_2024b}, together with its bootstrapped $95\%$ uncertainty region (black solid line and grey shaded region, respectively). A lognormal fit to these estimates yields $\mu=4.42(18)$ and $\sigma=0.76(9)$ (black dashed line). The green lines show the model results from the left panel but corrected for kinematic bias following the method of \citet{Disberg_2026}, so that they are comparable to observations. The right panel shows the distributions where the three \textit{Gaia} NSs with halo-like Galactic orbits are omitted from the data and the model results, for $\Mfid$ and $\Mroc$, only contain systems that experienced a CE. The lognormal fit to these curated observations (black dashed line) resulted in $\mu=4.18(14)$ and $\sigma=0.38(13)$. We also show simulated systems where we changed the CE prescription to the two-stage CE formalism of \citet{Hirai_2022}, in which the $\Mlowroc$ model yields systems in the \textit{Gaia} orbital-parameter range (green dash-dotted line).}
    \label{Fig_Gaia_Velocities}
\end{figure*}

Our results for the $\Mroc$ model, in which NSs obtain high natal kicks and a relatively small rocket kick, can explain the observed periods and eccentricities of the \textit{Gaia} NSs. However, we note that ruling out other models depends on several assumptions in the \lstinline{COMPAS} model, such as the common-envelope (CE) formalism. If we change the CE prescription to the two-stage CE formalism of \citet{Hirai_2022}, we find that the Blaauw kicks are sufficient to produce systems in the \textit{Gaia} sensitivity range and the results for $\Mroc$ and $\Mlowroc$ are practically indistinguishable. This is due to the fact that in the two-stage CE formalism the simulation contains a population of tight binaries with progenitor masses of ${\gtrsim}\,3M_{\odot}$, resulting in significant Blaauw kicks and thus post-SN orbital periods similar to the \textit{Gaia} NS periods even in the absence of strong natal kicks. We therefore argue that, although in our model the \textit{Gaia} NS periods and eccentricities can be explained by the $\Mroc$ model, the systematic model uncertainties make it difficult to draw strong conclusions about their natal kicks based on their orbits. 

\subsection{Gaia NS Velocities}
\label{sec3.2}
\noindent We also consider the systemic velocities of these binaries. In particular, we use the method of \citet{Disberg_2024b,Disberg_2025a} to constrain their systemic kicks by estimating the eccentricity of their Galactic trajectory based on their present-day velocity. We also fit lognormal distributions to these systemic kick estimates (see Appendix \ref{appC}), which should be compared to the model results. Since we use radial velocity estimates from \textit{Gaia} to estimate the present-day velocity vectors and do not need to assume isotropy, as discussed in Appendix \ref{appC}, the lognormal fits do not differ significantly from the median distribution of the kinematically constrained kick estimates. We use the method of \citet{Disberg_2026} to correct the model results for kinematic bias (i.e., the fact that systems with higher systemic kicks are less likely to be observed) in order to compare them to the observed kick distribution.

In Figure \ref{Fig_Gaia_Velocities} we show the systemic kick distributions predicted by our kick models, which are weighted by the \textit{Gaia} detection probability function of \citet{Baibhav_2026}. The figure shows that the systemic kicks for $\Mfid$ and $\Mroc$ are higher than the ones for $\Mlow$. This can be explained by the fact that the former two models have a contribution from tight pre-SN orbits (as shown in Figure \ref{Fig_Gaia_Progenitors}), and their higher pre-SN orbital velocities allow for higher systemic kicks without becoming unbound \citep{Mandel_2026}. We do not consider the effect of the rocket kick on the systemic velocity, meaning the only difference between $\Mfid$ and $\Mroc$ is the fact that the rocket kicks alter the post-SN orbits and therefore the observational weight of each system. In reality, isotropically distributed rocket kicks will add a scatter on the systemic kick distribution, but should not affect the median systemic kick distribution significantly.

In the central panel of Figure \ref{Fig_Gaia_Velocities} we also show the kinematically constrained kick estimates for the \textit{Gaia} NSs, together with the model distributions corrected for kinematic bias. We find that the model results predict significantly lower systemic kicks than the estimates for the observed \textit{Gaia} NSs. 

However, \citet{El-Badry_2024b} identify three systems that are on halo-like Galactic orbits (J0152-2049, J1432-1021, and J1739+4502) in their \textit{Gaia} NS sample. Indeed, we estimate the median kicks that are necessary to explain the eccentricity of their Galactic trajectories to be $680$, $710$, and $380$\,km\,s$^{-1}$, respectively. \citet{El-Badry_2024b} find that these three systems have a significantly lower metallicity than the other systems in their sample, and they state that this suggests their halo-like orbits are the result of them being part of an old stellar population. \citet{SchiebelbeinZwack_2026} argue that these systems might have been formed in a metal-poor dwarf galaxy that was accreted onto the Milky Way, which would mean that (1) their systemic velocities are not just affected by the NS natal kicks and (2) our systemic kick estimates listed above are invalid since these systems did not form in the Galactic thin disc on circular trajectories. Moreover, the two distinct populations that can produce post-SN systems in the detection range of \textit{Gaia}, from interacting and non-interacting binaries (see Figure \ref{Fig_Gaia_Progenitors}), have significantly different systemic kicks. In an attempt to reconcile the models and observations, we consider the systemic kick distribution of the \textit{Gaia} NSs where we omit the systems on halo-like orbits, and compare them to the model binaries that originate from tight, interacting binaries that went through a CE (which in practice corresponds to the systems with pre-SN orbital period $P_0<100\,$d, as shown in Figure \ref{Fig_Gaia_Progenitors}).

Figure \ref{Fig_Gaia_Velocities} shows that the systemic kick estimates of the \textit{Gaia} NSs on non-halo Galactic orbits are similar to the model predictions for $\Mfid$ and $\Mroc$ when omitting the systems that have wide pre-SN orbits. In our model, the orbits and systemic velocities of the \textit{Gaia} NSs can thus be explained by natal kicks similar to the ones of isolated NSs ($v_{\text{SSN}}=630$\,km\,s$^{-1}$) and a rocket kick of $30$\,km\,s$^{-1}$ ($\Mroc$). However, this conclusion depends on two critical assumptions.

\begin{figure*}
    \centering
    \includegraphics[width=18cm]{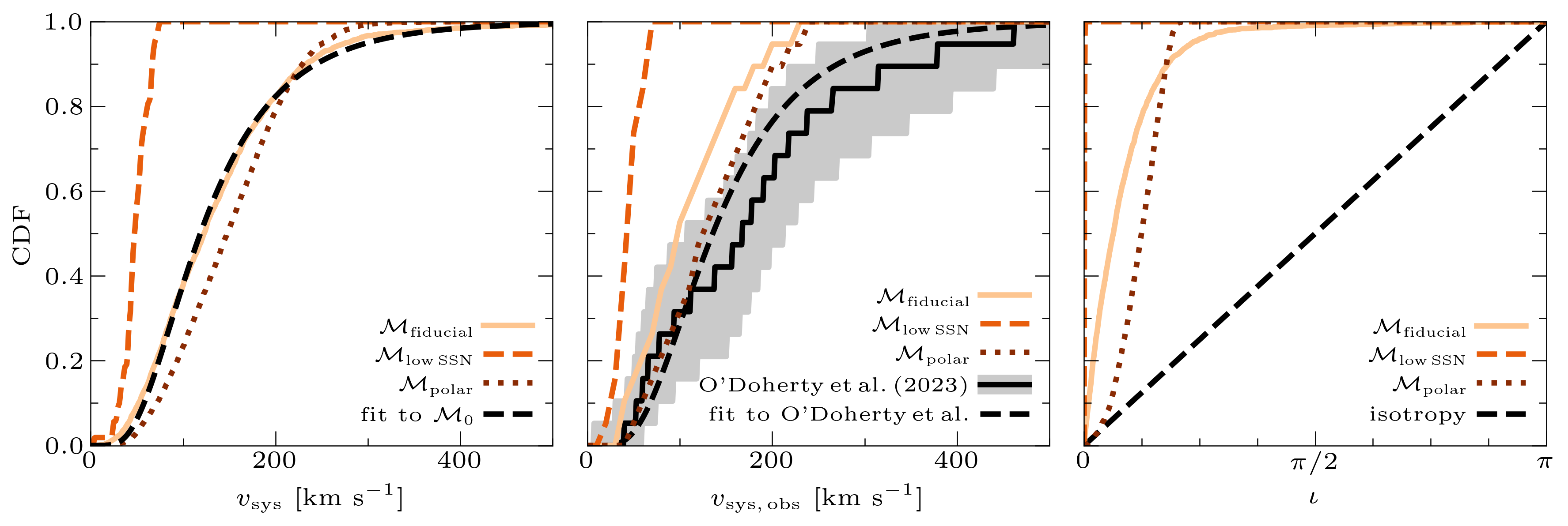}
    \caption{Systemic kicks of the LMXBs. The left panel shows the intrinsic kick distributions as predicted through our models $\Mfid$, $\Mlow$, and $\Mpol$ (light orange solid line, orange dashed line, and dark orange dotted line, respectively). We also fit a lognormal distribution to $\Mfid$ which results in $\mu=4.76$ and $\sigma=0.58$ (black dashed line). In the central panel we show the median distribution of the systemic kick estimates of \citet{ODoherty_2023}, together with its bootstrapped $95\%$ uncertainty region (black solid line and grey shaded region, respectively). A lognormal fit to these estimates yields $\mu=4.91(17)$ and $\sigma=0.54(14)$ (black dashed line). The orange lines show the model results from the left panel but corrected for kinematic bias following the method of \citet{Disberg_2026}, so that they are comparable to observations. The right panel shows the distribution of the orbital plane tilt $\iota$ induced by the SN (determined through Equation \ref{eq_iota}), together with an isotropic $\iota$ distribution (black dashed line).}
    \label{Fig_LMXB_Velocities}
\end{figure*}

Firstly, the default CE model in \lstinline{COMPAS} is the $\alpha{-}\lambda$ formalism \citep{Webbink_1984,DeKool_1990,COMPAS_2022a}, and this formalism determines the pre-SN orbits for the interacting binaries that go through a CE phase. However, as discussed in Section \ref{sec3.1}, if we instead use the two-stage CE formalism of \citet{Hirai_2022} we find that the Blaauw kicks become sufficient to produce systems in the \textit{Gaia} detection range, even with small natal kicks. In this case, the orbits for $\Mlowroc$ look indistinguishable from the $\Mroc$ results (cf.\ Figure \ref{Fig_Gaia_Orbits}). However, in the right panel of Figure \ref{Fig_Gaia_Velocities} we also show the systemic kicks of the $\Mlowroc$ results using the two-stage CE formalism, which are significantly lower than the results for the models with high natal kicks and seem difficult to reconcile with the observations.

Secondly, as discussed in Section \ref{sec3.1}, our method of determining systemic kicks neglects the effects of dynamical heating. As explained by \citet{Lacey_1985}, dynamical heating through velocity perturbations induced by gravitational interaction with structures in the Galaxy will cause a random walk in velocity space. This increases the peculiar velocities of objects on circular Galactic orbits, since peculiar velocities cannot be negative. However, if objects have already obtained a significant peculiar velocity, due to a kick, dynamical heating will add scatter to the velocity distribution but is unlikely to increase its median meaningfully. For this reason, the effects of dynamical heating will be greatest for $\Mlow$, which predicts the lowest systemic kicks. We also note that \citet{El-Badry_2024b} show that the velocities of (non-halo) \textit{Gaia} NSs overlap with the velocities of other \textit{Gaia} binaries in the thin and thick disc \citep[see also][]{Shariat_2026}.

In light of these caveats, it is difficult to decisively distinguish between high and low natal kicks or to conclude the presence of a rocket kick. Nevertheless, it appears that the current \lstinline{COMPAS} model can match the orbital properties and systemic velocities of \textit{Gaia} NSs if we assume high natal kicks combined with rocket kicks (through the $\Mroc$ model). Lastly, we note that \citet{Mapelli_2026} find that the black hole binaries observed by \textit{Gaia} can also be explained by natal kicks drawn from the kick distribution of \citet{Disberg_2025b}.

\section{Neutron Star Low-mass X-ray Binaries}
\label{sec4}
\noindent We also compare our model to the NS LMXB systemic kick estimates of \citet{ODoherty_2023}. These systems typically consist of an NS and a low-mass companion (${\lesssim}\,1M_{\odot}$) in a tight orbit $P\lesssim10\,$d \citep{Tauris_2006}, are relatively old \citep{Cowley_1987}, can reach vertical offsets from the Galactic disc of ${\gtrsim}\,1$\,kpc \citep{Jonker_2004}, and are observed through X-ray emission originating from mass transfer from the companion onto the NS \citep{VandenHeuvel_1975}. To estimate the NS LMXB systemic kicks, \citet{ODoherty_2023} collect a sample of 19 systems that have relatively accurate proper motion and distance estimates and then apply the method of (\citealt{Atri_2019}, see also \citealt{Zhao_2023}), which entails tracing the trajectories of these systems back through the Galaxy and estimating their peculiar velocities at disc crossings. This method assumes that the systems are formed in the Galactic thin disc, so \citet{ODoherty_2023} omitted systems associated with globular clusters, where LMXBs can form dynamically \citep{Pooley_2003}.

Similarly to \citet{Disberg_2026}, we compare the observed LMXBs to the simulated binaries that stay bound after the SN, have post-SN orbital periods shorter than $10$\,d, and have companions with masses below $1M_{\odot}$. In Figure \ref{Fig_LMXB_Velocities} we show the systemic kicks of these simulated systems for the kick models $\Mroc$, $\Mlow$, and $\Mpol$ (since we do not consider the effects of rocket kicks on systemic velocities). The higher natal kicks lead to higher systemic kicks, while polar kicks ($\Mpol$) cause a slightly higher median systemic kick than isotropic kicks ($\Mfid$). We note that for $\Mfid$, the pre-SN orbital periods are ${\lesssim}\,1$\,d.

We applied the method of \citet{Disberg_2026} to correct these intrinsic kick distributions for kinematic bias and compare them to the systemic kick estimates of \citet{ODoherty_2023}. As Figure \ref{Fig_LMXB_Velocities} shows, the models that assume high SSN natal kicks result in systemic kicks that are comparable to the observed distribution. The main difference is the fact that the observations contain kicks above ${\sim}\,250$\,km\,s$^{-1}$ whereas the models do not predict binaries with systemic kicks in this region. In part, this is caused by the projection effects present in the observational distribution that were caused by assuming isotropy to estimate the radial velocities \citep{ODoherty_2023}. The low SSN natal kicks ($\Mlow$), in contrast, result in systemic kicks significantly lower than observed. This difference is likely larger than model uncertainties.  We therefore conclude that we need high natal kicks to explain the observed LMXB systemic velocities.

\begin{figure*}
    \centering
    \includegraphics[width=18cm]{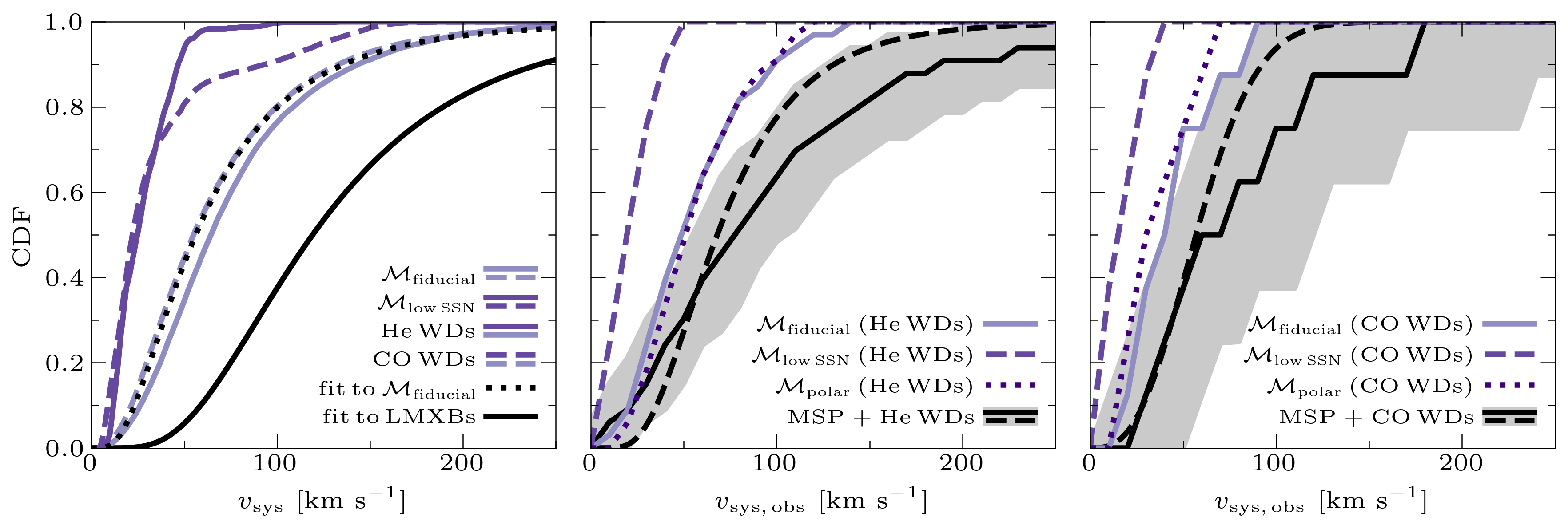}
    \caption{Systemic kicks of the NSWDs. The left panel shows the intrinsic kick distributions as predicted through our models $\Mfid$ and $\Mlow$ (light purple lines and purple lines, respectively). We separately show the model results for He WD progenitor companions (solid lines) and CO WD progenitor companions (dashed lines), and fit a lognormal distribution to the $\Mfid$ results of all NSWDs (of which ${\sim}\,85\%$ have CO WD companions), resulting in $\mu=4.03$ and $\sigma=0.69$. We also display the lognormal fit to the $\Mfid$ LMXB systemic kicks (black solid line, see also Figure \ref{Fig_LMXB_Velocities}). In the central panel we show the median distribution of the kinematically constrained kick estimates for the MSPs in our sample with He WD companions, together with its bootstrapped $95\%$ uncertainty region (black solid line and grey shaded region, respectively). A lognormal fit to these estimates yields $\mu=4.22(14)$ and $\sigma=0.51(13)$ (black dashed line). The purple lines show the model results for $\Mfid$, $\Mlow$, and $\Mpol$ (light purple solid line, purple dashed line, and dark purple dotted line, respectively) but corrected for kinematic bias following the method of \citet{Disberg_2026}. The right panel shows the model results compared to the systemic kick estimates for the MSPs in our sample with CO WD companions. Since this sample is not big enough to constrain a lognormal fit, we instead use a Maxwellian distribution to describe these estimates, which only has one parameter. The black dashed line shows the Maxwellian fit to the kick estimates, described by the scale parameter $\sigma=37(8)$\,km\,s$^{-1}$.}
    \label{Fig_NSWD_Velocities}
\end{figure*}

In Figure \ref{Fig_LMXB_Velocities} we also show the inclinations induced by the SN relative to the pre-SN orbits. The small natal kicks do not cause noticeable inclinations, because Blaauw kicks are restricted to the orbital plane. The higher natal kicks cause median inclinations of ${\sim}\,15\degree$. Constraining the kick directions to be polar increases the median inclination, but the largest inclinations are part of the model with isotropic kicks. This is due to the fact that the largest inclinations are caused by post-SN polar velocities, but the post-SN velocities are a combination of the natal kick and pre-SN velocity (as defined in Equation \ref{eq_v_prime_vec}), meaning that polar kicks do not lead to polar post-SN velocities.

\section{Neutron Star--White Dwarf Binaries}
\label{sec5}
\noindent The third observational sample that we compare our model to is NSWDs. These systems consist of an MSP \citep{Alpar_1982} with a WD companion, which is either a He WD or a CO WD. We curate this sample by selecting pulsars from the ATNF Pulsar Catalog v2.7.0 \citep{Manchester_2005} that have (1) a WD companion, (2) proper motion estimates, (3) parallax estimates with a fractional uncertainty less than $50\%$, (4) no association with a globular cluster, and (5) a rotational period below $50$\,ms. This selection resulted in $33$ MSPs with a He WD companion and $8$ MSPs with a CO companion. The classification of a pulsar companion as a He or CO WD can be based on mass estimates and binary orbital periods \citep[e.g.,][]{Lorimer_1995,Graikou_2017,Wang_2025} or optical detections of the WD \citep[e.g.,][]{Swiggum_2017,Stovall_2019,Zyuzin_2019}. It is thought that MSPs + He WD systems evolve from LMXBs whereas MSPs + CO WD systems evolve from intermediate-mass X-ray binaries \citep{Tauris_2012}. 

While the periods and eccentricities of NSWDs might harbor information about the NS natal kicks, their orbits are likely affected by post-SN binary interaction with the WD progenitor and are expected to be circularized. However, it is conceivable that in a certain number of systems the pre-SN mass transfer causes a mass ratio reversal, resulting in the WD being formed before the NS and the present-day orbits being comparable to our model predictions. Although there are indeed NSWDs in the ATNF Pulsar Catalog with orbits that are significantly eccentric, we find that the majority of these systems (1) are associated with a globular cluster where they might have formed dynamically, and (2) harbor a (recycled) MSP which is a sign of post-SN mass transfer. Because of this, we do not consider the NSWD binary orbits reliable constraints on the NS natal kicks and limit our analysis to their systemic kicks, which we do not expect to be significantly affected by the WD formation.

Similarly to the LMXBs and \textit{Gaia} NSs, we use the method of \citet{Disberg_2024b,Disberg_2025a} to kinematically constrain the NSWD systemic kicks, where we differentiate between He WDs and CO WDs. We compare these observational estimates to our model, where we select the simulated binaries that stay bound after the SN and have a post-SN orbital period ${\leq}\,10$\,d (assuming these systems go through an LMXB-like phase in which the pulsar is recycled). Then, we differentiate between systems in which the companion ends up as a He WD or a CO WD according to the \lstinline{COMPAS} model, which in practice is approximately equivalent to selecting companions with zero-age main-sequence (ZAMS) masses of ${<}\,2M_{\odot}$ and ${<}\,6M_{\odot}$, respectively. Approximately $85\%$ of the selected simulated systems have a CO WD companion (as opposed to a He WD), whereas the majority of observed systems have He WD companions. It is not clear whether this is due to an observational bias or a tension between the \lstinline{COMPAS} model and the observations.

In Figure \ref{Fig_NSWD_Velocities} we show the intrinsic systemic kick distributions of the simulated binaries. In general, the CO WD progenitors are more massive which means that the systems obtain lower systemic kicks, although this difference is relatively small. However, in these systems the SN progenitor also tends to be more massive compared to the He WD systems, leading to increased Blaauw kicks. This is the reason why for $\Mlow$ the CO WD systemic kicks exceed the He WD systemic kicks. We also show the fitted distribution to the $\Mfid$ LMXB kicks from Figure \ref{Fig_LMXB_Velocities}, which exceeds the $\Mfid$ NSWD kicks. This is likely due to the fact that we model the observed LMXBs as systems with companion masses below $1M_{\odot}$, which leads to higher systemic kicks \citep[see Appendix D of][]{Disberg_2026}. The NSWDs, and particularly the He WD progenitors, have likely gone through an LMXB phase as well, but these types of LMXBs are less likely to be observed due to the fact that for higher mass companions the mass transfer that causes the X-ray emission lasts a shorter amount of time \citep[e.g.,][]{Tauris_2006}. 

\begin{figure*}
    \centering
    \includegraphics[width=18cm]{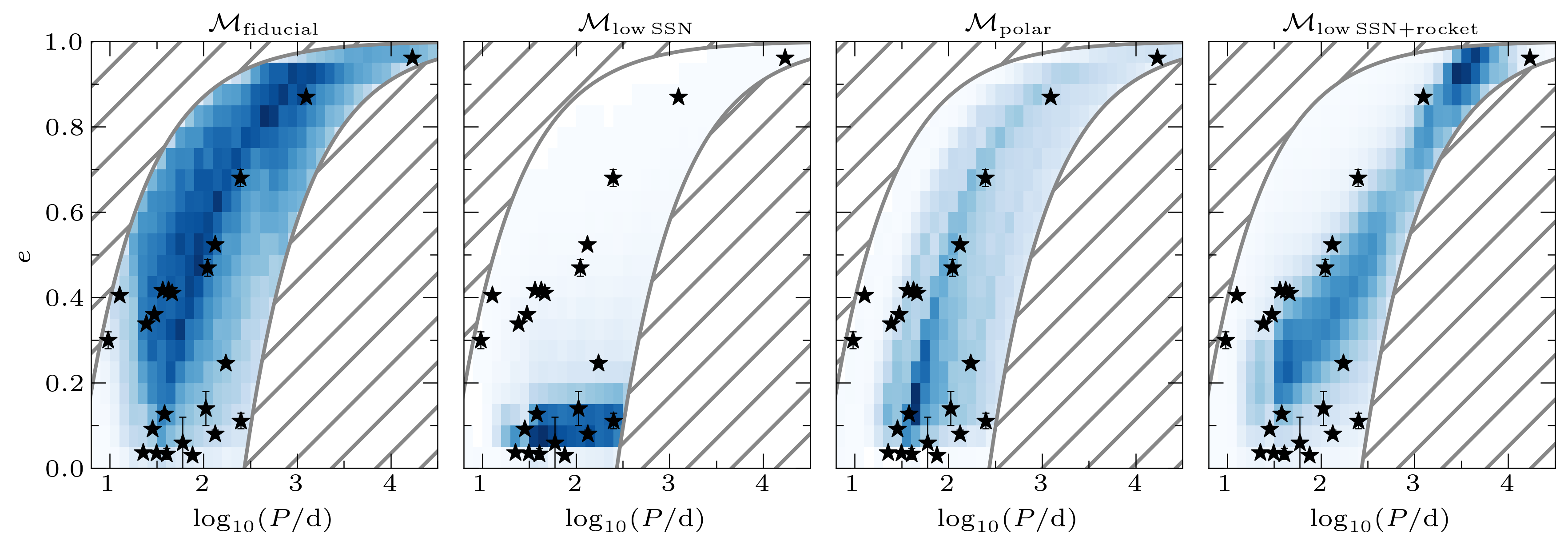}
    \caption{Binary orbits of the BeXBs (black stars) in the sample of \citet{Valli_2025}. The grey hashed region corresponds to periastron distances below $30R_{\odot}$ or above $450R_{\odot}$ (assuming stellar masses of $1.4M_{\odot}$ and $15M_{\odot}$), in which they argue BeXBs are not observable. The blue distributions show the density of the simulated post-SN binaries as a function of the orbital period $P$ and the eccentricity $e$. These distributions depend on the different kick models $\mathcal{M}$ (Table \ref{tab_models}), and are re-normalized for each panel.}
    \label{Fig_BeXB_Orbits}
\end{figure*}

Figure \ref{Fig_NSWD_Velocities} also shows the systemic kick estimates for the pulsar systems in our sample. The observed MSPs with He WD companions have slightly higher kick estimates than the CO WD counterparts, since they have lower mass companions and also have slightly larger observed distances. The model predictions for high natal kicks ($\Mfid$) are relatively consistent with the observations. Although the model predicts slightly lower systemic kicks, the difference is smaller than the expected model uncertainties. The results for small natal kicks are less consistent with the observations, but we note that for these systems it may be possible to reconcile the $\Mlow$ results and the NSWDs through model uncertainties or effects such as dynamical heating. Nevertheless, the high natal kicks in the $\Mfid$ model paint a consistent picture between the LMXBs and the NSWDs. We note that despite the high natal kicks the systemic kicks can be relatively small, with ${\sim}\,10\%$ below $25\,$km\,s$^{-1}$, which is compatible with observed MSPs with low systemic velocities \citep[e.g.,][]{Zhang_2025}. Lastly, for the NSWDs the polar kicks in $\Mpol$ are practically indistinguishable from the isotropic kicks in $\Mfid$.

\section{Neutron Star High-mass X-ray Binaries}
\label{sec6}
\noindent The orbits and systemic velocities of NS-harboring HMXBs (and in particular BeXBs), which have companions with masses above ${\sim}\,10M_{\odot}$, also provide clues about the NS natal kicks. Based on HMXB kinematic ages and associations with recent star formation regions \citep{Shtykovskiy_2005,Tauris_2006,Bodaghee_2012,Antoniou_2016,Vinciguerra_2020,Fortin_2022a}, and Be star lifetimes \citep{Fabregat_2000,McSwain_2005,Ekstrom_2008}, we assume that most of these systems are likely young enough that their present-day kinematic properties are still representative of their birth properties. Below, we analyze the orbital periods and eccentricities of BeXBs (Section \ref{sec6.1}), discuss their progenitor systems (Section \ref{sec6.2}), and consider HMXB systemic velocities (Section \ref{sec6.3}).

\subsection{BeXB Orbits}
\label{sec6.1}
\noindent Binary systems called Be X-ray binaries are NSs with rapidly rotating Be star companions. The decretion discs of Be stars \citep[e.g.,][]{Rivinius_2013} can interact with the NS, resulting in X-ray emission. \citet{Valli_2025} curate a BeXB sample, selecting systems with (1) a classical Be star that is not expected to be close to filling its Roche lobe, (2) evidence for the presence of an NS through X-ray pulsations, (3) a location within the Galaxy, and (4) accurate orbital parameter estimates. This selection resulted in 23 BeXB systems with period and eccentricity estimates. In their BeXB sample, \citet{Valli_2025} find two distinct groups: one group consists of low-eccentricity systems ($e<0.25$) and the other group contains eccentric systems that show a narrow relation between period and eccentricity \citep[cf.][]{Pfahl_2002c,Oller_2026}. We compare the BeXB observations with the model results of simulated binaries with companion masses between $10M_{\odot}$ and $20M_{\odot}$ \citep{Fortin_2022b}, that stay bound after the SN. Also, we follow \citet{Valli_2025} in only considering systems with periastron distances between $30R_{\odot}$ and $450R_{\odot}$, since they argue that outside this range the system is unlikely to produce observable X-rays from an interaction between the NS and the decretion disc. All selected binaries will have experienced mass transfer from the NS progenitor onto the Be star, plausibly responsible for spinning up the Be stars.

In Figure \ref{Fig_BeXB_Orbits}, we show the BeXB period and eccentricity observations together with the predictions for $\Mfid$, $\Mlow$, $\Mpol$, and $\Mlowroc$ models, where we note that the results for $\Mroc$ are similar to those for $\Mfid$. The figure shows that large SSN natal kicks ($\Mfid$) cannot explain the observations: the simulated binaries are scattered over the parameter space and do not align with the observed eccentricities. For example, we find that when randomly drawing $12$ objects from the model distribution, the probability of finding a period-eccentricity relation as tight as observed is negligibly low. However, we note that \citet{Rocha_2024} show that it is possible to constrain this distribution further by considering selection effects beyond the periastron limits. For lower natal kicks ($\Mlow$), the simulated results align well with the low-eccentricity BeXBs \citep[as also found by][]{Valli_2025}. In this model, the SN progenitors have a typical mass of $2.5M_{\odot}$ and are in systems with orbital periods mostly between $10\,$d and $300\,$d.

For polar kicks ($\Mpol$) the post-SN systems have a relatively narrowly peaked periastron distribution, but this is still difficult to reconcile with the observations. Similarly, rocket kicks ($\Mlowroc$) cause some wider orbits to become more eccentric, resulting in a period-eccentricity relation comparable to the observations, where we note that a slightly higher rocket kick of $\Delta v_{\text{roc}}=40\,$km\,s$^{-1}$ makes the distribution align better with the observations. However, there is still significantly more scatter in these results than in the observations, and if one considers a scatter in $\Delta v_{\text{roc}}$ instead of the single fixed value in our model, this would increase the eccentricity scatter even further. We therefore conclude that our model cannot explain the BeXBs with $e>0.25$ and their tight period-eccentricity relation.

In their analysis, \citet{Valli_2025} fit distributions of the pre-SN orbital period, mass loss, kick magnitude, and kick direction to the low-eccentricity and high-eccentricity BeXBs. They find that the most-likely progenitors for the low-eccentricity BeXBs obtain isotropic kicks with a mean of ${\sim}\,7\,$km\,s$^{-1}$ and lose approximately $1M_{\odot}$ in the SN, which aligns well with our $\Mlow$ model. Moreover, they show that the period-eccentricity relation for the high-eccentricity BeXBs can be reproduced by a relatively narrow kick distribution around ${\sim}\,100\,$km\,s$^{-1}$, where the kicks are almost perfectly polar relative to the pre-SN orbit, and a mass loss of ${\sim}\,3.5M_{\odot}$.

However, if the kick is truly one-dimensional (polar), as their model suggests, it is difficult to imagine what process would yield a bimodal kick distribution with sharply peaked kicks either aligned or anti-aligned with the orbital angular momentum. Moreover, they fix the Be star mass in their model to be $15M_{\odot}$, but in the observations the Be stars have masses approximately between $10M_{\odot}$ and $20M_{\odot}$ as inferred from their spectral types \citep{Fortin_2022b}, which increases the scatter in eccentricity as a function of period. Finally, \citet{Valli_2025} suggest that the ${\sim}\,100$\,km\,s$^{-1}$ kicks align with the lower mode in the bimodal kick distributions of \citet{Verbunt_2017} and \citet{Igoshev_2020}, but \citet{Disberg_2025b} show that the bimodality in these distributions is likely statistical and not physical.

\subsection{BeXB Progenitors}
\label{sec6.2}
\begin{figure}
    \centering
    \resizebox{\hsize}{!}{\includegraphics{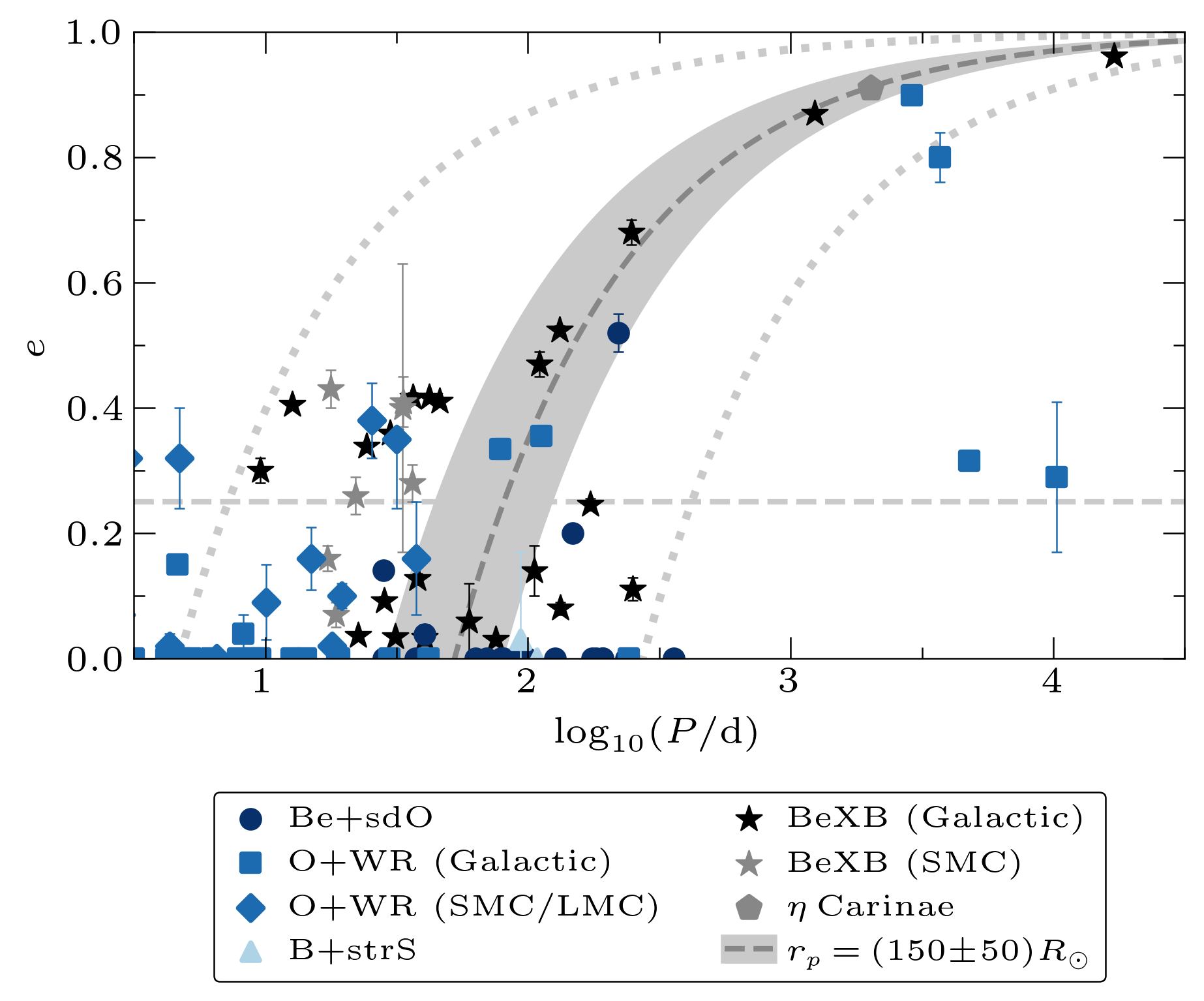}}
    \caption{Systems comparable to BeXB progenitors, combining the listed samples of \citet{Valli_2025} and \citet{VanSon_2026}: Be+sdO (dark blue circles), Galactic O+WR (blue squares), O+WR systems in the SMC and LMC (blue diamonds), and B+strS (light blue triangles), together with Galactic BeXBs \citep[black stars,][]{Valli_2025} and BeXBs in the SMC \citep[grey stars,][]{Coe_2015} and $\eta$ Carinae \citep[grey pentagon,][]{Grant_2020}. We also show constant periastron distances ($r_p$) of $(150\pm50)R_{\odot}$ (grey dashed line and light grey shaded region). The light grey dashed line at $e=0.25$ divides the low-eccentricity and the high-eccentricity BeXBs and the light grey dotted lines correspond to the minimum and maximum BeXB periastron distances of $30R_{\odot}$ and $450R_{\odot}$, respectively. For the periastron distances we assume stellar masses of $1.4M_{\odot}$ and $15M_{\odot}$.}
    \label{Fig_BeXB_Progenitors}
\end{figure}
\noindent One additional critical assumption in the analysis of \citet{Valli_2025} is that the pre-SN orbits are circular. In order to justify this assumption, they look at systems consisting of a massive star with a possible NS progenitor companion. In particular, they collect a sample of binaries consisting of a Be star with a type-O subdwarf companion \citep[Be+sdO,][]{Peters_2008,Peters_2013,Mourard_2015,Chojnowski_2018,Gies_2020,Shenar_2020,Harmanec_2022,Klement_2024,Muller-horn_2025}, an O star with a Wolf-Rayet companion \citep[O+WR,][]{North_2007,Chevrotiere_2011,Richardson_2021,Thomas_2021}, or a B star with a stripped-star companion \citep[B+strS,][]{Villasenor_2023,Ramachandran_2024}, since these systems are kinematically comparable to BeXB progenitors. They argue that it is justified to consider the pre-SN orbit to be circular since the majority of these systems show no sign of significant eccentricity. 

However, several of these systems do have significant eccentricity, comparable to the BeXBs. In Figure \ref{Fig_BeXB_Progenitors} we show the progenitor-like systems listed by \citet{Valli_2025}, where we expand their O+WR sample with the sample of \citet{VanSon_2026}, which contains Galactic systems as well as systems from the Large Magellanic Cloud (LMC) and Small Magellanic Cloud (SMC). Although most of the massive stars in this O+WR sample are indeed O-type stars, we note that two are early B-type. The figure shows that within the parameter space where BeXBs are observable (i.e., periastron distances below $450R_{\odot}$) there are several systems with $e>0.25$ and they align relatively well with the BeXB period-eccentricity relation. If there are BeXB progenitors that already follow this relation, one does not need natal kicks to explain the eccentric BeXBs. However, even though the eccentric progenitor systems in the BeXB parameter space seem to be relatively consistent with this relation, it is difficult to draw strong conclusions due to the low number of observations. Also, one might argue that the majority of these progenitor-like systems are circular while approximately half of the BeXBs are eccentric, but we note that observational biases for the progenitors and the BeXBs differ, complicating comparisons.

As Figure \ref{Fig_BeXB_Progenitors} shows, the eccentric progenitor systems and the BeXBs with $e>0.25$ and $\log_{10}(P/\text{d})\gtrsim2$ all have similar periastron distances of ${\sim}\,150R_{\odot}$, with the exception of WR 19 and WR 137 ($e<0.4$ and $P>1000\,$d) which lie outside the BeXB parameter space. We suggest that a possible merger history in these systems can cause pre-SN eccentricity, and that this may induce a periastron bias. According to this hypothetical explanation, a fraction of the BeXB progenitor systems are formed in triple systems, which can be unstable and cause two stars to merge. The resulting binary system, then, is likely to have a non-zero eccentricity, which remains relatively unaffected by the SN if the natal kicks are weak.

In our hypothesis, the Be stars in high-eccentricity BeXBs were thus created through stellar mergers. The B[e] star R4, located in the SMC, is an example of such a massive star that is thought to be formed through a stellar merger \citep{Pasquali_2000,Podsiadlowski_2006,Wu_2020}. In Figure \ref{Fig_BeXB_Progenitors} we show $\eta$ Carinae \citep{Grant_2020}, which is also theorized to have gone through a merger-in-a-triple \citep{Hirai_2021}, and this system indeed aligns well with the high-eccentricity BeXBs. The post-merger periapsis distribution may stem from the requirement for the third star to be sufficiently close to make the hierarchical triple unstable and drive the inner binary toward merger. However, if stellar mergers produce massive stars with strong magnetic fields \citep{Schneider_2019}, these may rapidly spin down, inconsistent with the Be phenomenon \citep{Rivinius_2013}. 

Although we pose the hypothetical merger-in-a-triple scenario as a possible explanation for the high-eccentricity BeXBs, we stress that the observational data on the progenitor systems are limited, making it difficult to draw decisive conclusions. However, the systems displayed in Figure \ref{Fig_BeXB_Progenitors} show that BeXB progenitors may have non-negligible pre-SN eccentricity, meaning that it is not obvious that the period-eccentricity relation is caused by NS kicks. Indeed, \citet{VanSon_2026} show that non-zero eccentricities are ubiquitous among post-mass-transfer binary systems \citep[see also][]{Parkosidis_2026}. Because of this, a kick model that can explain the high-eccentricity BeXBs may need to account for pre-SN eccentricity, and both our model and the model of \citet{Valli_2025} are thus likely insufficient to explain these systems. Nevertheless, in Appendix \ref{appD} we show that, if the period-eccentricity relation is present pre-SN, the low natal kicks in $\Mlow$ can preserve this relation post-SN, because the Blaauw kick does not alter the orbit significantly due to the massive companion star, while high natal kicks would destroy a pre-SN period-eccentricity relation (see the $\Mfid$ results in Figure \ref{Fig_BeXB_Orbits}). If the period-eccentricity relation is indeed present pre-SN, the $\Mlow$ model can thus explain the BeXBs in the sample of \citet{Valli_2025}.

\subsection{HMXB Velocities}
\label{sec6.3}
\noindent One way to distinguish between our $\Mlow$ model and the polar kick model of \citet{Valli_2025} is through the effects they have on the systemic velocities of BeXBs. In particular, the fact that \citet{Valli_2025} predict significantly more mass loss will increase the strength of the Blaauw kick and therefore the systemic kick of short-period systems will be greater. To estimate the magnitude of this effect, we consider a range of pre-SN orbital periods and apply polar natal kicks of $5\,$km\,s$^{-1}$ and $100\,$km\,s$^{-1}$ and then compute the post-SN period and the systemic kick, assuming an NS mass of $1.5M_{\odot}$, a mass loss of $3.5M_{\odot}$, a Be star mass of $15M_{\odot}$, and no pre-SN eccentricity. We also compute the systemic kicks for orbits where the pre-SN eccentricity follows our polynomial fit to the high-eccentricity BeXBs as shown in Appendix \ref{appD}, assuming the same stellar masses but a mass loss of $1.0M_{\odot}$ and a kick of $5\,$km\,s$^{-1}$.

\begin{figure}
    \centering
    \resizebox{\hsize}{!}{\includegraphics{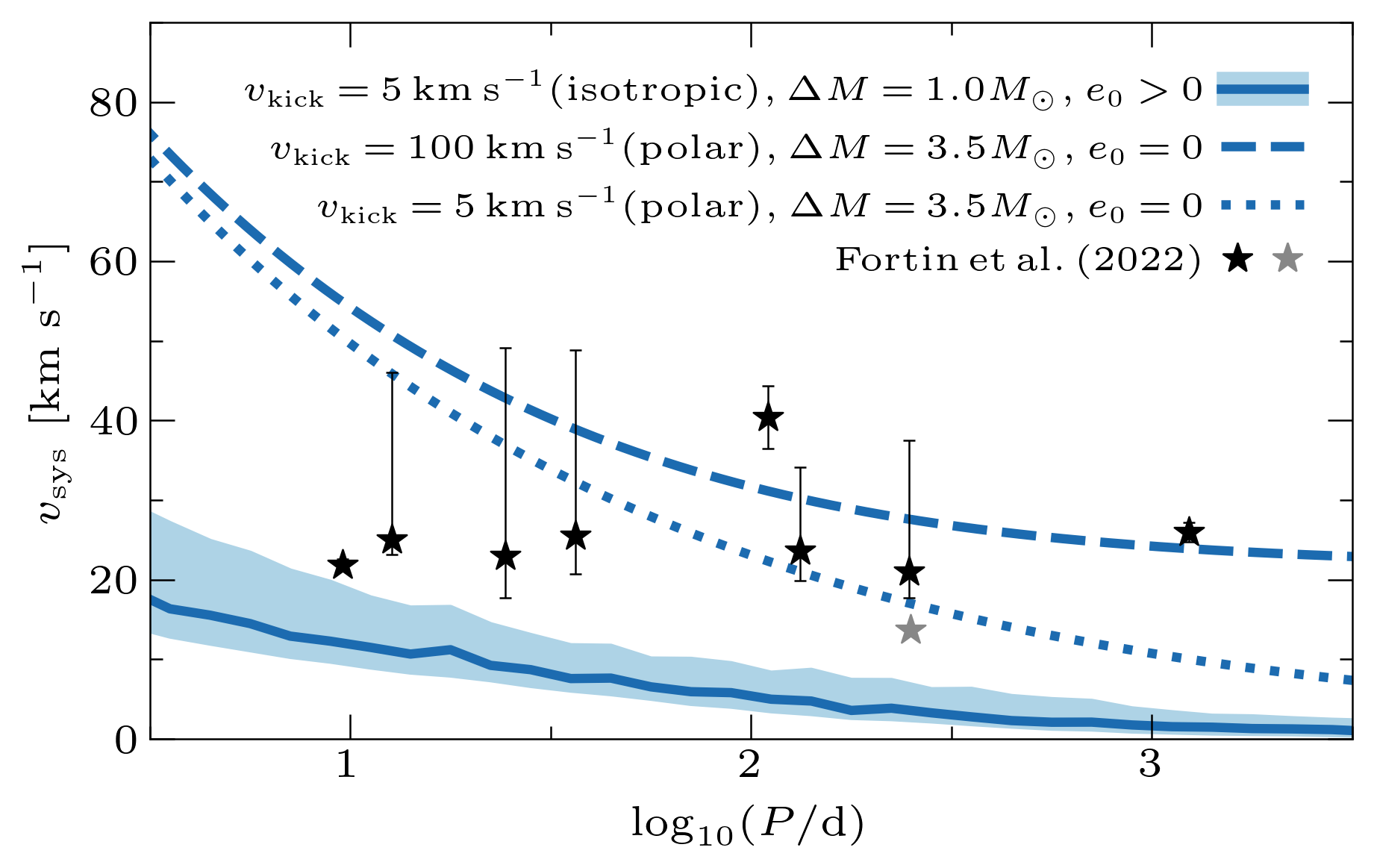}}
    \caption{Systemic velocity estimates of the BeXBs in the HMXB sample of \citet{Fortin_2022b}, as a function of the observed orbital period, where we make a distinction between high-eccentricity and low-eccentricity systems (black and grey stars, respectively) similarly to \citet{Valli_2025}. We show predictions for the systemic kicks of initially circular binaries ($e_0=0$) that experience a mass loss of $\Delta M=3.5M_{\odot}$ and polar natal kicks of $5$\,km\,s$^{-1}$ (dotted line) and $100$\,km\,s$^{-1}$ (dashed line), where the latter is similar to the model of \citet{Valli_2025}. These predictions assume an NS mass of $1.5M_{\odot}$ and a Be star mass of $15M_{\odot}$. We also compute the systemic kicks for systems with kicks of $5$\,km\,s$^{-1}$ and a mass loss of $\Delta M=1.0M_{\odot}$, where the pre-SN eccentricities follow our polynomial fit to the high-eccentricity BeXBs (see Appendix \ref{appD}), and display the median and $68\%$ confidence interval of the resulting distribution for a given period (blue solid line and light blue region, respectively).}
    \label{Fig_BeXB_Velocities}
\end{figure}

In Figure \ref{Fig_BeXB_Velocities} we show the relationship between post-SN orbital period and systemic kick for these different estimates, together with the systemic velocity estimates of \citet{Fortin_2022b}, as also done by \citet{Valli_2025}. The systemic velocities are dominated by the Blaauw kick due to the massive companions, and for shorter orbital periods the systemic kicks become increasingly sensitive to the mass loss. For $\Delta M=3.5M_{\odot}$, which is similar to the model of \citet{Valli_2025}, there is a steep increase in systemic kick towards shorter orbital periods. However, the systemic velocity estimates of the observed BeXBs show no evidence of this period dependence. Nevertheless, it is possible that there is indeed a significant amount of mass loss, but that in shorter-period systems the NS progenitor is more strongly stripped, negating the period-dependence of the systemic kick. The observed velocities exceed the predictions from the low-mass-loss model by ${\sim}\,10$--$30$\,km\,s$^{-1}$, but this might be accounted for by systematic uncertainties in the velocity estimates, for example due to the fact that the pre-SN systemic velocity is not necessarily zero \citep{Fortin_2022b}.

We therefore conclude that the high-eccentricity BeXB velocities show no evidence of the significant SN mass loss as proposed in the model of \citet{Valli_2025}. However, \citet{Valli_2025} show that if one considers scatter in their model predictions and extends the uncertainty of the velocity estimates by $15\,$km\,s$^{-1}$ in order to account for pre-SN velocity \citep{Fortin_2022b}, then the uncertainties overlap with the model. This means that it is difficult to decisively say that the observations are inconsistent with the strong Blaauw kicks predicted by \citet{Valli_2025}, due to the small sample size.

The velocities shown in Figure \ref{Fig_BeXB_Velocities} mainly correspond to the high-eccentricity BeXBs, but there is also one velocity estimate for a low-eccentricity BeXB. This system has a slightly lower velocity than the others, but the difference is small (${<}\,10$\,km\,s$^{-1}$) and there is only one low-eccentricity BeXB velocity estimate so this is not statistically significant (although \citealt{Shariat_2026} show that this difference is also visible in an expanded sample of BeXB systemic velocities). However, if we assume that high-eccentricity BeXBs have slightly higher systemic velocities, this could be compatible with the merger-in-a-triple scenario. It is conceivable that the stellar merger causes mass to be ejected from the system and if this happens anisotropically and on a short enough timescale, the system receives a Blaauw kick.

If a post-merger system consists of a $10M_{\odot}$ NS progenitor and a $10M_{\odot}$ B star in a $100\,$d orbit such that $v_{\text{orb}}\approx150\,$km\,s$^{-1}$, where ${\sim}\,10\%$ of the B star mass has been ejected (similar to the estimate of \citealt{Lombardi_2002} for gentle mergers in globular clusters) with a velocity equal to the post-merger $v_{\text{orb}}$, then the systemic kick will be on the order of $5-10$\,km\,s$^{-1}$ by conservation of linear momentum. This value will be lower if the mass is ejected more isotropically and higher if there is more mass ejected or the pre-SN relative velocity is greater than the post-merger $v_{\text{orb}}$ due to eccentricity. Because of this, a merger origin can in theory cause relatively high pre-SN systemic velocities, which might explain the velocity of ${\sim}\,40\,$km\,s$^{-1}$ of the BeXBs at $P\approx100\,$d.

\begin{figure}
    \centering
    \resizebox{\hsize}{!}{\includegraphics{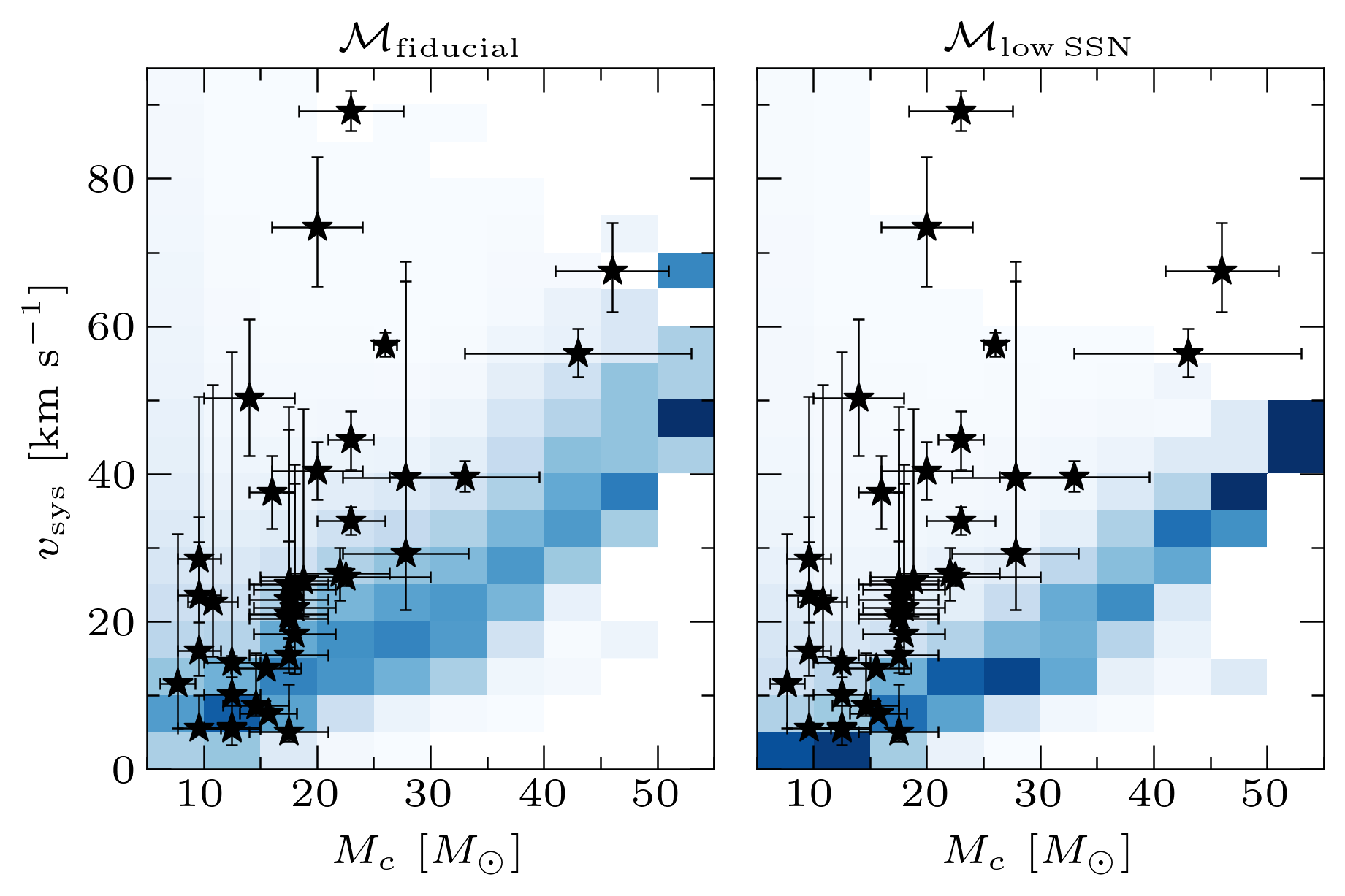}}
    \caption{Systemic velocity estimates of the HMXBs in the sample of \citet{Fortin_2022b}, as a function of the companion mass (black stars). This sample includes NS BeXBs as well as NSs in X-ray binaries with Oe-type companions and supergiant companions. The blue distribution shows the model results for $\Mfid$ and $\Mlow$, where the columns are normalized individually such that the distribution corresponds to $p(v_{\text{sys}}|M_{c})$. The model results show all simulated systems that survive the SN, but we note that these results are not sensitive to a limit on the post-SN orbital period.}
    \label{Fig_BeXB_v-M}
\end{figure}

We also consider the systemic velocities of other HMXBs, as estimated by \citet{Fortin_2022b}. Besides BeXBs, they also estimate the systemic velocities of X-ray binaries with Oe-type companions and supergiant companions. In Figure \ref{Fig_BeXB_v-M} we show their velocity estimates as a function of the companion mass, together with our model results for $\Mfid$ and $\Mlow$. The observed velocities increase for higher companion mass \citep[see also][]{Nuchvanichakul_2025}, due to the fact that more massive companions tend to have accreted mass from more massive primary stars, leading to more massive NS progenitor stars and thus more mass loss and a higher Blaauw kick. Indeed, our model results show that the systemic kicks of these systems with high-mass companions are dominated by the Blaauw kick (as also shown in Figure \ref{Fig_BeXB_Velocities}). The natal kicks present in $\Mfid$ only induce a relatively small scatter on the systemic kicks, as described in Equation \ref{eq_v_sys}. 

\begin{figure*}
    \centering
    \includegraphics[width=18cm]{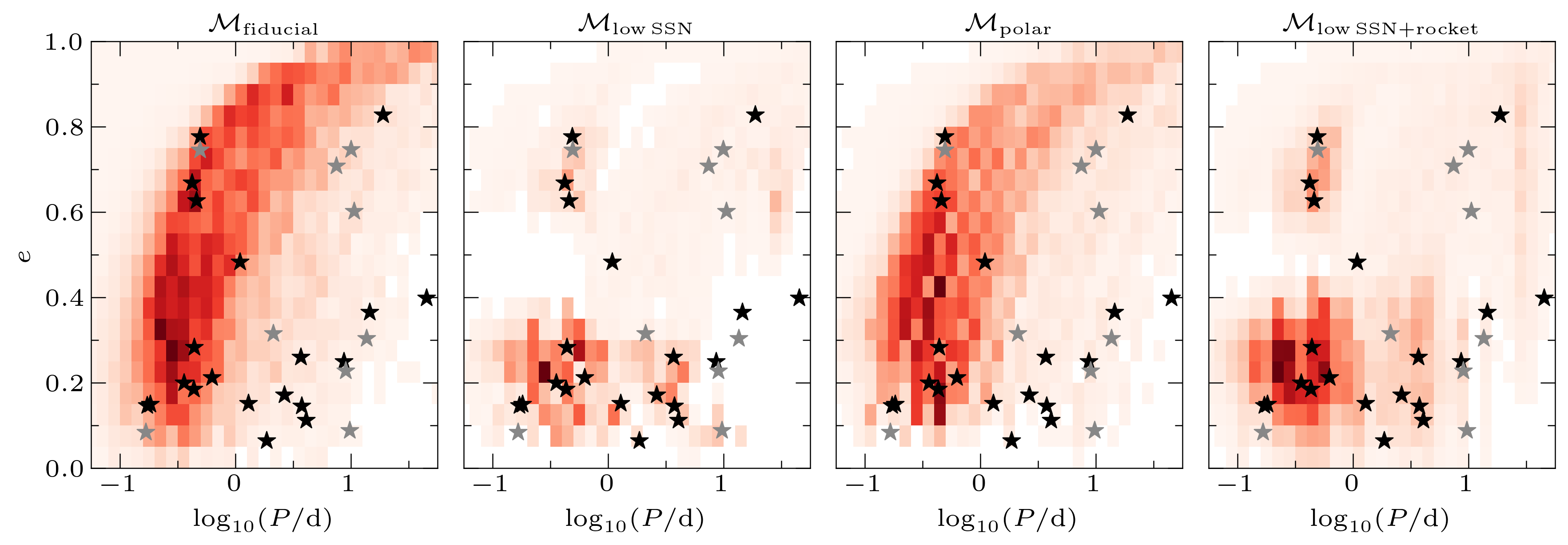}
    \caption{Binary orbits of the DNSs in the sample of \citet{Grichener_2026}, including confirmed DNSs (black stars) and DNSs that are either not confirmed to be DNSs or have a globular cluster association (grey stars). The periods and eccentricities are calculated by taking the observed values and integrating the orbits back in time to account for the effects of GW emission \citep[following][]{Peters_1964}, for a duration equal to the pulsar characteristic age. The red distributions show the density of the simulated post-second-SN binaries as a function of the orbital period $P$ and the eccentricity $e$, which are weighted by the observation probability defined in Equation \ref{eq_P_obs}. These distributions depend on the different kick models $\mathcal{M}$ (Table \ref{tab_models}), and are separately normalized in each panel.}
    \label{Fig_DNS_Orbits}
\end{figure*}

The results for $\Mfid$ and $\Mlow$ both show increased systemic kicks towards higher companion masses, similar to the observations. However, the observed velocities exceed the model predictions. In particular, there are two systems at $v_{\text{sys}}>70\,$km\,s$^{-1}$ \citep[with an additional system found by][]{Nuchvanichakul_2025} that seem difficult to explain with our models. This could indicate that \lstinline{COMPAS} is underestimating the mass loss in these systems, where a small change in mass loss affects the systemic kicks noticeably (as also shown in Figure \ref{Fig_BeXB_Velocities}). Moreover, it might be possible that some of these systems indeed have a merger-in-a-triple origin where they obtained a systemic kick due to asymmetric mass loss, or were dynamically ejected from the cluster in which they were formed \citep[e.g.,][]{Fujii_2011,Oh_2016}. These model uncertainties are likely greater than the scatter induced by natal kicks, making it difficult to detect the presence of natal kicks based on the systemic velocities and differentiate between $\Mfid$ and $\Mlow$.

\section{Double Neutron Stars}
\label{sec7}
\noindent The final sample we use in our comparison consists of DNSs, which have experienced two SNe in order to form two NSs. We use the DNS sample of \citet{Grichener_2026}, which combines the ATNF Catalog \citep{Manchester_2005}, the list of \citet{Chattaraj_2026}, and the works of \citet{Jacoby_2006}, \citet{Barr_2024}, \citet{Zhao_2024}, and \citet{Yang_2026}. In this sample there are $21$ DNSs and $8$ systems for which there is either a globular cluster association or discussion about whether it is actually a DNS or rather an NS with a WD or black hole companion. We compare our model to the observed DNS orbital periods and eccentricities (Section \ref{sec7.1}), systemic kicks (Section \ref{sec7.2}), and masses (Section \ref{sec7.3}).

\subsection{DNS Orbits}
\label{sec7.1}
\noindent We consider the binary orbits of the observed DNSs in the sample of \citet{Grichener_2026}, who determine the post-SN orbital periods and eccentricities by integrating the observed values back in time to correct for the effects of GW emission \citep[following the model of][]{Peters_1964}, for a duration equal to the characteristic ages of the observed pulsars. These characteristic ages, which are based on the spin-down of the observed pulsar, are considered by \citet{Maoz_2024} to be reasonable age estimates, and tend to be consistent with kinematic age estimates \citep[particularly for young pulsars, see e.g.,][]{Igoshev_2019,Disberg_2024b,Disberg_2025a}. For the resulting post-SN orbital properties, \citet{Grichener_2026} make a distinction between observed DNSs and systems for which there is either an association with a globular cluster or no consensus about the nature of the companion star. A globular cluster origin, for instance, makes comparison with our model difficult since in these environments dynamical interactions can affect binary properties.

We compare the observed DNSs to our model results. In particular, we consider the \lstinline{COMPAS} systems that already contain an NS and now experience a second SN, as shown in Figure \ref{Fig_BSE}. The first NS in these systems obtained a large kick following $\Mfid$, though this may not be the dominant channel for forming DNS if primaries can form NSs through low-kick ECSNe \citep[e.g.,][]{Vigna_2018}. In any case, we neglect the effect of the first SN on the DNS orbits since they are mostly determined by the second SN, which occurs after the orbit has been circularized by (Case BB) mass transfer onto the NS. We test this assumption by changing the pre-second-supernova parameters to results from a \lstinline{COMPAS} simulation with $v_{\text{NS}}=5\,$km\,s$^{-1}$ and $f_{\text{NS}}=0.45$, as defined in Equation \ref{eq_mu_sig}, and find no relevant differences in the results that would affect our conclusions.

After applying our kick models to these pre-second-SN orbits, we weight the resulting binary by an observation probability $p_{\text{obs}}$. This probability takes into account that DNSs (1) are less likely to be observed if they have shorter merger times, and (2) can only be observed when one (or both) of the NSs are visible as radio pulsars. This means that a DNS with merger time $\tau_{\text{gw}}$ and radio lifetime $\tau_{\text{radio}}$ is observable as a DNS with probability
\begin{equation}
    \label{eq_P_obs}
    p_{\text{obs}}\propto\min(\tau_{\text{gw}},\tau_{\text{radio}}),
\end{equation}
where we determine $\tau_{\text{gw}}$ as a function of the period, eccentricity, and masses through the model of \citet{Peters_1964}. However, the radio lifetime is difficult to estimate, and its correlation with the DNS properties is not well understood. We therefore tentatively choose $\tau_{\text{radio}}=500\,$Myr. We note that several DNSs have characteristic ages that exceed this value and $\tau_{\text{radio}}$ for MSPs is often expected to be several Gyr \citep{Belczynski_2010,Belczynski_2018}, but we choose the value of $500\,$Myr to emphasize the short-period systems in view of the systematic uncertainties in determining the observability of an MSP.  In any case, our main conclusions do not depend significantly on $p_{\text{obs}}$.

The distributions of the DNS periods and eccentricities, corrected for GW emission by \citet{Grichener_2026}, show interesting features. Most DNSs have eccentricities ${\lesssim}\,0.4$, but there are a few high-eccentricity systems (${\gtrsim}\,0.6$) with short periods (${<}\,1$\,d) or long periods (${>}\,10\,$d). This means that there is an apparent gap in the DNS eccentricities between ${\sim}\,0.4$ and ${\sim}\,0.6$ that has been a topic of discussion \citep[e.g.,][]{Vigna_2018,Andrews_2019,Andrews_2019b,Chattaraj_2026,Grichener_2026}. \citet{Andrews_2019} define three DNS sub-populations, containing the systems with (1) short periods and low eccentricities, (2) short periods and high eccentricities, or (3) long periods. They suggest that the short-period high-eccentricity DNSs might have formed dynamically in a globular cluster, and that this can explain their clustering in period-eccentricity space (but see \citealt{Ye_2019}). \citet{Chattaraj_2026}, in turn, show that a bifurcation between short- and long-period DNSs may correspond to different kinds of CE evolutionary channels.

\citet{Grichener_2026} show that the \citet{Mandel_2020} prescription for NS remnant masses results in a bimodal Blaauw kick distribution which may explain the observations. In particular, they note that the \citet{Mandel_2020} model, as displayed in Figure \ref{Fig_SSE}, has a non-monotonicity in the relation between pre-SN CO core mass and NS remnant mass characterized by a jump in expelled mass $M_{\text{CO}}-M_{\text{NS}}$ at $M_{\text{CO}}>3M_{\odot}$. This leads to a clear distinction between systems with $M_{\text{CO}}<3M_{\odot}$ that have small Blaauw kicks and systems with $M_{\text{CO}}>3M_{\odot}$ that obtain relatively large Blaauw kicks. As \citet{Grichener_2026} show, if the natal kicks in DNS progenitor systems are small and the Blaauw kicks dominate, the resulting orbits show a bimodal eccentricity distribution that looks similar to the observed Galactic DNSs.

\begin{figure}
    \centering
    \resizebox{\hsize}{!}{\includegraphics{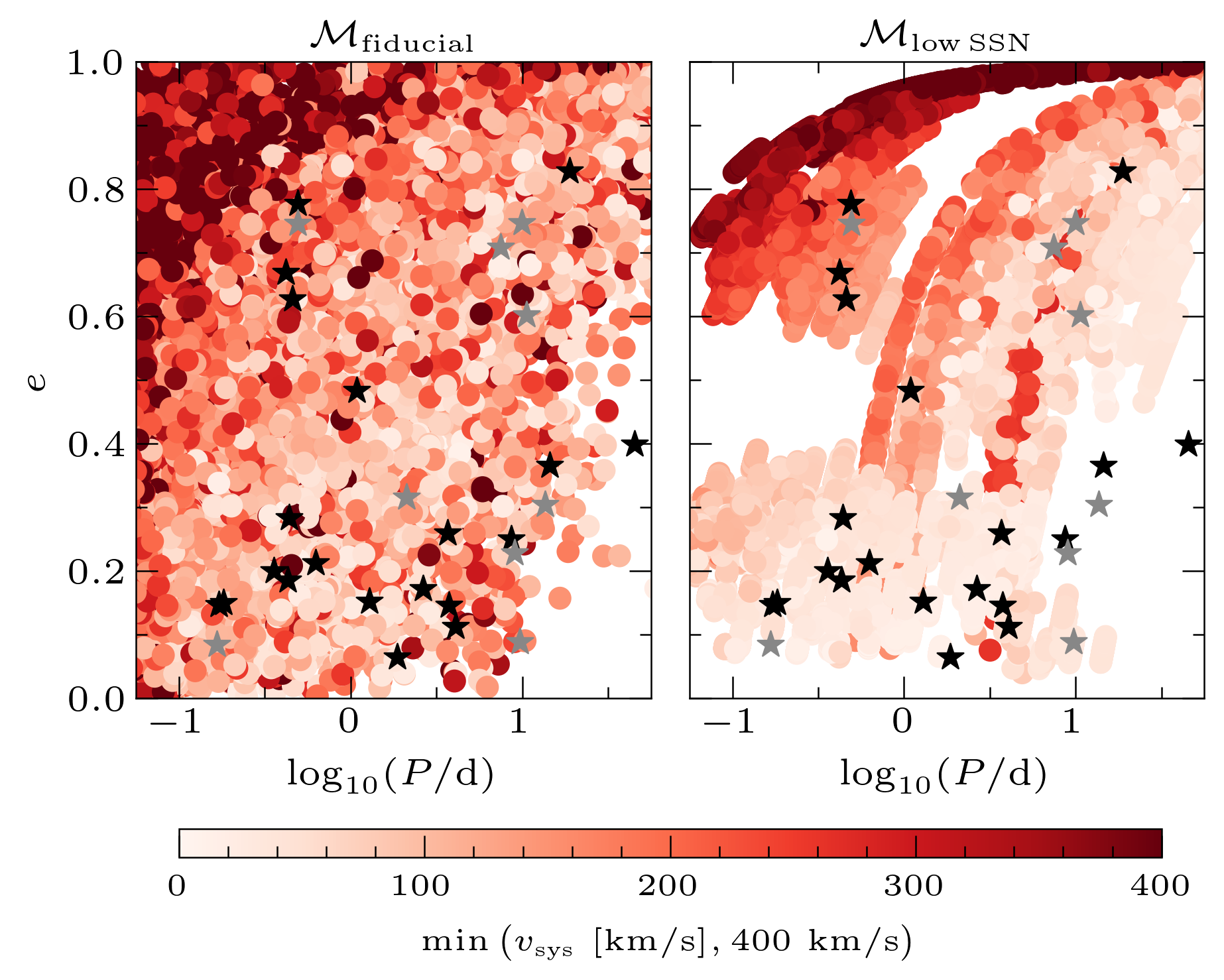}}
    \caption{Periods and eccentricities of all simulated DNSs, neglecting the probability of them being observable, where the color scale corresponds to their systemic kicks. The foreground/background position of the colored dots is assigned randomly. We truncate the velocity scale at $400\,$km\,s$^{-1}$ for visibility. The black stars correspond to the DNSs in the sample of \citet{Grichener_2026}, where the grey stars are limited-confidence/globular cluster DNSs.}
    \label{Fig_DNS_e-P-v}
\end{figure}

However, \citet{Grichener_2026} also find that the apparently bimodal DNS eccentricities are statistically consistent with a unimodal underlying distribution. Indeed, after integrating the orbit of the DNS J1208--5936 back in time to account for gravitational-wave emission, they find a post-SN eccentricity of $0.47$, which falls in the apparent eccentricity gap. 

\begin{figure*}
    \centering
    \includegraphics[width=18cm]{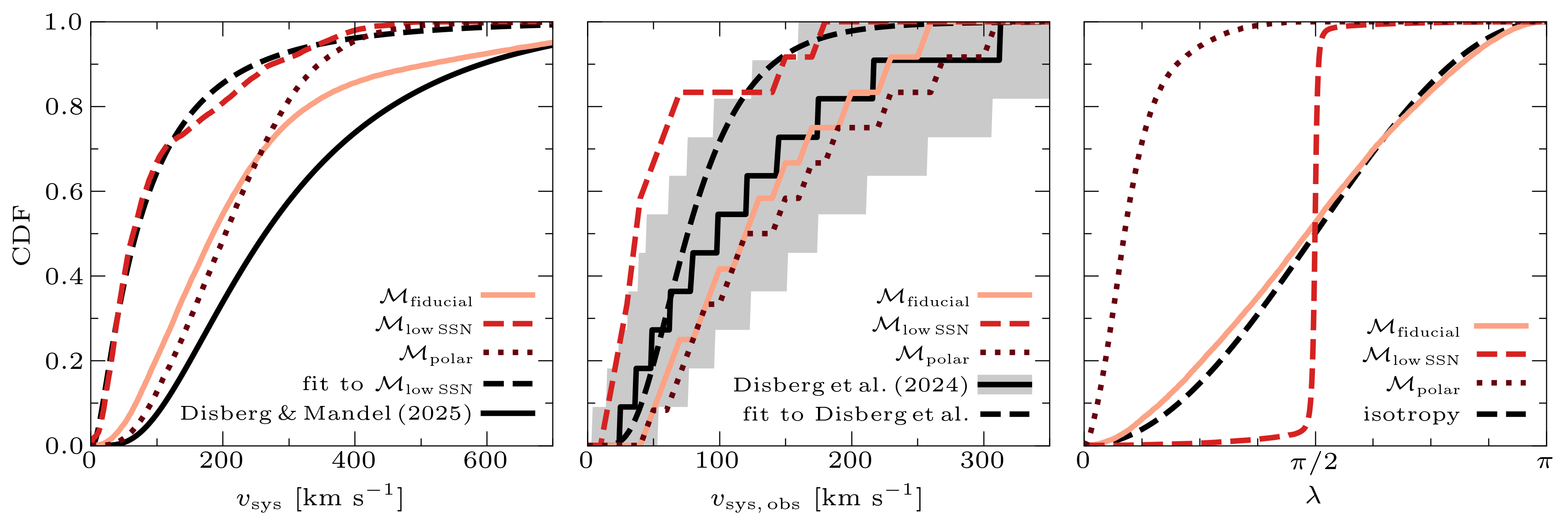}
    \caption{Systemic kicks of the DNSs. The left panel shows the intrinsic kick distributions as predicted through our models $\Mfid$, $\Mlow$, and $\Mpol$ (light red solid line, red dashed line, and dark red dotted line, respectively). We also fit a lognormal distribution to $\Mlow$ which results in $\mu=4.24$ and $\sigma=1.02$ (black dashed line), and display the natal kick distribution of \citet[][black solid line]{Disberg_2025b}. In the central panel we show the median distribution of the systemic kick estimates of \citet{Disberg_2024b}, together with its bootstrapped $95\%$ uncertainty region (black solid line and grey shaded region, respectively). A lognormal fit to these estimates yields $\mu=4.33(26)$ and $\sigma=0.48(24)$ (black dashed line). The red lines show the predicted systemic velocities for the DNSs in the sample of \citet{Disberg_2024b}, given their orbital periods and eccentricities (see, e.g., Figure \ref{Fig_DNS_e-v}). The right panel shows the distributions of the angle $\lambda$ between the post-SN orbital angular momentum and the systemic kick (determined through Equation \ref{eq_dlambda}), together with an isotropic $\lambda$ distribution (black dashed line).}
    \label{Fig_DNS_Velocities}
\end{figure*}

In Figure \ref{Fig_DNS_Orbits} we show the period and eccentricity distributions of the simulated \lstinline{COMPAS} DNSs for $\Mfid$, $\Mlow$, $\Mpol$, and $\Mlowroc$, weighted by $p_{\text{obs}}$ as defined in Equation \ref{eq_P_obs}. For high natal kicks, in $\Mfid$ and $\Mpol$, the DNSs follow a broad distribution that shows little structure \citep[as also found by, e.g.,][]{Mandel_2021}, where we note that the results for $\Mroc$ are similar to those for $\Mfid$. However, for low natal kicks ($\Mlow$) the simulated DNSs follow a distribution very similar to the observations. As a result of the bimodal Blaauw kicks as described by the \citet{Mandel_2020} model, there are DNSs with (1) low eccentricities (${\lesssim}\,0.4$) and progenitor CO core masses below $3M_{\odot}$, (2) short periods (${\lesssim}\,1\,$d), high eccentricities (${\gtrsim}\,0.6$), and progenitor core masses above $3M_{\odot}$ \citep[see also][]{Grichener_2026}, and (3) long periods (${\gtrsim}\,10\,$d) and high eccentricities (${\gtrsim}\,0.6$). The first two groups align relatively accurately with the observed DNSs, with the notable exception of J1208--593 at $e=0.47$. The third group contains systems with SN progenitor CO core masses between $2M_{\odot}$ and $3M_{\odot}$, leading to moderate Blaauw kicks. In this parameter space there are also systems with non-zero pre-SN eccentricity (as shown in Figure \ref{Fig_BSE}). There are several long-period systems that seem to lie outside of the model distribution.

\citet{Grichener_2026} show that the \citet{Mandel_2020} remnant mass prescription can explain the DNS periods and eccentricities if the natal kicks are small, and in our model we find that the $\Mlow$ model indeed aligns well with the observations. This means that Blaauw kicks can be sufficient to explain the DNS eccentricities \citep[contrasting with the bimodal DNS kicks proposed by][]{Beniamini_2016}. They pose the caveats, however, that there may be a tension between this model and (1) the masses of the observed DNSs (as we discuss in Section \ref{sec7.3}), and (2) the observed spin-orbit misalignment angles of ${\sim}\,20\degree$ for the DNS B1913+16 \citep{Kramer_1998,Weisberg_2002} and ${\sim}\,25\degree$ for B1534+12 \citep{Thorsett_2005}. Lastly, the results for $\Mlowroc$ are similar to the ones for $\Mlow$, since the rocket kick applies a relatively small scatter in the DNS eccentricities due to their high orbital velocities. It is therefore difficult to detect the presence of a rocket kick in the observed DNS sample.

\subsection{DNS Velocities}
\label{sec7.2}
\noindent The systemic velocities of the Galactic DNSs that have distance and proper motion estimates are known to not deviate significantly from their LSR \citep{Gaspari_2024a}. \citet{Disberg_2024b} kinematically constrained the systemic kicks of $11$ DNSs with no globular cluster associations, finding a typical systemic kick velocity of ${\sim}\,50$--$100\,$km\,s$^{-1}$. We find one additional DNS in the ATNF Catalog \citep{Manchester_2005} with proper motion and distance estimates and no globular cluster association, J2150+3427 \citep{Wu_2023,Wu_2025}, but we find that its Galactic trajectory is difficult to constrain since inverting its velocity vector leads to a distribution of trajectories where the vast majority escape the Galaxy.

In Figure \ref{Fig_DNS_e-P-v} we show the systemic kicks of the simulated binaries for the $\Mfid$ and $\Mlow$ models. If the natal kicks are small, as in $\Mlow$, both the eccentricity and the systemic kick are dominated by the Blaauw kick, resulting in a clear correlation between eccentricity and systemic kick (particularly for short periods). If the natal kicks are large, in contrast, the systems are scattered over the parameter space and there is no clear correlation between eccentricity and systemic kick. In the comparison between our models and the observed kicks, we must account for this possible correlation between the binary orbit and its systemic kick. We therefore compare observed DNSs with systemic kick estimates against the average systemic kicks among the simulated binaries with similar periods and eccentricities.

In Figure \ref{Fig_DNS_Velocities} we show the systemic kicks of all simulated DNSs, dependent on the different kick models. For the high natal kicks in $\Mfid$ the difference between the systemic kicks and the natal kicks of \citet{Disberg_2025b} is relatively small, but for the low natal kicks in $\Mlow$ the systemic kicks are significantly lower. In the central panel of the figure, we compare the lognormal distribution fitted to the systemic kicks of observed DNSs \citep{Disberg_2024b} with the systemic kick distributions of simulated DNSs with periods and eccentricities similar to the observed ones. For high natal kicks the model predicts systemic kicks similar to the median observed kick, but these kicks are higher than the lognormal fit. The lognormal fit is a better representation of the observed distribution than the bootstrapped confidence interval since the kick posteriors of \citet{Disberg_2024b} are asymmetrically skewed to higher values due to projection effects as a result of assuming isotropy to describe the radial velocity components \citep[see Appendix \ref{appC} and also][]{Disberg_2025b}. The low natal kicks produce systemic kicks more similar to the lognormal fit and therefore seem to be somewhat more aligned with the observations, but because of the width of the bootstrapped confidence interval the comparison does not allow a decisive choice between the natal kick models.

\begin{figure}
    \centering
    \resizebox{\hsize}{!}{\includegraphics{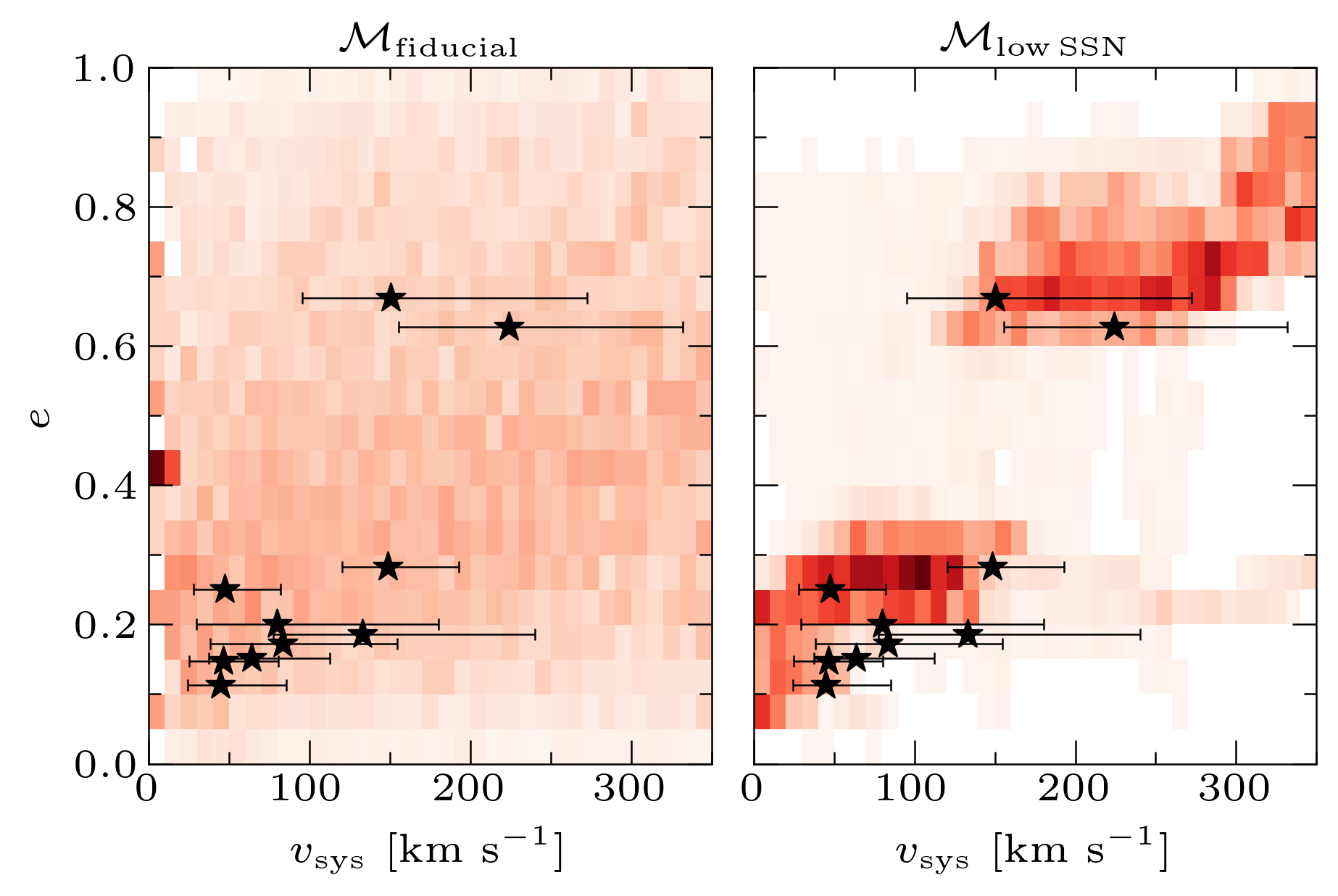}}
    \caption{Eccentricities of the short-period (${<}\,10\,$d) DNSs in the sample of \citet{Disberg_2024b}, as a function of systemic kick. The black stars correspond to the median systemic kick estimate, and the error-bars show the $68\%$ confidence intervals. The red distributions show the model results with post-SN periods ${<}\,10\,$d for $\Mfid$ and $\Mlow$, where the columns are normalized individually such that the distribution corresponds to $p(e|v_{\text{sys}})$.}
    \label{Fig_DNS_e-v}
\end{figure}

Moreover, Figure \ref{Fig_DNS_Velocities} shows the distributions of the angle $\lambda$ between the post-SN orbital angular momentum and the systemic kick, as defined in Equation \ref{eq_dlambda}. For the $\Mfid$ results, with high isotropic natal kicks, the $\lambda$ distribution is similar to an isotropic distribution. If these natal kicks are polar ($\Mpol$), the systemic kicks are significantly more aligned with the orbital angular momentum. However, for the small natal kicks in $\Mlow$ the systemic kicks are dominated by the Blaauw kicks which are restricted to the pre-SN orbital plane, meaning almost all simulated DNS systems obtain a systemic kick perpendicular to the orbital angular momentum. This is relevant for analyzing SGRB offsets from host galaxies \citep[e.g.,][]{Levan_2007,Fong_2022,Gaspari_2025,Skobe_2026}, since the jets of GRBs resulting from DNS mergers are thought to be aligned with the orbital angular momentum. If this is true and the DNS systemic kicks are dominated by Blaauw kicks, then the projected offsets of SGRBs from their host galaxies are approximately equal to the true offsets. \citet{Mandel_2026} derive a formalism for the maximum offset a DNS can obtain from its host galaxy before merging, assuming high natal kicks, but in Appendix \ref{appE} we show that their results also hold for low natal kicks.

Even though the $\Mlow$ model seems to be more consistent with the lognormal fit to the observed kick distribution in Figure \ref{Fig_DNS_Velocities}, it is difficult to make decisive conclusions due to systematic uncertainties both in the \lstinline{COMPAS} model and the observed systemic kick estimates. As an additional way to distinguish between high and low natal kicks, we use the correlation between systemic kick and eccentricity for low natal kicks, as displayed in Figure \ref{Fig_DNS_e-P-v}. In Figure \ref{Fig_DNS_e-v} we show the DNS eccentricities predicted by our models $\Mfid$ and $\Mlow$ for a given systemic kick magnitude, where we restrict this analysis to systems with periods below $10\,$d to avoid comparison with long-period systems for which the correlation shown in Figure \ref{Fig_DNS_e-P-v} is less clean. The figure indeed shows that for $\Mlow$ one expects systems with larger systemic kicks to be more eccentric, whereas for $\Mfid$ the natal kicks break this correlation. However, in comparing our models to the systemic kick estimates of \citet{Disberg_2024b} we find that the eccentricities of the DNSs in their sample are consistent with the correlation found for $\Mlow$: the systems with higher systemic kicks are more eccentric. Although it is difficult to draw strong conclusions due to the limited sample size, these systems appear to show evidence of systemic kicks that are dominated by Blaauw kicks. We note that particularly for the low-eccentricity systems the observational kick estimates slightly exceed the model predictions, similarly to the distributions in Figure \ref{Fig_DNS_Velocities}, but it is not obvious whether this is due to model uncertainties or systematic uncertainties in the observational estimates.

\begin{figure}
    \centering
    \resizebox{\hsize}{!}{\includegraphics{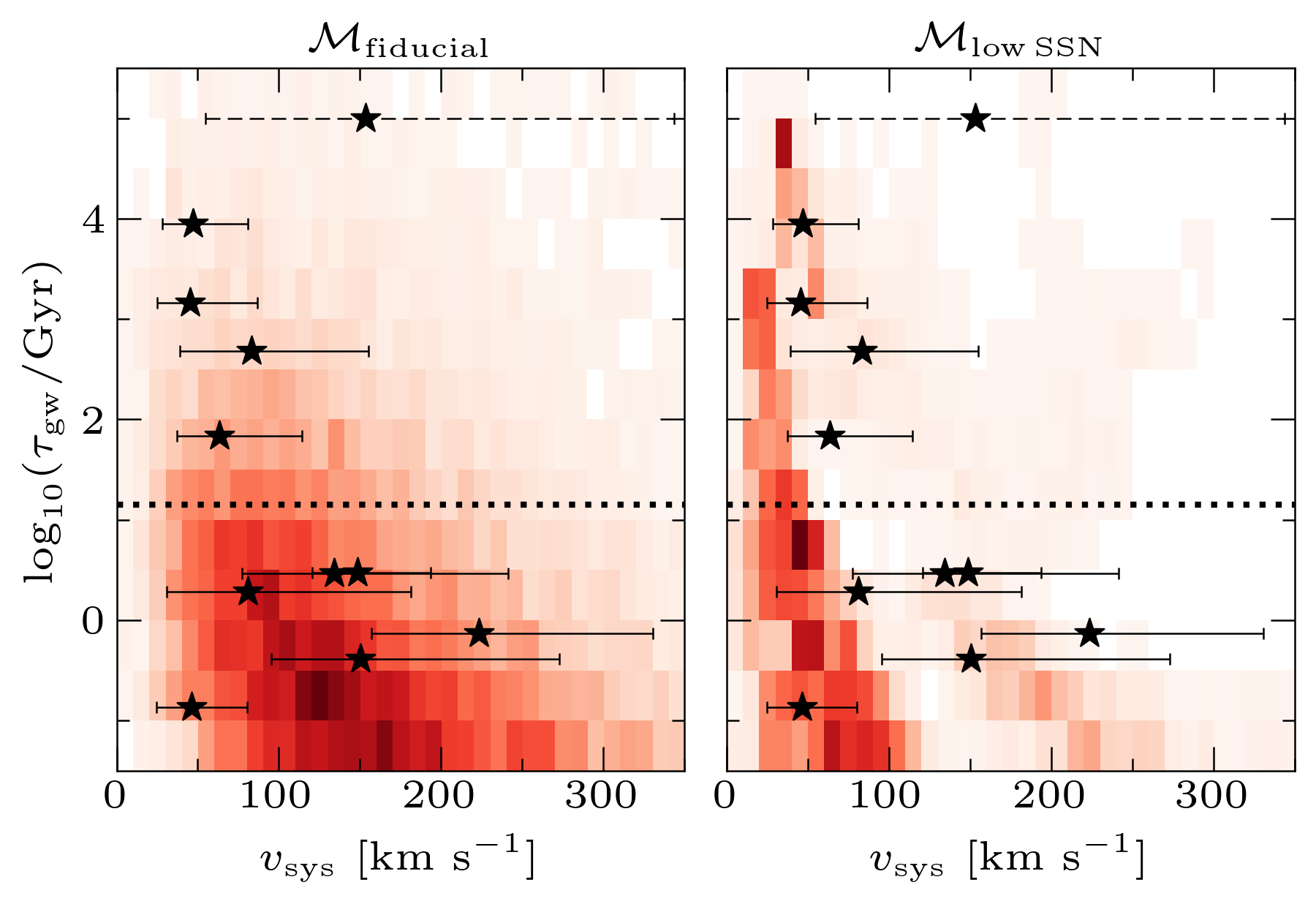}}
    \caption{Times between formation and merger of the DNSs in the sample of \citet{Disberg_2024b}, as a function of systemic kick (black stars), where we highlight the DNS J1930--1852 with a dashed errorbar since it has a poorly constrained Galactic trajectory. The merger times are determined by adding the present-day merger time of the DNSs and their characteristic ages. The red distributions show the model results for $\Mfid$ and $\Mlow$. The dotted line shows the Hubble time.}
    \label{Fig_DNS_t-v}
\end{figure}

In Figure \ref{Fig_DNS_Velocities} we show the intrinsic systemic kick distributions for the DNSs in our models, but in order to compare these distributions to observations it is necessary to also take into account that the systemic kick correlates with the merger time of the system. If natal kicks are high, wide binaries are disrupted so only tight binaries with short merger times survive, while if the natal kicks are low, wide binaries with long merger times can survive but end up with small systemic kicks. We determine the merger times of the DNSs in our model and compare them to the sum of the merger time and the characteristic age of the Galactic DNSs. In Figure \ref{Fig_DNS_t-v} we show the merger times as a function of systemic kicks, for the $\Mfid$ and $\Mlow$ models. For both models the systems with higher systemic kicks also have shorter merger times and the observed DNSs indeed follow this trend, making it difficult to distinguish between the models.

While it is true that for $\Mlow$ the observed systemic kicks slightly exceed the model predictions for DNSs with a given merger time, this difference can be explained by, for example, non-zero pre-SN systemic velocities or the asymmetric kick posteriors of \citet{Disberg_2024b} that result from assuming isotropy. In particular the system with the longest merger time, J1930--1852, significantly exceeds the model prediction, but we note that in the analysis of \citet{Disberg_2024b} the Galactic trajectory and therefore the systemic kick of this system is poorly constrained and could be compatible with a low systemic kick similar to the model prediction. Lastly, we note that the kick models differ in the merger time distributions they predict \citep[see also, e.g.,][]{Beniamini_2024}, which may be relevant for DNS merger rates and thus the chemical enrichment history of the Milky Way \citep[e.g.,][]{DeDonder_2004,Cavallo_2021,Greggio_2021,Kobayashi_2023,Chattaraj_2026a,Fishbach_2026}. Large kicks reduce the total DNS formation rate by disrupting more DNSs, but may have less effect on the DNS merger rate because ultimately merging DNSs are already tight at second SN and less sensitive to disruption by kicks.

\subsection{DNS Masses}
\label{sec7.3}
\begin{figure}
    \centering
    \resizebox{\hsize}{!}{\includegraphics{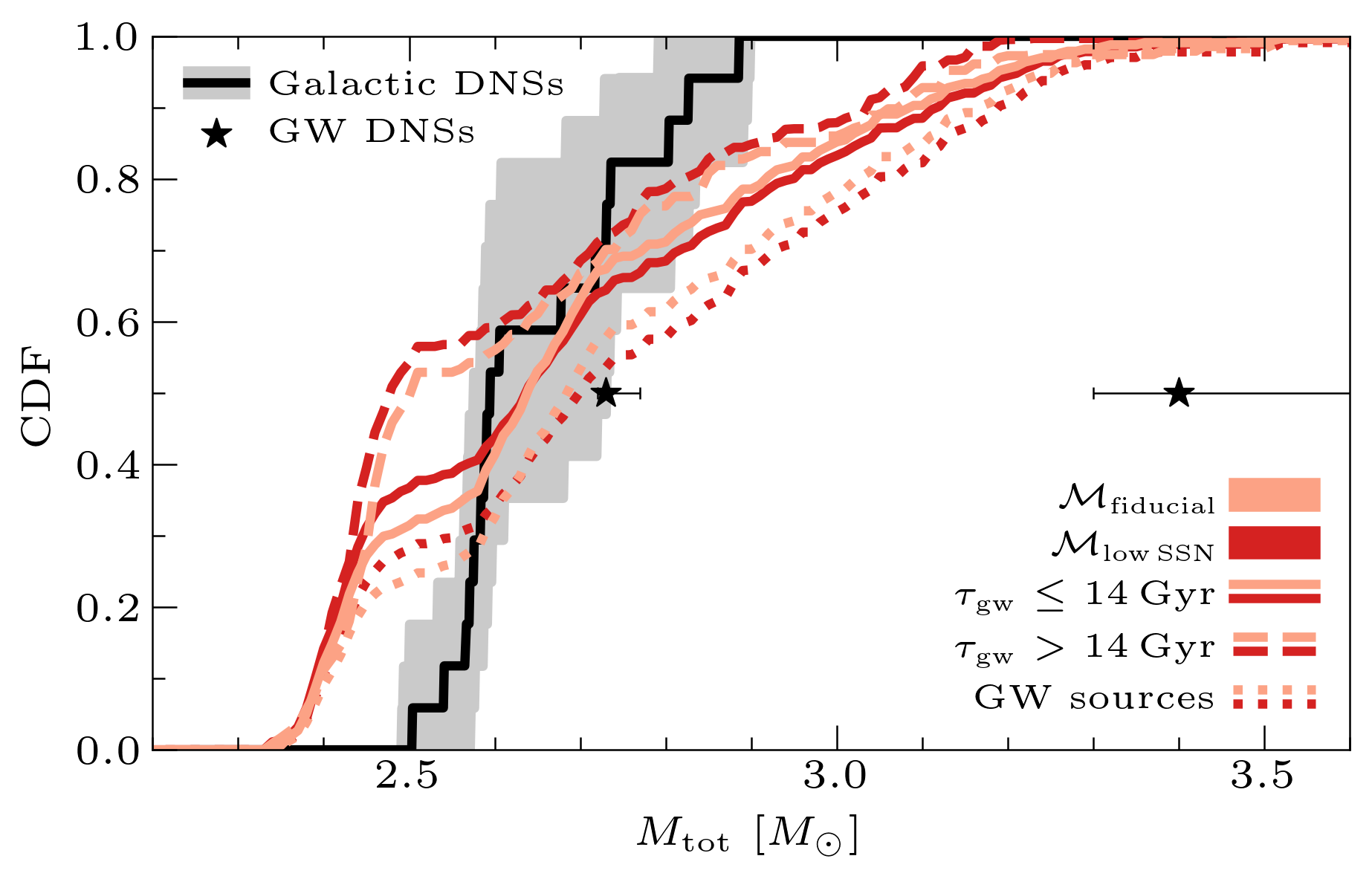}}
    \caption{Total masses of the Galactic DNSs in the sample of \citet{Grichener_2026}, where we omit the limited-confidence/globular cluster systems in their sample. We show the median distribution and the bootstrapped $95\%$ confidence interval (black solid line and grey shaded region, respectively). We further show the masses of the two DNSs detected through GWs: GW170817 with $M_{\text{tot}}=2.73^{+0.04}_{-0.01}M_{\odot}$ \citep{Abbott_2019} and GW190425 with $M_{\text{tot}}=3.4^{+0.3}_{-0.1}M_{\odot}$ \citep{Abbott_2020}. We also show the total masses of the DNSs in the $\Mfid$ (light red lines) and $\Mlow$ models (red lines), which we divide into systems that merge within $14\,$Gyr (solid lines) and systems with longer merger times (dashed lines). And we show the mass distribution of possible GW sources, where we weight the simulated DNSs with merger times below $14\,$Gyr by $M_{\text{ch}}^{5/2}$.}
    \label{Fig_DNS_Masses}
\end{figure}
\noindent The total masses of the observed Galactic DNSs can be constrained relatively accurately and compared to our model predictions. We therefore take the total DNS masses listed by \citet{Grichener_2026}, where we omit the limited-confidence/globular cluster DNSs in their sample, and determine their median distribution and the $95\%$ confidence interval. Then, we compare these observed masses to our model results for $\Mfid$ and $\Mlow$, making a distinction between systems with merger times below and above $14\,$Gyr. We also show the distribution of masses that can be compared to GW observations, in particular the two observed GW DNSs GW170817 \citep{Abbott_2019} and GW190425 \citep{Abbott_2020}. Since the distance to which a DNS merger can be observed through GWs scales with $M_{\text{ch}}^{5/6}$, where $M_{\text{ch}}=(M_1M_2)^{3/5}/(M_1+M_2)^{1/5}$ is the chirp mass of a binary with stellar masses $M_1$ and $M_2$, the cosmic volume in which these mergers can be observed scales with $M_{\text{ch}}^{5/2}$. We therefore construct the mass distribution of possible GW sources by taking the simulated DNSs with merger times below $14\,$Gyr and weighting them by $M_{\text{ch}}^{5/2}$.

\begin{figure}
    \centering
    \resizebox{\hsize}{!}{\includegraphics{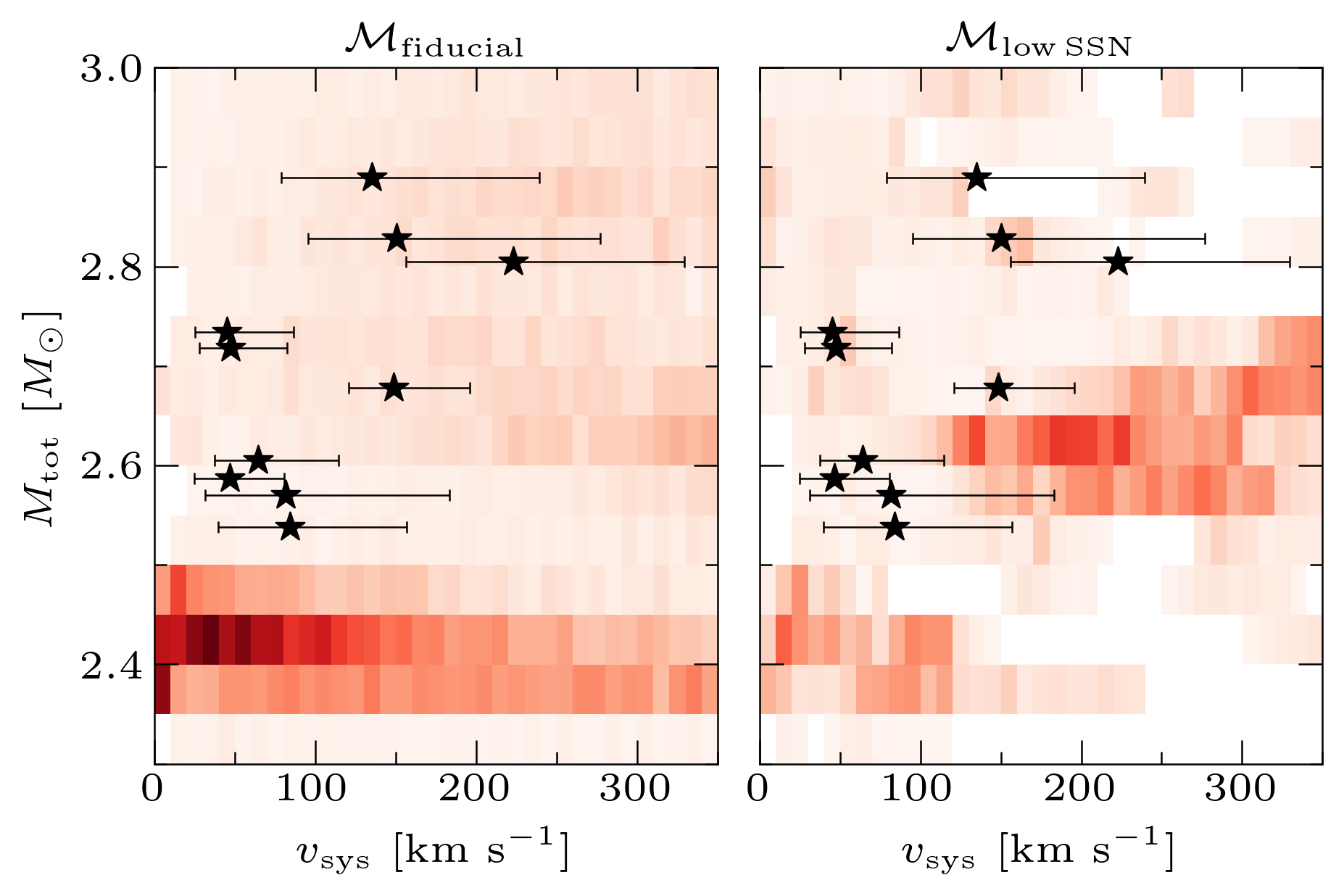}}
    \caption{Total masses of the short-period (${<}\,10\,$d) DNSs in the sample of \citet{Disberg_2024b}, as a function of systemic kick (black stars). The red distributions show the model results with post-SN periods ${<}\,10\,$d, for $\Mfid$ and $\Mlow$, where the columns are normalized individually such that the distribution corresponds to $p(M_{\text{tot}}|v_{\text{sys}})$.}
    \label{Fig_DNS_M-v}
\end{figure}

In Figure \ref{Fig_DNS_Masses} we show the total mass distribution of the Galactic DNSs compared to our model results. The observed masses are tightly concentrated between $2.5M_{\odot}$ and $2.8M_{\odot}$, whereas the model results have a larger spread. Indeed, \citet{Mandel_2021} find that there is tension between the masses of Galactic DNSs and the \lstinline{COMPAS} predictions through the \citet{Mandel_2020} model. However, it is not straightforward to determine whether this difference is caused by the remnant mass prescription of \citet{Mandel_2020} as shown in Figure \ref{Fig_SSE}, or by \lstinline{COMPAS} assumptions concerning case BB mass transfer onto the NS, for example. Nevertheless, it appears as if the \citet{Mandel_2020} model is underpredicting the masses of NSs from progenitors with CO core masses below $2M_{\odot}$, while overpredicting the masses of NSs from progenitors with CO core masses between $2M_{\odot}$ and $3M_{\odot}$. We note that these latter NSs are not required to explain the observed periods and eccentricities of the Galactic DNSs \citep{Grichener_2026}. Nevertheless, the GW DNS GW190425 has been determined to have a relatively high mass of $3.4^{+0.3}_{-0.1}M_{\odot}$ \citep{Abbott_2020}, and while this can be reconciled with our models, there is a noticeable difference with the Galactic DNSs \citep[see also][]{Nair_2025,Nair_2026}. In part, this might be explained by observational biases. \citet{Chu_2025}, for example, argue that there may be a correlation between the radio lifetime of a DNS pulsar and its mass.

Although the different kick models predict similar DNS mass distributions (Figure \ref{Fig_DNS_Masses}), they differ in the relationship between mass and systemic kick. In Figure \ref{Fig_DNS_M-v} we show this relationship for $\Mfid$ and $\Mlow$. While high natal kicks ($\Mfid$) predict DNSs of approximately the same mass for systemic kicks in the range of the observations, low natal kicks ($\Mlow$) show a trend in which the more massive DNSs tend to obtain larger kick velocities. The observed DNS masses exceed the model predictions by ${\sim}0.2$--$0.3M_{\odot}$, but they do appear to agree with this trend, although we stress again that it is difficult to draw strong conclusions due to the limited sample size.

\section{Discussion}
\label{sec8}
\begin{figure*}
    \centering
    \includegraphics[width=18cm]{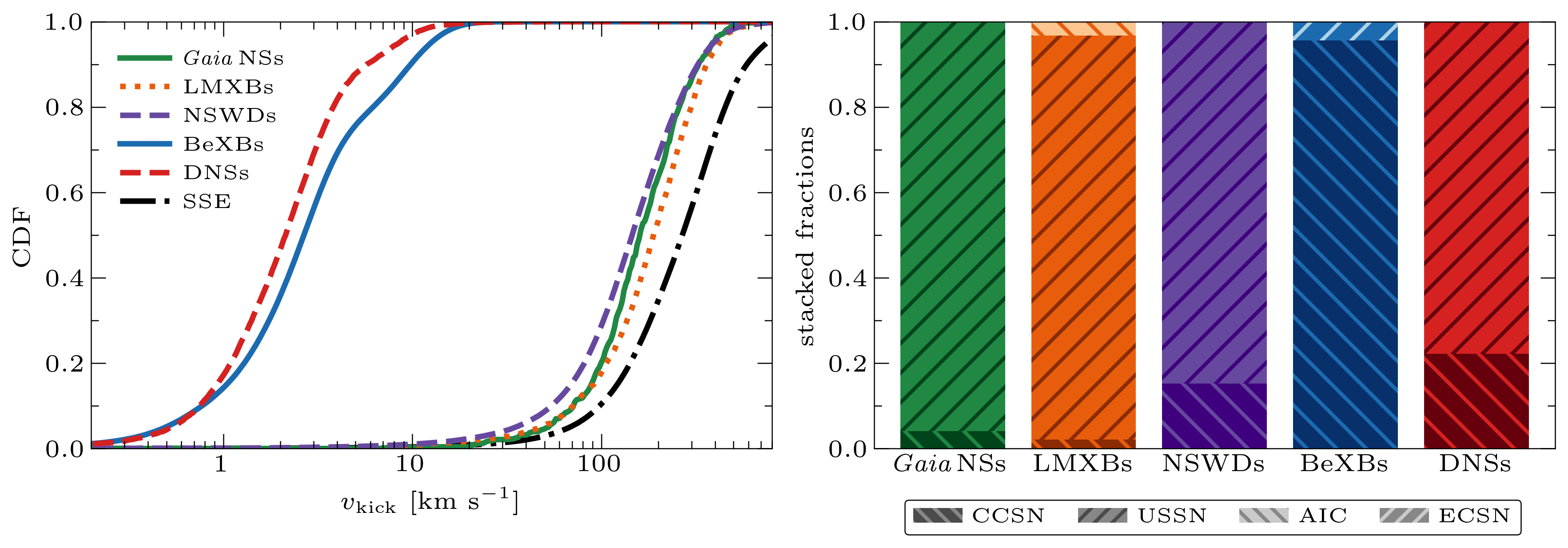}
    \caption{The natal kicks (left panel) and \lstinline{COMPAS}-recorded SN types (right panel) in our preferred models of the \textit{Gaia} NSs (green, $\Mroc$), LMXBs (orange, $\Mfid$), NSWDs (purple, $\Mfid$), BeXBs (blue, $\Mlow$), and second-born NSs in DNSs (red, $\Mlow$). In the left panel we also show the natal kicks for an SSE model (black dash-dotted line) calibrated to the results of \citet{Disberg_2025b} through the model of \citet{Mandel_2020}, as also displayed in Figure \ref{Fig_Calibration}. In the right panel, where we display the SN type fractions for each binary type, we neglect fractions of ${<}\,1\%$.}
    \label{Fig_Comparison}
\end{figure*}
\noindent Based on the comparison between the kick models and observations for the \textit{Gaia} NSs, LMXBs, NSWDs, BeXBs, and DNSs, we discuss the different natal kick magnitudes that appear to explain the different kinds of binary systems (Section \ref{sec8.1}). Moreover, we note several caveats that complicate our analysis (Section \ref{sec8.2}). 
\subsection{Kick Magnitudes}
\label{sec8.1}
\noindent We find that combining the pre-SN binary parameters from \lstinline{COMPAS} with natal kicks calibrated to the velocities of isolated pulsars \citep{Disberg_2025b} through the model of \citet{Mandel_2020} can explain the observed properties of the \textit{Gaia} NSs in the sample of \citet{El-Badry_2024b}, the LMXBs in the sample of \citet{ODoherty_2023}, and the NSWDs in the ATNF Catalog \citep{Manchester_2005}, where we note that in order to match the \textit{Gaia} NS eccentricities we also need a rocket kick of ${\sim}\,30\,$km\,s$^{-1}$. In contrast, our best explanations for the BeXBs in the sample of \citet{Valli_2025} and the DNSs in the sample of \citet{Grichener_2026} require natal kicks that are significantly reduced.

We plot the natal kick velocities in the preferred kick models for the different kinds of binary systems in Figure \ref{Fig_Comparison}. Since the natal kicks depend on the pre-SN CO core mass in the \citet{Mandel_2020} model, they are affected by the evolutionary histories and therefore not necessarily equal for different kinds of systems. The figure shows that the \textit{Gaia} NSs, LMXBs, and NSWDs receive similar natal kicks in $\Mfid$, which are slightly lower than the SSE kicks calibrated to the results of \citet{Disberg_2025b}. Meanwhile, BeXBs and DNSs have similar natal kicks of ${\lesssim}\,10\,$km\,s$^{-1}$ in $\Mlow$. This bifurcation is also shown in Table \ref{tab_results}, in which we summarize how well our different kick models fit the observations. In this table we show our preferred kick model for each kind of system, but also list the other models that can fit the observations, for example because our model results are not sensitive to kick directions or the presence of a rocket kick. In general, we do not include polar kicks or rocket kicks in our preferred model unless these are necessary to reconcile model and observation. Lastly, in Table \ref{tab_distributions} we list the lognormal fits to the intrinsic systemic kicks in our preferred model and to the observations. While the table shows the fitted distribution parameters, we also list the individual kick estimates for the \textit{Gaia} NSs and NSWDs in Appendix \ref{appF}.

\definecolor{mygreen}{HTML}{208843}
\definecolor{myorange}{HTML}{e85d0c}
\definecolor{myred}{HTML}{d52221}
\begin{table*}
\centering
\caption{Summary of how well our considered kick models ($\mathcal{M}$, defined in Table \ref{tab_models}) fit the observed orbits (i.e., periods and eccentricities) and velocities \citep[i.e., systemic kicks as estimated through the method of][]{Disberg_2024b,Disberg_2025a}.\label{tab_results}}
\hspace{-29mm}\begin{tabular}{lc|c|c|c!{\vrule width 1.2pt}c|c|}
\cline{1-7}\\[-11pt]\cline{1-7}\\[-13pt]
 & & \multicolumn{3}{c!{\vrule width 1.2pt}}{natal kicks calibrated to isolated NSs} & \multicolumn{2}{c|}{significantly reduced} \\[-2pt]
  & & \multicolumn{3}{c!{\vrule width 1.2pt}}{\citep{Disberg_2025b}} & \multicolumn{2}{c|}{natal kicks (${\lesssim}\,10\,\text{km}\,\text{s}^{-1}$)} \\
 & & $\Mfid$ & $\Mpol$ & $\Mroc$ & $\Mlow$  & $\Mlowroc$\\[2pt]\hline\\[-13pt]
 \multirow{4}{*}{\textit{Gaia} NSs} & \multirow{2}{*}{orbits} & \multirow{2}{*}{\textcolor{myred}{poor fit}} & \multirow{2}{*}{\textcolor{myred}{poor fit}} & \multirow{2}{*}{\textcolor{mygreen}{\textbf{good fit}}} & \multirow{2}{*}{\textcolor{myred}{poor fit}} & \textcolor{myorange}{for different}\\[-2pt]
  & & & & & & \textcolor{myorange}{CE formalism}$^{\text{(a)}}$ \\[2pt]\cline{2-7}\\[-13pt]
 & \multirow{2}{*}{velocities} & \textcolor{mygreen}{for CE} & \textcolor{mygreen}{for CE} & \textcolor{mygreen}{\textbf{for CE}} & \textcolor{myorange}{requires} & \textcolor{myorange}{requires} \\[-2pt]
  & & \textcolor{mygreen}{systems}$^{\text{(b)}}$ & \textcolor{mygreen}{systems}$^{\text{(b)}}$ & \textcolor{mygreen}{\textbf{systems}}$^{\text{(b)}}$ & \textcolor{myorange}{dyn.\ heating}$^{\text{(c)}}$ & \textcolor{myorange}{dyn.\ heating}$^{\text{(c)}}$\\[2pt]\hline\\[-13pt]
 \multirow{2}{*}{LMXBs} & \multirow{2}{*}{velocities} & \multirow{2}{*}{\textcolor{mygreen}{\textbf{good fit}}} & \multirow{2}{*}{\textcolor{mygreen}{good fit}} & \multirow{2}{*}{\textcolor{mygreen}{good fit}} & \multirow{2}{*}{\textcolor{myred}{poor fit}} & \multirow{2}{*}{\textcolor{myred}{poor fit}}\\[-2pt]
 & & & & & & \\[2pt]\hline\\[-13pt]
 \multirow{2}{*}{NSWDs} & \multirow{2}{*}{velocities} & \multirow{2}{*}{\textcolor{mygreen}{\textbf{good fit}}} & \multirow{2}{*}{\textcolor{mygreen}{good fit}} & \multirow{2}{*}{\textcolor{mygreen}{good fit}} & \textcolor{myorange}{requires} & \textcolor{myorange}{requires}\\[-2pt]
 & & & & & \textcolor{myorange}{dyn.\ heating}$^{\text{(c)}}$ & \textcolor{myorange}{dyn.\ heating}$^{\text{(c)}}$ \\[2pt]\specialrule{1.2pt}{0pt}{0pt}\\[-13pt]
 \multirow{2}{*}{HMXBs} & \multirow{2}{*}{orbits} & \multirow{2}{*}{\textcolor{myred}{poor fit}} & \multirow{2}{*}{\textcolor{myred}{poor fit}} & \multirow{2}{*}{\textcolor{myred}{poor fit}} & \textcolor{mygreen}{\textbf{fits low-$e$}} & \multirow{2}{*}{\textcolor{myred}{poor fit}} \\[-2pt]
 & & & & & \textcolor{mygreen}{\textbf{systems}}$^{\text{(d)}}$ & \\[2pt]\hline\\[-13pt]
 \multirow{4}{*}{DNSs} & \multirow{2}{*}{orbits} & \multirow{2}{*}{\textcolor{myred}{poor fit}} & \multirow{2}{*}{\textcolor{myred}{poor fit}} & \multirow{2}{*}{\textcolor{myred}{poor fit}} & \multirow{2}{*}{\textcolor{mygreen}{\textbf{good fit}}} & \multirow{2}{*}{\textcolor{mygreen}{good fit}} \\[-2pt]
 & & & & & & \\[2pt]\cline{2-7}\\[-13pt]
 & \multirow{2}{*}{velocities} & \multirow{2}{*}{\textcolor{myorange}{subopt.\ fit}$^{\text{(e)}}$} & \multirow{2}{*}{\textcolor{myorange}{subopt.\ fit}$^{\text{(e)}}$} & \multirow{2}{*}{\textcolor{myorange}{subopt.\ fit}$^{\text{(e)}}$} & \multirow{2}{*}{\textcolor{mygreen}{\textbf{good fit}}} & \multirow{2}{*}{\textcolor{mygreen}{good fit}} \\[-2pt]
 & & & & & & \\[2pt]\hline
\end{tabular}
\tablecomments{\footnotesize The bold vertical line divides the models with relatively strong natal kicks and the models with weak natal kicks. The boldface labels correspond to our preferred models. We do not consider the impact of rocket kicks on systemic kicks, since they would only induce a relatively small scatter, but we list them here for completeness. (a) We can only reproduce the \textit{Gaia} NSs with low natal kicks if we use the two-stage CE formalism of \citet{Hirai_2022}. (b) The velocities of the \textit{Gaia} NSs on non-halo Galactic trajectories coincide with our predictions for systems that experienced a CE. (c) For models that predict systemic kicks significantly lower than the observed kick estimates, we note that dynamical heating is required to reconcile model and observation, unless we consider the difference sufficiently large to label the model a poor fit. (d) The model with low natal kicks can reproduce the low-eccentricity BeXBs; we propose that the orbits of the high-eccentricity BeXBs might be affected by non-zero pre-SN eccentricities. (e) We do not consider the predicted DNS systemic kicks for high natal kicks to be in significant tension with their observed velocities, but consider the fit suboptimal since one would expect the observed kick posteriors to slightly exceed the predicted ones.}
\end{table*}
\begin{table*}
\centering
\caption{Preferred models for the different systems considered in this work and the corresponding intrinsic and observed systemic kick distributions, described by the lognormal parameters $\mu$ and $\sigma$ as defined in Appendix \ref{appC}.\label{tab_distributions}}
\hspace{-29mm}\begin{tabular}{l|cccc|ccc|c|c}
\hline\hline\\[-13pt]
systems & \multicolumn{4}{c|}{intrinsic systemic kicks} & \multicolumn{3}{c|}{observed systemic kicks} & Fig. & sample\\
 & preferred & \multirow{2}{*}{$\mu$} & \multirow{2}{*}{$\sigma$} & median & \multirow{2}{*}{$\mu$} &\multirow{2}{*}{$\sigma$} & median & & reference \\[-2pt]
 & model & & & $\left[\text{km}\,\text{s}^{-1}\right]$ & & & $\left[\text{km}\,\text{s}^{-1}\right]$ & & \\[2pt]\hline\\[-13pt]
isolated NSs$^{\text{(a)}}$ & $\Mfid$ & $5.60$ & $0.68$ & $266$ & $5.60(12)$ & $0.68(10)$ & $266$ & \ref{Fig_Calibration} & (1)\\
\textit{Gaia} NSs$^{\text{(b)}}$ & $\Mroc$ & $4.60$ & $0.55$ & $88$ & $4.18(14)$ & $0.38(13)$ & $65$ & \ref{Fig_Gaia_Velocities} & (2)\\
LMXBs & $\Mfid$ & $4.76$ & $0.58$ & $117$ & $4.91(17)$ & $0.54(14)$ & $136$ & \ref{Fig_LMXB_Velocities} & (3)\\
NSWDs$^{\text{(c)}}$ & $\Mfid$ & $4.03$ & $0.69$ & $56$ & $4.22(14)$ & $0.51(13)$ & $68$ & \ref{Fig_NSWD_Velocities} & (4)\\
HMXBs & $\Mlow$ & -- & -- & -- & -- & -- & -- & \ref{Fig_BeXB_v-M} & (5,\,6)\\
DNSs & $\Mlow$ & $4.24$ & $1.02$ & $69$ & $4.33(26)$ & $0.48(24)$ & $76$ & \ref{Fig_DNS_Velocities} & (7,\,8)\\
[2pt]\hline
\end{tabular}
\tablecomments{\footnotesize The intrinsic fits are least-squares fits to the simulated distributions and the observed fits were made by maximizing the likelihood as described in Appendix \ref{appC}, where the values in parentheses are half the widths of the $68\%$ credible intervals. The median values are determined for the distributions normalized between $0$ and $1000$ km s$^{-1}$. For the (young) isolated NSs the intrinsic kick distribution is equal to the observed one, and we do not estimate the kick distributions of HMXBs but list them here for completeness. (a) See the analysis of \citet{Disberg_2025b}. (b) Considering simulated systems that experienced a CE and observed systems on non-halo Galactic trajectories, as discussed in Section \ref{sec3}. (c) The observed kick distribution is valid for MSPs with He WD companions, for CO WD companions the observed kicks can be described by a Maxwellian distribution with $\sigma=37(8)$\,km\,s$^{-1}$, as shown in Figure \ref{Fig_NSWD_Velocities}. References: (1) \citet{Disberg_2025b}; (2) \citet{El-Badry_2024b}; (3) \citet{ODoherty_2023}; (4) ATNF catalog \citep{Manchester_2005}; (5) \citet{Valli_2025}; (6) \citet{Fortin_2022b}; (7) \citet{Grichener_2026}; (8) \citet{Disberg_2024b}.}
\end{table*}

The alignment between our results for models with high natal kicks and the observed \textit{Gaia} NSs, LMXBs, and NSWDs strengthens our confidence in the kick distribution of \citet{Disberg_2025b} and its application to the \lstinline{COMPAS} binaries through the model of \citet{Mandel_2020}. The fact that our preferred models for HMXBs and DNSs contain significantly reduced natal kicks could be a sign that the NSs in these systems are formed through different mechanisms, such as USSNe, ECSNe, and AIC. These events are thought to be more symmetric and lead to significantly reduced natal kicks (see, e.g., \citealt{Pfahl_2002b,Podsiadlowski_2004,VandenHeuvel_2004,VandenHeuvel_2010,Tauris_2015,Mandel_2020,Willcox_2021,Stevenson_2022,Richardson_2023,Tauris_2023}, and references therein, or the review of \citealt{Popov_2025}), which might be compatible with our $\Mlow$ model with natal kicks of ${\lesssim}\,10$\,km\,s$^{-1}$. In particular, \citet{Muller_2018} simulate a USSN and indeed find a kick velocity of less than $10\,$km\,s$^{-1}$ \citep[see also, e.g.,][]{Suwa_2015}. However, \citet{Muller_2019} consider several USSN models and find that for some models the kicks can exceed $100\,$km\,s$^{-1}$, meaning it is difficult to constrain USSN kicks theoretically \citep[see also the discussion in][]{Grichener_2026}.

Since one might expect mechanisms such as USSNe, ECSNe, or AIC to produce reduced kicks, it is tempting to link our $\Mlow$ model to these mechanisms and $\Mfid$ to CCSNe that result in high kicks. However, in Figure \ref{Fig_Comparison} we show what fraction of the SNe in the progenitor systems considered in our analysis are labeled as either a CCSN, USSN, ECSN, or AIC by \lstinline{COMPAS}, and there is no clear link between our kick models and these labels. For instance, we find that the \textit{Gaia} NSs, LMXBs, and NSWDs can be explained by high natal kicks, but in \lstinline{COMPAS} the SNe in these systems are mainly USSNe. Likewise, we argue that at least part of the BeXB sample of \citet{Valli_2025} can be explained by small natal kicks, but \lstinline{COMPAS} labels the corresponding SNe as CCSNe. This means that, if one trusts the SN labels in the current \lstinline{COMPAS} model, the causal relationship between progenitor systems, SN mechanism, and kick magnitude is not easily constrained by the observed NS-harboring binaries.

The question remains what differences between the considered systems can explain the significant difference in natal kicks. One contrast between (1) the \textit{Gaia} NSs, LMXBs and NSWDs, and (2) the HMXBs and DNSs consists of the ZAMS masses of the companions, where the latter likely have more massive companions than the other systems. Even though the companion masses affect the binary evolution of the systems, it is not clear whether they affect the kick velocities of the primary in a way that can explain the difference between $\Mfid$ and $\Mlow$. There may be differences in the star formation process of an equal-mass high-mass binary versus an unequal-mass binary, for example in the stars' rotational profiles \citep[though see][]{Burrows_2026}. Moreover, \citet{Shariat_2026} make a similar (but observation-based) argument that some NSs obtain kicks of ${\lesssim}\,10\,$km\,s$^{-1}$, but they attribute these low kicks to different classes of systems (i.e.\ \textit{Gaia} NSs and low-eccentricity BeXBs), highlighting the model-dependence of inferring the natal kicks of NSs in binary systems.

We note that (1) if the natal kicks of some NSs in binary systems are indeed significantly reduced, these NSs are more likely to stay bound to their companions \citep[e.g.,][]{Igoshev_2021,Willcox_2021} so it is possible these NSs would not contribute to the population of isolated NSs that forms the basis for the kick distribution of \citet{Disberg_2025b}, and (2) the effects of observational selection biases affecting the orbital properties of radio pulsars in binary systems are likely relatively small \citep{Drummond_2026}. Moreover, there have been a significant number of NSs observed in globular clusters with typical escape velocities of a few tens of km\,s$^{-1}$ \citep[e.g.,][]{Heinke_2005}, which may be in tension with the typical kick velocities of isolated pulsars \citep[e.g.,][]{Katz_1975,Sigurdsson_1995,Drukier_1996,Davies_1998,Pfahl_2002,Sigurdsson_2003,Podsiadlowski_2004,Ivanova_2008,Willcox_2021,Ding_2025}, and our proposed models containing significantly reduced natal kicks for some classes of NSs in binary systems may be relevant for resolving this possible tension.

\subsection{Caveats}
\label{sec8.2}
\noindent Our analysis allows for constraining the kick magnitudes of NSs in binaries by considering the observed binary orbits, systemic velocities, and dependencies such as the DNS eccentricity-systemic kick relation shown in Figure \ref{Fig_DNS_e-v}. However, we find that it is difficult to constrain the presence of a rocket kick (i.e., $\Delta v_{\text{roc}}=30\,$km\,s$^{-1}$ in $\Mroc$ and $\Mlowroc$) or the kick direction (i.e., the polar kicks in $\Mpol$), since these have relatively small effects on most of the considered binaries. Although the rocket kicks are needed to reconcile the simulation with the observed \textit{Gaia} NSs \citep[cf.][]{Hirai_2024,Baibhav_2026}, this is difficult to verify in other systems (see also Table \ref{tab_results}). We do note that the low-eccentricity BeXBs are difficult to explain with rocket kicks, as shown in Figure \ref{Fig_BeXB_Orbits}, which could perhaps imply a correlation between natal kick and rocket kick magnitudes (e.g.\ due to the dependence of the rocket kick on the NS birth spin and magnetic field complexity), but might also mean that no NSs receive rocket kicks and (1) the post-SN \textit{Gaia} NS eccentricities are reduced by a different mechanism or (2) the binary evolution of the \textit{Gaia} NSs (e.g., their CE phase) is poorly described by \lstinline{COMPAS}.

Moreover, there are several caveats to the methodology we used to infer the systemic kicks of old NS binaries whose current velocities are impacted by motion through the Galactic potential \citep{Disberg_2024b,Disberg_2025a}. Firstly, it assumes a static, smooth Galactic potential, while the potential evolves over time \citep{Amend_2025} and includes structures such as giant molecular clouds \citep{Spitzer_1951,Spitzer_1953}. Secondly, we use the method of \citet{Disberg_2026} to correct our simulated systemic kicks for kinematic bias, which is caused by the fact that different kinds of systems are observable at different distances, but there are approximations in this method that add uncertainty to our results \citep[see the discussion in][]{Disberg_2026}. Lastly, for systems without radial velocity estimates we assume that their velocity vectors are isotropically distributed in their LSR, but if there is an alignment between the natal kick and the NS spin \citep[e.g.,][]{Noutsos_2013,Biryukov_2025} this can cause us to slightly overestimate the true velocities \citep{Mandel_2023}. Moreover, if such an alignment exists, a rocket kick will be aligned with the natal kick and thus non-negligibly increase the systemic kicks.

Additionally, we apply our kick models to pre-SN binaries obtained with the default \lstinline{COMPAS} model. This has the benefit that the pre-SN properties are informed by physical theory and were not fitted to match specific systems, adding confidence to the general picture regarding NS kicks that is constructed by our analysis. However, the \lstinline{COMPAS} model makes assumptions and approximations that increase the uncertainty of our results. Nevertheless, a partial exploration of the parameter space (e.g., exploring the impact of CE formalism and metallicity) suggests that our results are relatively robust. Moreover, a non-negligible fraction of the observed systems could be triple systems instead of binaries, which could affect the evolutionary history as well as measurements of system properties \citep[e.g.,][]{Ransom_2014,Burdge_2024,Freire_2026}.

We use the \citet{Mandel_2020} model to determine the NS mass and the corresponding kick magnitude, based on the pre-SN CO core mass. The non-monotonic relation between pre-SN mass and NS mass in this model can explain the apparent bimodality in DNS eccentricities \citep[see also][]{Grichener_2026}. However, there are alternative models for determining NS masses that might alter our results. For example, \citet{Schneider_2021} argue that NS progenitor masses might be divided into two separate ranges, with a \textquotedblleft black hole island\textquotedblright\ in between \citep[see also][]{OConnor_2011,Ugliano_2012,Ertl_2016,Muller_2016,Sukhbold_2016,Kresse_2021,Maltsev_2025}. This model might be able to explain apparent structures in the binary black hole mass distribution as observed through GWs \citep{Disberg_2023,Schneider_2023,Laplace_2025,Willcox_2025,Galaudage_2026}. Such a model may also provide a non-monotonic NS mass prescription that might explain the bimodal DNS eccentricities; our conclusions about NS kick magnitudes are likely relatively robust despite possible uncertainties in the \citet{Mandel_2020} model.

\section{Conclusions}
\label{sec9}
\noindent We have formulated NS kick models and applied them to simulated \lstinline{COMPAS} binaries. This allowed for comparison between the theoretical models and observations of NS-harboring binaries, where we considered their binary orbits (i.e., periods and eccentricities) and their kinematically constrained systemic kicks (Section \ref{sec2}). In particular, we compared our models to observational samples of \textit{Gaia} NSs (Section \ref{sec3}), LMXBs (Section \ref{sec4}), NSWDs (Section \ref{sec5}), HMXBs and in particular BeXBs (Section \ref{sec6}), and DNSs (Section \ref{sec7}). Although there are several caveats (Section \ref{sec8}), we conclude the following.
\begin{enumerate}
    \item We can relatively accurately reproduce the binary orbits and systemic kicks of the considered systems, if we apply high natal kicks, calibrated to the velocities of young isolated pulsars, to the progenitors of \textit{Gaia} NSs, LMXBs, and NSWDs, and significantly reduced natal kicks (i.e., ${\lesssim}\,10$\,km\,s$^{-1}$) to the progenitors of HMXBs and DNSs. The data are consistent with a unimodal kick distribution for isolated NSs and NSs that are formed with low-mass companions (i.e., \textit{Gaia} NSs, LMXBs, and NSWDs), which is significantly suppressed for NSs that are formed with high-mass companions (i.e., HMXBs, and DNSs).
    \item For the \textit{Gaia} NSs, we need to additionally apply rocket kicks of ${\sim}\,30\,$km\,s$^{-1}$ to match the observations by randomizing the post-SN eccentricities \citep[see also][]{Hirai_2024,Baibhav_2026}. However, most other systems are not sensitive to rocket kicks, preventing us from making decisive conclusions about the presence of a rocket kick. We note that the low-eccentricity BeXBs are difficult to explain with a rocket kick, which could mean that only NSs with a significant natal kick receive rocket kicks. In order to match the systemic kick estimates, we need to select the \textit{Gaia} NS progenitors that experienced a CE and omit the three observed systems that have a low metallicity and are on halo-like orbits \citep{El-Badry_2024b}.
    \item For the HMXBs, we cannot reproduce the high-eccentricity BeXBs, which show a tight period-eccentricity relation \citep{Valli_2025}. However, we note that observed systems that are kinematically comparable to BeXB progenitors, as listed by \citet{Valli_2025} and \citet{VanSon_2026}, show pre-SN eccentricities that are broadly compatible with this period-eccentricity relation. The non-zero pre-SN eccentricity is difficult to explain through isolated binary evolution. We therefore hypothesize that perhaps the scenario of a merger in a chaotic triple system can help explain the high-eccentricity BeXBs \citep[see, e.g.,][]{Hirai_2021}, where their pre-SN eccentricities were conserved in the absence of significant natal kicks. 
    \item For the DNSs, we find that combining the non-monotonic NS mass prescription of \citet{Mandel_2020} with small natal kicks results in a bimodal Blaauw kick distribution and therefore bimodal eccentricities \citep[as also found by][]{Grichener_2026}. This is supported by the observed DNS eccentricities. Moreover, we find that the Blaauw kicks are compatible with the DNS systemic kick estimates of \citet{Disberg_2024b}, where particularly the apparent correlation between eccentricity and systemic kick provides tentative evidence for Blaauw kicks being dominant over natal kicks. Thus, the observed Galactic DNSs can be explained relatively well without significant natal kicks.
\end{enumerate}
\noindent It is not straightforward to explain the reduced HMXB and DNS natal kicks based on binary evolution and SN physics. Although part of the answer to this question lies in the fact that SN mechanisms such as USSNe, ECSNe, and AIC are thought to have small natal kicks, it is difficult to reconcile this with the simulated \lstinline{COMPAS} binaries. Besides this, questions that remain concern aspects of our analysis such as (1) the low-metallicity \textit{Gaia} NSs, (2) the high-eccentricity BeXBs, (3) the high-velocity HMXBs, and (4) the DNS masses. These might, in turn, be at least partially explained by (1) a metal-poor dwarf galaxy that was accreted by the Milky Way \citep[e.g.,][]{SchiebelbeinZwack_2026}, (2) pre-SN eccentricity \citep[see also][]{VanSon_2026}, (3) dynamical ejections from clusters \citep[e.g.,][]{Fujii_2011,Oh_2016}, and (4) a different treatment of Case BB mass transfer \citep[see also][]{Grichener_2026}. The fact that these systems differ from our \lstinline{COMPAS} predictions is therefore not necessarily due to the natal kicks.

Apart from these remaining open questions, the observed NS-harboring binaries can be relatively well explained by a kick model in which NSs that are formed with low-mass companions (i.e., \textit{Gaia} NSs, LMXBs, and NSWDs) obtain kicks similar to young isolated NSs, and NSs that are formed with high-mass companions (i.e., HMXBs and DNSs) obtain significantly reduced natal kicks, providing a relatively consistent picture for the kick velocities of NSs in binary systems.

\newpage
\begin{acknowledgments}
\noindent We thank the referee for comments and suggestions that have helped improve this work. We are also thankful to Gijs Nelemans for insightful comments on the manuscript, and Matthew Bailes, Adam Burrows, Adam Deller, Aldana Grichener, Yuri Levin, Takashi Moriya, Bernhard Müller, and Tomer Shenar for useful discussions. We acknowledge support from the Australian Research Council (ARC) Centre of Excellence for Gravitational-Wave Discovery (OzGrav) through project number CE230100016. R.H. acknowledges support from the RIKEN Special Postdoctoral Researcher Program for junior scientists. The data from the \lstinline{COMPAS} simulation used in our analysis is publicly available at \href{https://gwlandscape.org.au/}{https://gwlandscape.org.au/}.This work has made use of data from the European Space Agency (ESA) mission \textit{Gaia} \citep[][\url{https://www.cosmos.esa.int/gaia}]{Gaia_2016,Gaia_2023}, processed by the \textit{Gaia} Data Processing and Analysis Consortium (DPAC, \url{https://www.cosmos.esa.int/web/gaia/dpac/consortium}). Funding for the DPAC has been provided by national institutions, in particular the institutions participating in the \textit{Gaia} Multilateral Agreement. We have also used pulsar data from the ATNF Pulsar Catalog \citep[][\url{https://www.atnf.csiro.au/research/pulsar/psrcat/}]{Manchester_2005}.
\end{acknowledgments}
\vspace{-4mm}
\noindent\software{\lstinline{ASTROPY} \citep{Astropy_2013,Astropy_2018,Astropy_2022}, \lstinline{COGSWORTH} \citep{Wagg_2025a,Wagg_2025b}, \lstinline{COMPAS} \citep{COMPAS_2022a,COMPAS_2022b,COMPAS_2025}, \lstinline{GALPY} \citep{Bovy_2015}, \lstinline{MATPLOTLIB} \citep{Hunter_2007}, \lstinline{NUMPY} \citep{Harris_2020}, \lstinline{SCIPY} \citep{Virtanen_2020}.}

\clearpage
\appendix
\section{Rocket-induced Mergers}
\label{appA}
\noindent In Section \ref{sec2.3} we describe how a rocket kick affects a post-SN binary system, through the formalism of \citet{Hirai_2024} in which the rocket kick induces oscillations in the eccentricity of the binary. However, in this formalism the eccentricities can theoretically reach values of $1$, after which they oscillate back to lower values. This is not physical, since an eccentricity of ${\sim}\,1$ implies a periastron distance of ${\sim}\,0$, which would result in a stellar collision/merger. We therefore determine the maximum rocket kick a binary can obtain before a merger is induced.

In order to determine the rocket kick that is necessary to reach a certain critical eccentricity, we define the function $f(\tilde{v}_{\text{roc}})=(e'/e)^2$, which is explicitly given in Equation \ref{eq_e_roc}, where $e$ and $e'$ are the pre- and post-rocket eccentricities, respectively. Through Equation \ref{eq_e_roc} we invert this function, resulting in 
\begin{equation}
    \label{eq_f_inv_roc}
    f_{\pm}^{\text{inv}}\left(e',n\right)=\dfrac{2}{3}\left[\tan^{-1}\left(\dfrac{A\pm\sqrt{A^2+B^2-\varepsilon^2}}{\varepsilon+B}\right)+\pi n\right],
\end{equation}
where $\varepsilon=\left(e'/e\right)^2+B-1$, $n$ is an integer, and the parameters $A$ and $B$ only depend on the rocket angles $\theta_r$ and $\phi_r$, as defined in Section \ref{sec2.3}. This function describes the value of $\tilde{v}_{\text{roc}}$ that results in a post-rocket eccentricity $e'$ for a given $e$, $\theta_r$, and $\phi_r$. However, since the oscillations are periodic, this function has either infinite values corresponding to different $n$ if $e'$ lies within the oscillation range, or has no (real) values if $e'$ lies outside the oscillation range (resulting in $A^2+B^2-\varepsilon^2<0$).

We are interested in determining the first time $f(\tilde{v}_{\text{roc}})$ reaches a certain value, which corresponds to $\tilde{v}_{\text{roc}}=f^{\text{inv}}_{-}(e',n)$ since $df(\tilde{v}_{\text{roc}})/d\tilde{v}_{\text{roc}}\geq0$ at $f^{\text{inv}}_-(e',n)$ and $df(\tilde{v}_{\text{roc}})/d\tilde{v}_{\text{roc}}\leq0$ at $f^{\text{inv}}_+(e',n)$. Moreover, if $n=0$ then $f^{\text{inv}}_{-}(e',0)$ describes the root of $f(\tilde{v}_{\text{roc}}) - e' = 0$ that is closest to $\tilde{v}_{\text{roc}}=0$, but this might occur for $\tilde{v}_{\text{roc}}<0$ which is not physical. In this case, the first intersection for non-zero $\tilde{v}_{\text{roc}}$ occurs at $n=1$. The critical rocket kick magnitude $\tilde{v}_{\text{roc},\,\text{crit}}$ for which a binary with initial eccentricity $e$ reaches a certain critical eccentricity $e'_{\text{crit}}$ therefore equals
\begin{equation}
    \label{eq_v_crit}
    \tilde{v}_{\text{roc},\,\text{crit}}=\left\{\begin{matrix}f^{\text{inv}}_-(e'_{\text{crit}},0)&\text{if }f_-^{\text{inv}}(e'_{\text{crit}},0)>0;\ e<e'_{\text{crit}}\hfill\\f^{\text{inv}}_-(e'_{\text{crit}},1)&\text{if }f_-^{\text{inv}}(e'_{\text{crit}},0)<0;\ e<e'_{\text{crit}}\hfill\\\infty&\text{if }\varepsilon^2>A^2+B^2;\ e<e'_{\text{crit}}\hfill\\0&\text{if }e\geq e'_{\text{crit}}\hfill\end{matrix}\right..
\end{equation}
This means that the post-rocket eccentricity for a given rocket kick $\tilde{v}_{\text{roc}}$ is given by $(e'/e)^2=f(\min(\tilde{v}_{\text{roc}};\tilde{v}_{\text{roc},\,\text{crit}}))$. Moreover, we model the critical eccentricity as the value for which the periastron distance of the binary orbit is less than the sum of the stellar radii, or $a(1-e'_{\text{crit}})\leq R_1+R_2$, since this results in a collision/merger. In this work we omit simulated systems in which a rocket kick makes the binary reach a critical eccentricity, but we note that this only occurs for a small fraction of systems.

\section{Isolated Neutron Stars}
\label{appB}
\renewcommand{\thefigure}{B}
\begin{figure}
    \centering
    \resizebox{\hsize}{!}{\includegraphics{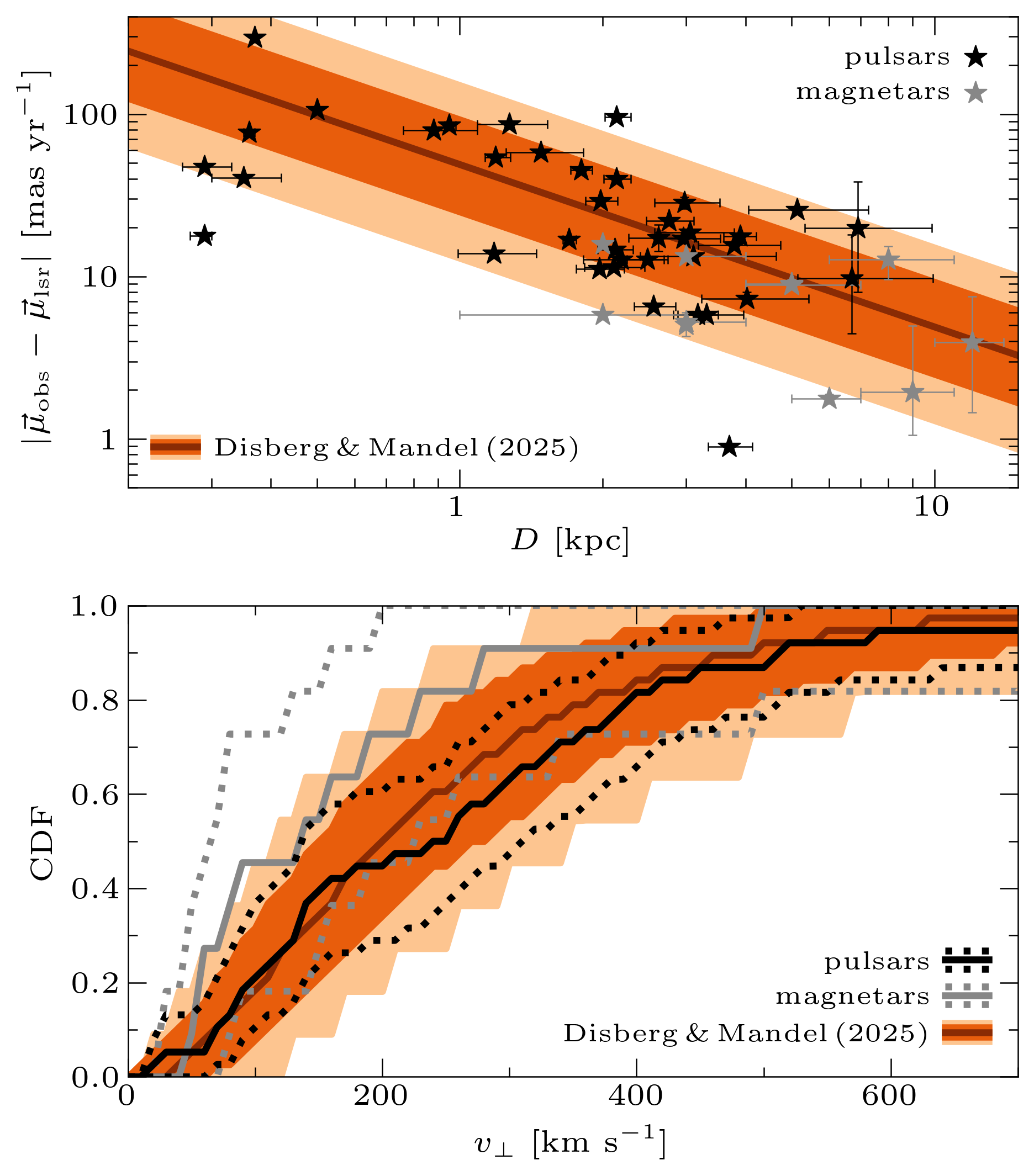}}
    \caption{Proper motions and transverse velocities of young pulsars (black) from the sample of \citet{Disberg_2025b} and the magnetars (grey) in the sample of \citet{Chrimes_2026}. The top panel shows the difference between the observed proper motion of an object and the proper motion of its LSR, as a function of distance. The error bars show the $68\%$ confidence interval, and the orange shaded regions correspond to the lognormal kick distribution of \citet{Disberg_2025b} projected on the sky assuming isotropy, where we show the median distribution, the $68\%$ confidence interval, and the $95\%$ confidence interval (dark orange line, orange shaded region, and light orange shaded region, respectively). The bottom panel shows the bootstrapped distributions of the transverse velocities of the pulsars and magnetars, including their median distributions (solid lines) and the $95\%$ confidence intervals (dotted lines). We also show the bootstrapped $95\%$ confidence regions that are sampled from the \citet{Disberg_2025b} distribution, with sample sizes equal the number of pulsars (orange shaded region) or the number of magnetars (light orange shaded region).}
    \label{Fig_INS}
\end{figure}
\noindent In Figure \ref{Fig_Calibration} we show our calibration of the \citet{Mandel_2020} natal kick prescription to the kick estimates of \citet{Disberg_2025b}, who determined the peculiar velocities of young isolated pulsars. The figure shows that the calibrated kicks align with the lognormal fit of \citet{Disberg_2025b}. In order to verify that this lognormal distribution is indeed a valid description of the natal kicks of isolated NSs, we compare it to the proper motions and transverse velocities of these objects. In particular, we consider their observed proper motion $\mu_{\text{obs}}$ and correct it for the proper motion of their LSR $\mu_{\text{lsr}}$, such that $|\vec{\mu}_{\text{obs}}-\vec{\mu}_{\text{lsr}}|$ equals their peculiar proper motion. We determine the proper motions of the pulsars in the sample of \citet{Disberg_2025b} that are younger than $10\,$Myr and have parallax estimates with a fractional uncertainty less than $25\%$, where we assume that their LSR is moving in a circular Galactic orbit with a velocity following the Milky Way potential of \citet{McMillan_2017}. The distances are estimated based on the parallax following the formalism of \citet{Verbunt_Cator_2017} and \citet{Verbunt_2017}. The peculiar proper motions, as a function of the distance $D$ between the object and the Sun, encode the kick velocities, and in order to quantify this better we also determine the peculiar transverse velocities $v_{\perp}=D |\vec{\mu}_{\text{obs}}-\vec{\mu}_{\text{lsr}}|$.

Moreover, we consider the proper motions of the magnetars in the sample of \citet{Chrimes_2026}. Magnetars are NSs with very high magnetic fields that decay over $\ll\,1\,$Myr, meaning their present-day peculiar velocities approximately equal their natal kicks. \citet{Chrimes_2026} argue that there is tentative evidence for a dearth of high-velocity magnetars compared to the natal kick distribution of \citet{Disberg_2025b}. We combine the proper motions and distance estimates listed by \citet{Chrimes_2026} with sky locations from literature \citep{Kaplan_2002,Israel_2003,Patel_2003,Gotthelf_2004,Camilo_2007,DeLuca_2009,Rea_2009,Israel_2016,Camilo_2018,Blumer_2020,Heyl_2024}, and determine their peculiar transverse velocities similarly to the young pulsars.

In Figure \ref{Fig_INS} we show the peculiar proper motions of the young pulsars as a function of their distance. We also show the median distribution and the $68\%$ and $95\%$ confidence intervals of the proper motions predicted by the kick distribution of \citet{Disberg_2025b}, which are determined by sampling from their lognormal distribution, projecting them on the sky assuming isotropy to give $v_{\perp}$, and converting them into (peculiar) proper motions through $\mu=v_{\perp}/D$. The proper motions of the pulsars and the lognormal model decrease for larger distances similarly, and the model encompasses most pulsars as well as all magnetars. There are two pulsars outside the $95\%$ confidence interval, but this is not statistically significant.

In the bottom panel of Figure \ref{Fig_INS} we show the bootstrapped confidence intervals of the peculiar transverse velocities $v_{\perp}$ for the pulsars and the magnetars. We also show the confidence intervals for the kick distribution of \citet{Disberg_2025b}, produced by taking samples from the lognormal model with sizes either equal to the number of pulsars or to the number of magnetars. The figure shows that the transverse velocities of the young pulsars are described well by the lognormal distribution. Moreover, although the median magnetar velocities are indeed slightly lower than the pulsar velocities, there is significant overlap in the uncertainty intervals and these velocities are also adequately described by the lognormal distribution. We find no statistically significant difference between the magnetar velocities of \citet{Chrimes_2026} and the natal kick distribution of \citet{Disberg_2025b}.

\section{Kinematic Constraints}
\label{appC}
\noindent In our analysis we employ the method of \citet{Disberg_2024b} to kinematically constrain the systemic kicks of dynamically old objects (as described in Section \ref{sec2.6}), and apply this method to our samples of \textit{Gaia} NSs, LMXBs, NSWDs, and DNSs. For \textit{Gaia} NSs the radial velocities are constrained, but for the other systems we estimate the radial velocity components by assuming the total systemic velocity vector to be isotropically distributed in the LSR of the system \citep[for more details see][]{Disberg_2024b,Disberg_2025a}. In order to estimate the transverse component of the systemic velocity vector we combine the observed proper motions with distance estimates. For distance estimates based on dispersion measure (DM) we assume an uncertainty of $20\%$ \citep{Ding_2024}, whereas for distance estimates based on parallax we use the formalism of \citet{Verbunt_Cator_2017} and \citet{Verbunt_2017}. Where available, we use parallax distances as opposed to DM distances since these are likely more accurate \citep[e.g.,][]{Deller_2009}.

We sample $100$ velocity vectors based on the transverse and radial velocity distributions, taking into account observational uncertainties. We use the sampled velocity vectors to trace back the Galactic trajectories of the object using \lstinline{GALPY}\footnote{\url{http://github.com/jobovy/galpy}} \citep{Bovy_2015} and the Milky Way potential of \citet{McMillan_2017}, and estimate their Galactic eccentricity. The posterior systemic kick distribution, then, follows from the Galactic eccentricity estimates and the relationship between Galactic eccentricity and systemic kick of \citet{Disberg_2025a}. For each Galactic eccentricity value, we sample one kick velocity from the corresponding posterior, resulting in $100$ kick estimates for each individual object.

This method makes two critical assumptions. The first assumption is that the pre-SN binary is on a circular Galactic trajectory at a Galactic height of $Z=0$\,kpc, where it receives an instantaneous kick which is responsible for its Galactic eccentricity. This assumption is valid because NS progenitors are massive stars which are born in the Galactic thin disc at heights of $Z\lesssim50$\,pc with velocity dispersions relative to their LSR of ${\lesssim}\,10$\,km\,s$^{-1}$ \citep{Glazebrook_2013}. For DNSs the assumption of only one instantaneous systemic kick does not hold, since two SNe are required for formation, but this is unlikely to affect the kick estimates due to small systemic velocities imparted by the first SN in a relatively wide binary, and the fact that the observed DNS peculiar velocities are relatively small \citep{Disberg_2024b,Gaspari_2024a}. This also ignores the possibility of formation in globular clusters (see, e.g., \citealt{Andrews_2019}, who suggested that some field DNSs were ejected from clusters, although \citealt{Ye_2019} argue that at least the merger rate of DNSs from globular clusters is relatively low). The second assumption is that after the kick the systemic velocity is only affected by migration through the (smooth) Galactic potential. However, dynamical heating due to gravitational interaction with structures such as giant molecular clouds or the Galactic bar can change the systemic velocity \citep{Spitzer_1951,Spitzer_1953}. Nevertheless, in the Galactic disc dynamical heating only increases the velocity dispersion of the oldest stars to ${\lesssim}\,50$\,km\,s$^{-1}$ \citep{Aumer_2016}.

In order to compare the kinematically constrained kick estimates to the model, we determine their confidence intervals through bootstrapping. If we have a sample of $N$ objects, we randomly draw $N$ objects from the sample (with replacement) and for each object randomly select one of their kick estimates. We repeat this $10^3$ times, and each time create a CDF of the resulting distribution. The resulting sample of $10^3$ CDFs allows for determining the median CDF and the $95\%$ confidence interval.

However, the radial velocity of objects for which these are not constrained (e.g., by \textit{Gaia}) is estimated by assuming isotropy and this causes the posterior kick distribution to be skewed towards higher velocities \citep[see appendix B in][]{Disberg_2025b}. These asymmetric kick posteriors will cause the bootstrapped intervals to overestimate the contribution of high velocities, to a certain degree. In order to correct for this, we fit a lognormal distribution to the kinematically constrained kick estimates, defined as
\renewcommand{\theequation}{C1}
\begin{equation}
    \label{eq_lognormal}
    p(v|\mu,\sigma)=\dfrac{1}{v\sigma\sqrt{2\pi}}\exp\left(-\dfrac{(\ln v-\mu)^2}{2\sigma^2}\right),
\end{equation}
where $v$ is given in km\,s$^{-1}$ and the parameters $\mu$ and $\sigma$ are dimensionless. If we have $N$ objects in a sample with $S$ kick estimates per object and the kick estimates are described by $\{v\}$ where $\{v\}_{n,i}$ is the $i$th kick estimate for the $n$th object, then we determine $\mu$ and $\sigma$ through the maximum value of their likelihood \citep[e.g.,][]{Mandel_2019}
\renewcommand{\theequation}{C2}
\begin{equation}
    \label{eq_likelihood}
    \mathcal{L}(\mu,\sigma|\{v\}\,)=\prod_{n=1}^{N}\dfrac{1}{S}\sum_{i=1}^{S}p(\{v\}_{n,i}|\mu,\sigma).
\end{equation}
This fit corrects for possible asymmetry in the individual kick estimates, which is why the lognormal model of \citet{Disberg_2025b} differs from the median of their bootstrapped uncertainty interval (as shown in Figure \ref{Fig_Calibration}). Throughout this work, we normalize the displayed kick distributions between $0$ and $1000$\,km\,s$^{-1}$.

\section{Eccentricity Conservation}
\label{appD}
\renewcommand{\thefigure}{D}
\begin{figure}
    \centering
    \resizebox{\hsize}{!}{\includegraphics{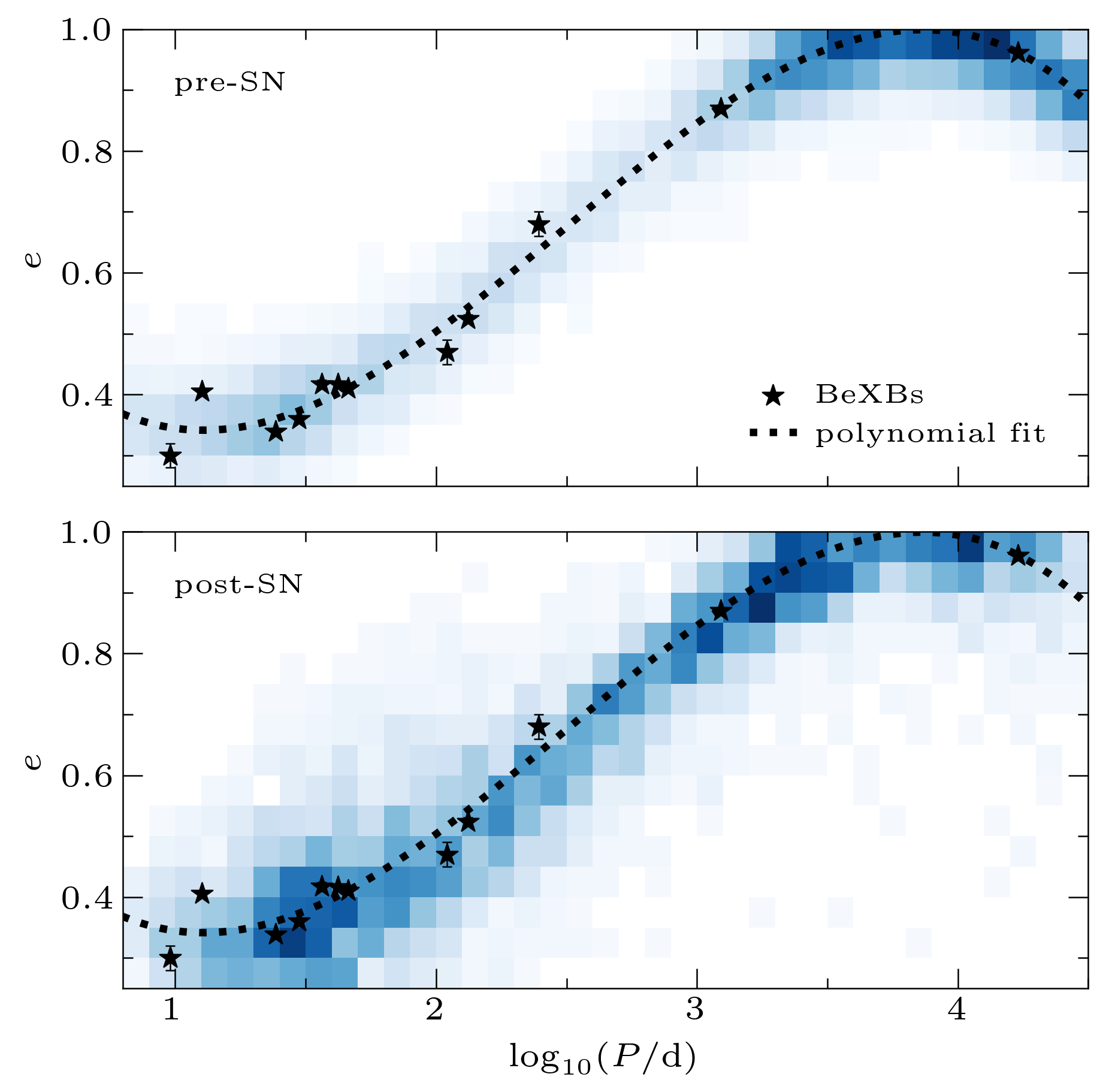}}
    \caption{Conservation of a BeXB period-eccentricity relation if the NS receives a small natal kick (following the $\Mlow$ model). We fit a third-degree polynomial to the high-eccentricity BeXBs (black dotted line), and resample the pre-SN eccentricity of the simulated BeXB progenitors from this polynomial with an added Gaussian scatter with a standard deviation of $0.05$. In the top panel the blue distribution corresponds to the pre-SN eccentricities, and in the bottom panel it shows the post-SN binaries after applying kicks through the $\Mlow$ model.}
    \label{Fig_BeXB_Test}
\end{figure}
\noindent In Section \ref{sec6} we discuss the binary orbits of the BeXBs in the sample of \citet{Valli_2025}, who find a tight correlation between period and eccentricity. After noting that pre-SN systems that may be kinematically comparable to BeXB progenitors might also be consistent with this relationship (see Figure \ref{Fig_BeXB_Progenitors}), we hypothesize that a merger-in-a-triple scenario may explain the period-eccentricity relation of high-eccentricity BeXBs. In order to show that such a correlation can survive an SN, we fit a third-degree polynomial to describe eccentricity as a function of period for these high-eccentricity systems, and change the pre-SN eccentricity of our simulated BeXB progenitors from zero to this polynomial fit (adding a Gaussian scatter of $0.05$). The polynomial is described by $e=\sum_{n{=}0}^3c_n\log_{10}(P/\text{d})^n$ where the fitted values of the coefficients $c_n$ are $(c_0,c_1,c_2,c_3)\simeq(0.7461,0.8070,0.4684,{-}0.0628)$. Then, we compute the post-SN binary orbits through the $\Mlow$ kick model. In Figure \ref{Fig_BeXB_Test} we show the pre- and post-SN periods and eccentricities, and the figure shows that a tight period-eccentricity relationship can survive post-SN, where mainly the Blaauw kicks add a relatively small scatter to the eccentricities. The fact that the companions are massive (${\geq}\,10M_{\odot}$) means that the orbits are relatively robust to changes due to Blaauw kicks. If there is a tight period-eccentricity relationship in progenitor systems, this can survive the SN and be detectable in BeXBs.

\section{Offsets}
\label{appE}
\renewcommand{\thefigure}{E}
\begin{figure}
    \centering
    \resizebox{\hsize}{!}{\includegraphics{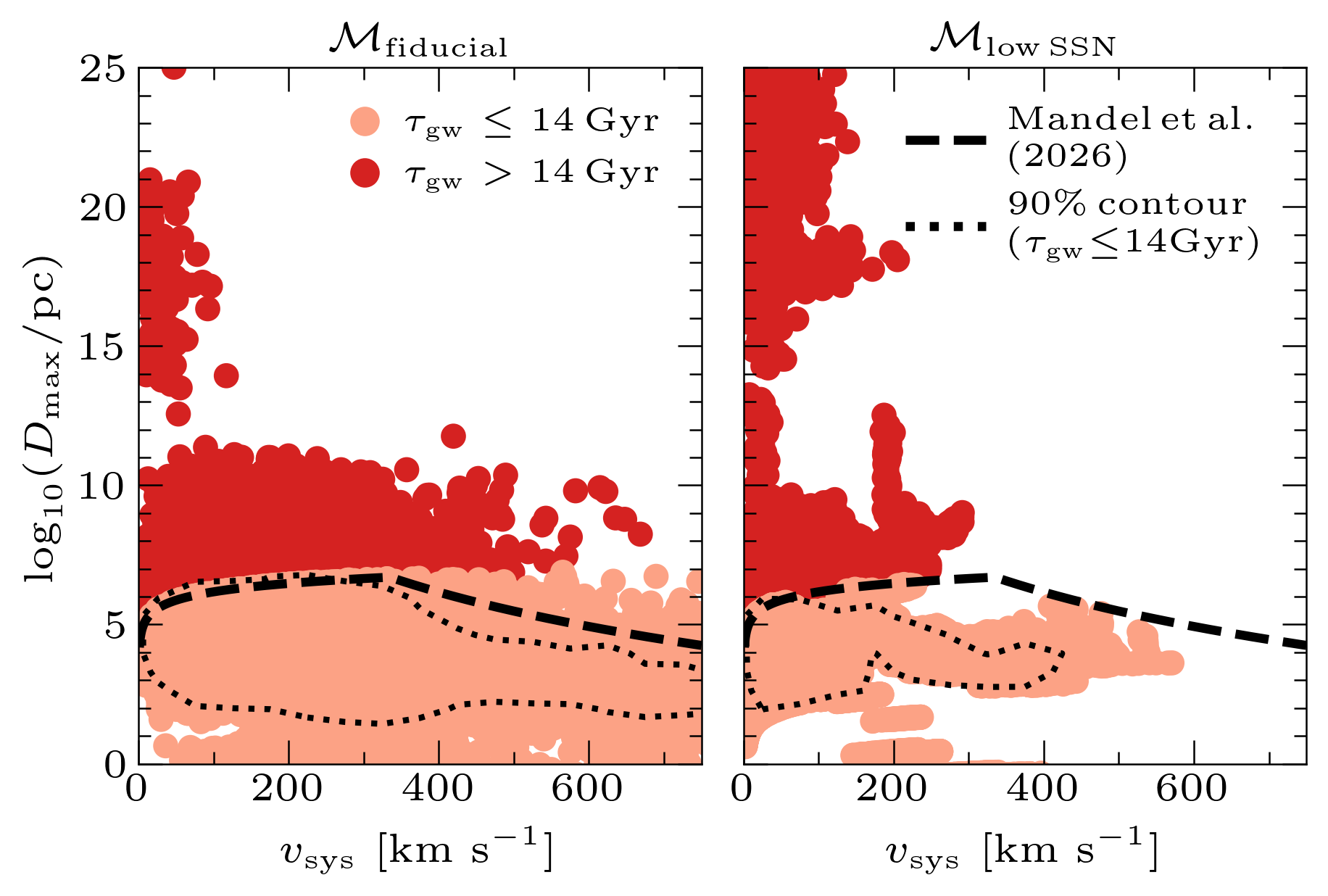}}
    \caption{Maximum offsets our simulated DNSs can obtain in the absence of a galactic potential before merging, determined through $D_{\max}=v_{\text{sys}}\tau_{\text{gw}}$ for $\tau_{\text{gw}}\leq14\,$Gyr (light red) and $\tau_{\text{gw}}>14\,$Gyr (red). The black dashed line corresponds to the upper limit estimate of \citet{Mandel_2026} as given by Equation \ref{eq_D_max}, and the dotted line shows the $90\%$ contour of the $\tau_{\text{gw}}\leq14\,$Gyr density distribution.}
    \label{Fig_DNS_Offsets}
\end{figure}
\noindent In Figure \ref{Fig_DNS_t-v} we show the merger times of the simulated DNSs in \lstinline{COMPAS} as a function of systemic kick, assuming reduced natal kicks as described by the $\Mlow$ model. This is relevant for DNS mergers, which may produce observable SGRBs, and the offsets from their host galaxies. \citet{Mandel_2026} argue that the DNSs that obtain large systemic kicks are unlikely to have long merger times, whereas low systemic kicks are less likely to be able to escape the gravitational potential of the host galaxy. Because of this, there is a limit on the offsets that DNSs can obtain. They describe the relationship between natal kick, systemic kick, and merger time, and estimate the maximum offsets of DNSs with merger times less than a Hubble time in absence of a galactic potential as
\renewcommand{\theequation}{E1}
\begin{equation}
    \label{eq_D_max}
    D_{\max}=5\,\text{Mpc}\cdot\min\left[\dfrac{v_{\text{sys}}}{335\,\text{km}\,\text{s}^{-1}},\left(\dfrac{v_{\text{sys}}}{335\,\text{km}\,\text{s}^{-1}}\right)^{-7}\right]
\end{equation}
In Figure \ref{Fig_DNS_Offsets} we show this upper limit estimate, together with the simulated offsets $D_{\max}=v_{\text{sys}}\tau_{\text{gw}}$ for the $\Mfid$ and $\Mlow$ models. The figure shows that the upper limit of \citet{Mandel_2026} is indeed a good description for high natal kicks, with ${>}\,90\%$ of the systems having a lower $D_{\max}$. However, for low natal kicks even more systems are consistent with the theoretical upper limit, with not a single simulated DNS that will merge within a Hubble time exceeding the estimate of \citet{Mandel_2026}.

\section{Systemic Kick Estimates}
\label{appF}
\noindent In our analysis, we consider the systemic kicks of several classes of NSs. In particular, we use kick estimates from the literature for LMXBs \citep{ODoherty_2023}, HMXBs \citep{Fortin_2022b}, and DNSs \citep{Disberg_2024b}. In Table \ref{tab_dns} we list the median and 68\% confidence intervals of the DNS systemic kick estimate posteriors of \citet{Disberg_2024b}. Moreover, we use the method of \citet{Disberg_2024b,Disberg_2025a} to estimate the systemic kicks of \textit{Gaia} NSs and NSWDs. We show the distributions of these estimates in Figures \ref{Fig_Gaia_Velocities} and \ref{Fig_NSWD_Velocities}, respectively, but in Table \ref{tab_gaia} we list the properties of the \textit{Gaia} NS sample and in Table \ref{tab_nswd} we list the properties of the NSWDs sample, together with our systemic kick estimates for the individual systems.

\begin{table*}
\centering
\caption{Systemic kick estimates of \citet{Disberg_2024b} for DNSs.\label{tab_dns}}
\hspace{-10mm}\begin{tabular}{l|c}
\hline\hline\\[-13pt]
system & $v_{\text{sys}}$\\
 & [km\,s$^{-1}$] \\\hline
J0737-3039 & $50^{+30}_{-20}$ \\
B1913+16 & $150^{+120}_{-60}$ \\
J1913+1102 & $130^{+110}_{-50}$ \\
J0509+3801 & $220^{+110}_{-60}$ \\
J1756-2251 & $80^{+100}_{-50}$ \\
B1534+12 & $150^{+40}_{-30}$ \\
J1829+2456 & $60^{+50}_{-20}$ \\
J1411+2551 & $80^{+80}_{-40}$ \\
J0453+1559 & $50^{+40}_{-20}$ \\
J1518+4904 & $50^{+30}_{-20}$ \\
J1930-1852 & $150^{+190}_{-100}$ \\\hline
\end{tabular}
\tablecomments{\footnotesize For the observed properties see \citet{Disberg_2024b} and \citet{Grichener_2026}.}
\end{table*}
\begin{table*}
\centering
\caption{Observed properties of the \textit{Gaia} NSs in the sample of \citet{El-Badry_2024b}, including the \textit{Gaia} IDs, right ascension (RA), declination (dec), proper motion in RA and dec ($\mu_{\alpha}$ and $\mu_{\delta}$, respectively), parallax ($\pi$), radial velocity ($v_r$), and systemic kick estimate ($v_{\text{sys}}$).\label{tab_gaia}}
\begin{tabular}{l|c|c|c|c|c|c|c}
\hline\hline\\[-13pt]
system & RA & dec & $\mu_{\alpha}$ & $\mu_{\delta}$ & $\pi$ & $v_r$ & $v_{\text{sys}}$\\
 & [deg] & [deg] & [mas\,yr$^{-1}$] & [mas\,yr$^{-1}$] & [mas] & [km\,s$^{-1}$] & [km\,s$^{-1}$] \\\hline
2995961897685517312 & 88.47(4) & -13.83(5) & 3.46(6) & 6.51(6) & 2.68(6) & 47.4(18) & $40^{+21}_{-10}$ \\
6481502062263141376 & 314.49(5) & -47.70(4) & -5.11(6) & 1.19(5) & 1.92(7) & -8(4) & $20^{+20}_{-0}$ \\
5820382041374661632 & 238.394(17) & -68.78(2) & -2.913(20) & -10.05(3) & 1.40(3) & -54(4) & $40^{+30}_{-10}$ \\
1871419337958702592 & 315.62(5) & 37.06(6) & 5.20(7) & 1.42(8) & 1.45(8) & 15(4) & $30^{+30}_{-0}$ \\
5530442371304583168 & 115.57(4) & -47.82(5) & -13.78(5) & 23.04(6) & 0.97(5) & 36(6) & $100^{+30}_{-20}$ \\
5136025521527939072 & 28.21(9) & -20.82(7) & 155.88(13) & -34.65(9) & 1.83(12) & 51(3) & $690^{+50}_{-261}$ \\
4922744974687373312 & 0.86(5) & -56.08(5) & 31.49(6) & -25.83(6) & 2.44(7) & 53(5) & $130^{+30}_{-20}$ \\
1434445448240677376 & 263.40(7) & 58.15(7) & -11.41(9) & 21.52(9) & 1.28(8) & -22(8) & $110^{+30}_{-20}$ \\
3494029910469026304 & 177.72(9) & -22.06(7) & -15.36(12) & -4.78(8) & 1.31(12) & 28(3) & $80^{+30}_{-20}$ \\
1694708646628402176 & 222.33(8) & 69.32(8) & -26.94(11) & 20.25(11) & 1.54(9) & -67(6) & $140^{+30}_{-30}$ \\
4637171465304969216 & 34.46(6) & -75.70(5) & 2.10(7) & -4.83(6) & 1.30(6) & 58(5) & $90^{+30}_{-20}$ \\
5580526947012630528 & 99.95(4) & -36.93(5) & -1.68(5) & -4.42(6) & 1.06(5) & -9(2) & $60^{+30}_{-10}$ \\
1350295047363872512 & 264.98(5) & 45.04(5) & 6.89(7) & 53.63(7) & 0.74(5) & -265(3) & $380^{+80}_{-101}$ \\
2426116249713980416 & 9.05(14) & -9.54(11) & -8.23(18) & -10.85(16) & 1.24(16) & 26(3) & $60^{+30}_{-20}$ \\
6328149636482597888 & 218.09(14) & -10.37(14) & -0.36(19) & -91.82(19) & 1.21(15) & 140(5) & $700^{+80}_{-260}$ \\
1058875159778407808 & 162.25(4) & 65.80(5) & -0.67(5) & -12.26(6) & 1.26(6) & 60(8) & $40^{+30}_{-10}$ \\
1801110822095134720 & 326.35(17) & 28.62(15) & 31.1(3) & -18.3(3) & 4.2(3) & -38(3) & $70^{+30}_{-10}$ \\
2397135910639986176 & 341.20(8) & -22.60(8) & 20.99(9) & -6.41(8) & 1.20(11) & 8(4) & $80^{+40}_{-10}$ \\
1028887114002082432 & 126.15(7) & 52.91(6) & 4.79(8) & 8.62(8) & 1.18(9) & -48(3) & $80^{+30}_{-20}$ \\
465093354131112960 & 37.53(6) & 59.85(7) & 12.51(7) & -15.43(8) & 2.61(9) & 4(2) & $40^{+30}_{-10}$ \\
1007185297091149824 & 98.69(3) & 62.93(3) & 5.43(4) & 2.60(3) & 0.74(4) & -50(8) & $60^{+40}_{-10}$ \\\hline
\end{tabular}
\tablecomments{\footnotesize The systems with $v_{\text{sys}}>300$\,km\,s$^{-1}$ correspond to the low-metallicity \textit{Gaia} NSs that are on halo-like Galactic trajectories. We note that the systemic kick estimates differ from the present-day peculiar velocities listed in table 4 of \citet{Shariat_2026}.}
\end{table*}
\begin{table*}
\centering
\caption{Observed properties of the NSWDs in our sample taken from the ATNF Pulsar Catalog v2.7.0 \citep{Manchester_2005}, including the system name, right ascension (RA), declination (dec), proper motion in RA and dec ($\mu_{\alpha}$ and $\mu_{\delta}$, respectively), parallax ($\pi$), and systemic kick estimate ($v_{\text{sys}}$).\label{tab_nswd}}
\begin{tabular}{l|c|c|c|c|c|c|c}
\hline\hline\\[-13pt]
system & WD & RA & dec & $\mu_{\alpha}$ & $\mu_{\delta}$ & $\pi$ & $v_{\text{sys}}$\\
 & & [deg] & [deg] & [mas\,yr$^{-1}$] & [mas\,yr$^{-1}$] & [mas] & [km\,s$^{-1}$] \\\hline
J0337+1715 & He & 54.4326079(5) & 17.2541189(6) & 4.8(5) & -4.4(5) & 0.7(3) & $70^{+50}_{-20}$ \\
J0437-4715 & He & 69.316872576(6) & -47.252783951(5) & 121.4429(5) & -71.4715(5) & 6.43(4) & $140^{+100}_{-60}$ \\
J0613-0200 & He & 93.433232924(9) & -2.01312150(2) & 1.837(2) & -10.359(6) & 1.00(5) & $95^{+45}_{-25}$ \\
J0614-3329 & He & 93.543117554(10) & -33.498368186(12) & 0.58(9) & -1.92(12) & 1.5(4) & $50^{+30}_{-10}$ \\
J0621+1002 & CO & 95.34214317(13) & 10.0440931(4) & 3.27(9) & -1.1(3) & 0.74(14) & $60^{+50}_{-20}$ \\
J0751+1807 & He & 117.78814720(2) & 18.12735717(14) & -2.700(10) & -13.27(7) & 0.85(4) & $120^{+60}_{-30}$ \\
J1012+5307 & He & 153.13932687(5) & 53.11729541(4) & 2.624(3) & -25.487(4) & 1.17(2) & $160^{+90}_{-40}$ \\
J1022+1001 & CO & 155.741665(3) & 10.031332(8) & -21.4(15) & -11(4) & 1.16(8) & $90^{+80}_{-50}$ \\
J1045-4509 & He & 161.45910336(4) & -45.16502896(4) & -6.09(2) & 5.18(3) & 1.7(7) & $20^{+30}_{-10}$ \\
J1125-6014 & He & 171.480177481(16) & -60.235224590(8) & 11.106(14) & -13.038(14) & 0.6(3) & $120^{+50}_{-30}$ \\
J1216-6410 & He & 184.03050144(3) & -64.169213051(12) & -7.87(8) & 2.57(8) & 1.1(5) & $20^{+40}_{-10}$ \\
J1400-1431 & He & 210.1541817(6) & -14.5297348(17) & 17(2) & -55(6) & 3.6(11) & $80^{+70}_{-40}$ \\
J1421-4409 & He & 215.33733404(3) & -44.15127166(4) & -11.6(4) & -7.9(8) & 1.2(4) & $160^{+140}_{-60}$ \\
J1455-3330 & He & 223.94990332(3) & -33.51289084(6) & 7.850(10) & -1.98(4) & 1.30(10) & $50^{+30}_{-10}$ \\
J1545-4550 & He & 236.483106078(16) & -45.84375766(2) & -0.48(2) & 2.37(5) & 0.45(14) & $70^{+40}_{-30}$ \\
J1600-3053 & He & 240.216262928(9) & -30.89705350(3) & -0.943(3) & -6.920(10) & 0.72(2) & $60^{+50}_{-20}$ \\
J1614-2230 & CO & 243.65211475(8) & -22.5086934(4) & 3.86(11) & -32.2(6) & 1.54(10) & $120^{+80}_{-40}$ \\
J1640+2224 & He & 250.069771123(16) & 22.40245113(2) & 2.102(4) & -11.333(7) & 0.73(6) & $80^{+30}_{-20}$ \\
J1643-1224 & He & 250.909009161(19) & -12.41629785(9) & 5.966(15) & 3.88(7) & 1.10(10) & $50^{+30}_{-10}$ \\
J1713+0747 & He & 258.4563907697(15) & 7.793744379(3) & 4.9215(8) & -3.920(2) & 0.880(10) & $30^{+30}_{-20}$ \\
J1738+0333 & He & 264.72485994(2) & 3.553018519(8) & 7.036(5) & 5.070(10) & 0.59(5) & $120^{+61}_{-30}$ \\
J1741+1351 & He & 265.379769744(7) & 13.862254413(10) & -8.98(2) & -7.42(2) & 0.60(10) & $120^{+110}_{-40}$ \\
J1751-2857 & He & 267.88619516(2) & -28.96293822(20) & -7.38(4) & -4.5(4) & 1.1(4) & $70^{+50}_{-20}$ \\
J1757-5322 & CO & 269.31314563(3) & -53.37405615(4) & -2.48(8) & -10.03(16) & 1.2(3) & $60^{+50}_{-20}$ \\
J1802-2124 & CO & 270.52222664(4) & -21.4010228(13) & -0.85(10) & 4.8(0) & 1.2(6) & $30^{+30}_{-10}$ \\
J1804-2717 & He & 271.08805446(4) & -27.2920381(8) & -0.00012(9) & -16.9(4) & 1.1(3) & $100^{+50}_{-20}$ \\
J1811-2405 & He & 272.83272651(3) & -24.0884540(11) & 0.53(6) & 0.0(0) & 0.72(15) & $10^{+10}_{-10}$ \\
J1853+1303 & He & 283.488825925(12) & 13.062235740(19) & -1.62(2) & -2.96(4) & 0.53(7) & $20^{+40}_{-10}$ \\
B1855+09 & He & 284.401626217(10) & 9.721442158(18) & -2.670(3) & -5.428(6) & 0.89(6) & $40^{+30}_{-10}$ \\
J1904+0553\_P & He & 286.07009513(17) & 5.8982257(2) & -5.5(6) & -6.2(17) & 4.5(17) & $120^{+120}_{-50}$ \\
J1909-3744 & He & 287.4476345132(12) & -37.737384372(3) & -9.5230(10) & -35.780(5) & 0.860(10) & $340^{+60}_{-80}$ \\
J1910+1256 & He & 287.540422896(17) & 12.94040385(3) & 0.24(2) & -7.10(3) & 0.36(6) & $70^{+90}_{-40}$ \\
J1911-1114 & He & 287.95534304(13) & -11.2395781(8) & -13.70(10) & -10.2(4) & 0.37(14) & $120^{+50}_{-40}$ \\
J1918-0642 & He & 289.700115424(10) & -6.70971197(3) & -7.124(5) & -5.96(2) & 0.75(7) & $80^{+50}_{-20}$ \\
J1933-6211 & CO & 293.385056816(17) & -62.196305232(13) & -5.620(10) & 11.09(3) & 0.7(2) & $110^{+60}_{-30}$ \\
J2017+0603 & He & 304.34460510(2) & 6.05154554(4) & 2.21(8) & 0.15(19) & 0.4(2) & $30^{+30}_{-10}$ \\
J2043+1711 & He & 310.837005635(6) & 17.191361107(8) & -5.703(10) & -10.841(16) & 0.72(7) & $80^{+40}_{-30}$ \\
J2145-0750 & CO & 326.46024803(2) & -7.83847452(6) & -9.49(2) & -9.09(6) & 1.40(8) & $60^{+30}_{-20}$ \\
J2222-0137 & CO & 335.52498722(5) & -1.62105004(13) & 44.707(5) & -5.403(12) & 3.723(13) & $60^{+40}_{-21}$ \\
J2234+0611 & He & 338.596137875(8) & 6.191301758(19) & 25.30(2) & 9.71(5) & 1.03(4) & $130^{+50}_{-20}$ \\
J2317+1439 & He & 349.28848492(2) & 14.65868362(4) & -1.305(14) & 3.45(3) & 0.60(8) & $70^{+50}_{-30}$ \\\hline
\end{tabular}
\end{table*}

\clearpage
\bibliography{references}{}
\bibliographystyle{aas_v7}

\end{document}